\documentclass[journal,10pt]{IEEEtran}
\usepackage{cite}
\usepackage{amsmath,amssymb,amsfonts}
\usepackage{algorithmic}
\usepackage{algorithm}
\usepackage{graphicx,color}
\usepackage{textcomp}
\usepackage{xcolor}
\usepackage{hyperref}
\usepackage{placeins}
\usepackage[caption=false]{subfig}
\usepackage{comment}
\usepackage{flushend}
\usepackage{comment}
\usepackage{cancel}

\usepackage{algorithm,algorithmic}
\def\BibTeX{{\rm B\kern-.05em{\sc i\kern-.025em b}\kern-.08em
    T\kern-.1667em\lower.7ex\hbox{E}\kern-.125emX}}
\AtBeginDocument{\definecolor{tmlcncolor}{cmyk}{0.93,0.59,0.15,0.02}\definecolor{NavyBlue}{RGB}{0,86,125}}

\begin{document}

\title{A Reinforcement Learning Framework for Resource Allocation in Uplink Carrier Aggregation in the Presence of Self Interference}

\author{Jaswanth Bodempudi$^{\dagger}$, Batta Siva Sairam$^{\ddagger}$, Madepalli Haritha$^{*}$, Sandesh Rao Mattu$^{**}$, and \\ Ananthanarayanan Chockalingam$^{*}$ 
\\ $^{\dagger}$ DRDO-Defence Electronics Research Laboratory, Hyderabad 500005, India
\\ $^{\ddagger}$ Qualcomm India Private Limited, Hyderabad 500081, India
\\ $^{*}$ Department of ECE, Indian Institute of Science, Bangalore 560012, India
\\ $^{**}$ Department of Electrical and Computer Engineering, Duke University, Durham, NC 27708, USA
}

\maketitle

\begin{abstract}
To meet the ever-increasing demand for higher data rates in mobile networks across generations, many novel schemes have been proposed in the standards. One such scheme is carrier aggregation (CA). Simply put, CA is a technique that allows mobile networks to combine multiple carriers to increase data rate and improve network efficiency. On the uplink, for power constrained users, this translates to the need for an efficient resource allocation scheme, where each user distributes its available power among its assigned uplink carriers. Choosing a good set of carriers and allocating appropriate power on the carriers is of paramount importance for good performance. Another factor that is critical to obtaining good performance is how well the degradation caused by the harmonic/intermodulation terms generated by the user's transmitter non-linearities is handled. Specifically, for example, if the carrier allocation is such that a harmonic of a user's uplink carrier falls on the downlink frequency of that user, it leads to a self coupling-induced sensitivity degradation of that user's downlink receiver. Considering these factors, in this paper, we model the uplink carrier aggregation problem as an optimal resource allocation problem with the associated constraints of non-linearities induced self interference (SI). This involves optimization over a discrete variable (which carriers need to be turned on) and a continuous variable (what power needs to be allocated on the selected carriers) in dynamic environments, a problem which is hard to solve using traditional methods owing to the mixed nature of the optimization variables and the additional need to consider the SI constraint in the problem. Therefore, in this paper, we adopt a reinforcement learning (RL) framework involving a compound-action actor-critic (CA2C) algorithm for the uplink carrier aggregation problem. We propose a novel reward function that is critical for enabling the proposed CA2C algorithm to efficiently handle SI. The CA2C algorithm along with the proposed reward function learns to assign and activate suitable carriers in an online fashion. Numerical results demonstrate that the proposed RL based scheme is able to achieve higher sum throughputs compared to naive schemes. The results also demonstrate that the proposed reward function allows the CA2C algorithm to adapt the optimization both in the presence and absence of SI. 
\end{abstract}


\maketitle

\section{Introduction}
\label{sec:sec1}
\IEEEPARstart{E}{ach} new generation of mobile communication system/standard pushes the limit on achievable data rate, spectral efficiency, and energy efficiency. A straightforward approach to achieve increased data rate is to increase the bandwidth proportionately. However, the available bandwidth (allotted spectrum) is limited and acquiring additional frequency bands has regulatory and cost considerations. Therefore, efficient techniques and schemes are required to achieve target data rates through judicious use of frequency resources. One such scheme is carrier aggregation (CA) \cite{CA_3GPP}. CA, as the name implies, combines multiple carriers to transmit data to/from users. In other words, instead of transmitting data on one  carrier, multiple carriers are used to transmit data simultaneously, and this offers increased data rates. CA can be applied on the downlink, i.e., base station (BS) to user equipment (UE), as well as on the uplink (UE to BS). Several works on CA reported in the literature consider CA on the downlink \cite{DL_CA_1},\cite{DL_CA_2},\cite{DL_CA_3},\cite{DL_CA_4}. CA involves the following two basic components. First, the activation of carriers, i.e., in addition to the assigned carrier called the primary component carrier (PCC), secondary component carriers (SCC) need to be activated for data transmission. Second, the power allocation strategy, i.e., the distribution of available transmit power among the PCC and activated SCCs. Since the UEs are often small hand-held devices, the transmit power constraints in uplink CA are relatively more critical compared to those in downlink CA. Therefore, optimal allocation of transmit power and frequency resources in uplink CA is crucial.  
 
The resource allocation problem in uplink CA can be posed as a constrained optimization problem, which can be broken down into two sub-problems. First, choosing the appropriate SCCs to be activated from the available SCCs. Second, choosing the transmit power on the PCC and SCCs, while meeting the maximum power constraints. The first sub-problem can be thought of as an optimization over a binary vector of length equal to the number of SCCs, where a `1' (or `0') in a location implies that the corresponding SCC is active (or inactive). The number of resource blocks (time-frequency slices) to be allotted to each user on each of the PCC and SCCs also needs to be considered in the optimization. Therefore, the first sub-problem is discrete in nature. The second sub-problem involves assigning transmit powers on the PCC and activated SCCs (obtained from first sub-problem) such that the sum of all the transmit powers is within the maximum allowed transmit power. This sub-problem is continuous in nature. Therefore, the uplink CA problem is a joint optimization problem over discrete and continuous variables cast in a dynamic environment, which is a difficult problem to solve using conventional methods \cite{MINLP},\cite{MINLP_1},\cite{opt_1},\cite{opt_2}.

Machine learning (ML) based methods are being increasingly adopted in wireless communications \cite{ce1}, \cite{dd1}, \cite{ra1}, thanks to the major advancements in ML hardware and software. These methods find use in situations where conventional methods are hard to develop or their performance fall short. More recently, reinforcement learning (RL) algorithms \cite{ref3} are being developed and used for various tasks in wireless research \cite{ref1}, \cite{ref2}, \cite{stat_1}, \cite{stat_2}. RL based algorithms are especially helpful when obtaining closed-form expressions for the optimization problem is tedious or is not feasible. These algorithms typically involve an agent which takes an action that affects the environment, and the consequence of the action is fed back to the agent using a reward/regret strategy, through which the algorithm learns in an ``online'' fashion. This self-learning ability of the RL algorithms has led to its widespread adoption.

Some solution approaches to uplink CA problem have been proposed in the literature \cite{UL_CA_1}, \cite{UL_CA_2}, \cite{stat_1}, \cite{ref1}. The work in \cite{UL_CA_1} proposes an optimal joint component carrier (CC) selection and resource allocation to maximize average throughput of users with quality of service (QoS) constraint in terms of delay, using a integer linear programming approach. The optimization problem is relaxed as an equivalent linear programming form with guaranteed binary solution. The work in \cite{UL_CA_2} considers resource allocation in the context of LTE-advanced systems with CA and dual-cluster scheduling. Based on a derived threshold, a UE is configured either with multiple CCs and/or dual-cluster scheduling. While the earlier approaches were conventional, the works in \cite{stat_1} and \cite{ref1} consider RL based approaches. The work in \cite{stat_1} proposes a joint power-sharing and carrier aggregation (JPSCA) algorithm to simultaneously optimize SCC activation and uplink power levels. The work in \cite{ref1} proposes a centralized optimization approach where the aim is to simultaneously minimize average delay and power consumption. 

An important issue that has not been considered in CA optimization problems reported in the literature so far is the effect of non-linearity of the transmitter power amplifier (PA). When multiple aggregated carriers are fed at the input of the PA, undesired harmonic and intermod terms are generated at the output of the non-linear PA \cite{book1},\cite{book2}. This is a cause for concern because it is possible that, say, a harmonic of one of the carriers at the output on the uplink (UL) coincides with the frequency of the downlink (DL) carrier. This self-interference (SI) due to such harmonic/intermod output causes degradation in the sensitivity of the UE receiver. The solution approaches in the literature described above do not consider this degrading effect of SI in their optimization framework.

In this paper, because of the mixed (continuous and discrete) nature of the optimization variables involved, we take an RL based approach involving a compound-action actor-critic (CA2C) algorithm \cite{stat_2} for the uplink CA problem with an objective to maximize the sum throughput of the network. There are two approaches to incorporate the CA2C algorithm. The first approach is to use a centralized framework as in \cite{ref1, stat_1}. The second approach is a decentralized approach wherein the CA2C algorithm is moved to edge nodes in close proximity to the UEs \cite{xu2023edge,edge2,edge3}. In this paper, we use the centralized approach. Decentralized approach can be investigated as future work. The proposed RL framework factors in the role played by the non-linearity induced SI at the UE in the optimization framework while choosing the SCCs, which is a new contribution compared to past works. A naive and straightforward approach to address the SI issue is to simply avoid those combinations of the UL and DL carriers in the allocation that will cause the SI. We call this as ``hard avoidance (HA)'' approach. Though simple, a drawback with this approach is the loss in throughput due to inefficient use of the available carrier resources. So, instead of putting such a hard restriction on the SI-inducing UL-DL carrier combinations, we allow these combinations and let the RL framework to optimize the UE transmit power allocation in such a way to maximize the sum throughput while keeping the receiver sensitivity degradation within acceptable levels. We call this proposed approach as ``soft avoidance (SA)'' approach. To realize such an RL framework, we come up with a novel reward function that strongly penalizes high levels of SI, and goes easy on the penalty when SI levels are low/moderate. Using this reward function, the CA2C algorithm learns to assign and activate suitable SCCs in an online fashion. Simulation results demonstrate that the proposed RL based framework is able to achieve higher sum throughputs compared to naive schemes. The results also show that the proposed reward function allows the CA2C algorithm to adapt the optimization both in the presence and absence of SI.   
\begin{figure}
\hspace{5mm}
\includegraphics[width=8.5cm,height=5.5cm]{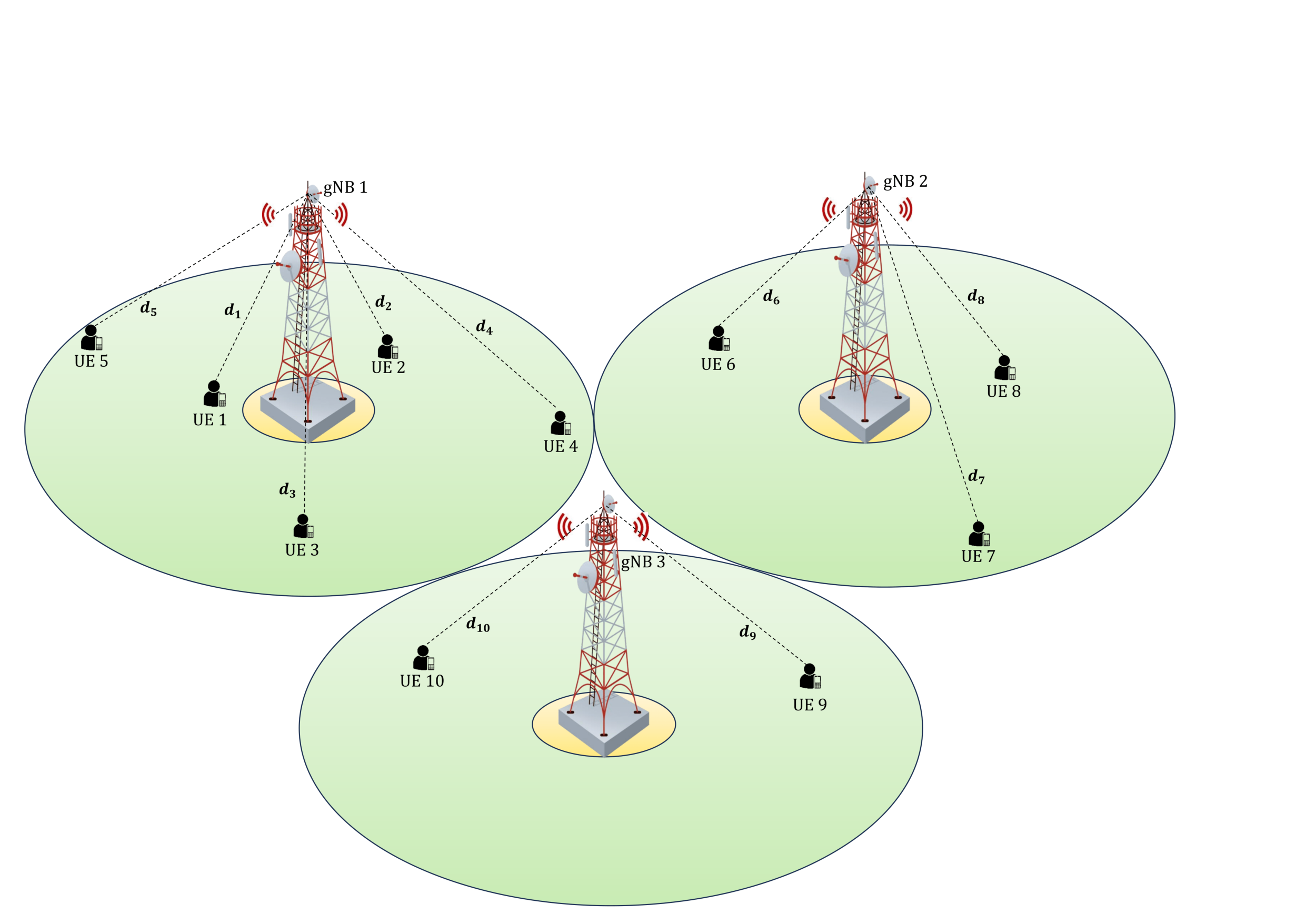}
\vspace{0cm}
\caption{Network environment.}
\label{network}
\vspace{-4mm}
\end{figure}

The rest of the paper is organized as follows. The network model is presented in Sec. \ref{sec:sec2}. The problem formulation and the rationale for adopting an RL approach is discussed in Sec. \ref{sec:sec3}. The RL approach using CA2C algorithm is presented in Sec. \ref{sec:sec3a}. The proposed RL framework is presented in Sec. \ref{sec:sec4}. Results and discussions are presented in Sec. \ref{sec:sec5}. Conclusions are presented in Sec. \ref{sec:sec6}.

\section{Network Model} 
\label{sec:sec2}
In this section, we introduce the network model including the network scenario and notations, the throughput model for the UE, and the SI model due to UE PA non-linearity.

\begin{table}
\centering
\begin{tabular}{|c|l|}
\hline
\textbf{Notation}        & \textbf{Description}         \\ \hline \hline
$N_\text{g}$                    & Number of gNBs in the network \\ \hline
$\mathcal{B}$            & Set of gNBs   \\ \hline
$M$                      & Number of CCs in the network \\ \hline
$\mathcal{M}$            & Set of CCs   \\ \hline
$N_j$                    & Number of RBs in CC $j$   
 \\ \hline
$\mathcal{N}_j$          & Set of RBs in CC $j$        \\ \hline
$K$                      & Number of UEs in the network \\ \hline
$\mathcal{K}$            & Set of UEs in the network   \\ \hline
$K_\text{b}$                    & Number of UEs associated to $\text{gNB}_\text{b}$ \\ \hline
$\mathcal{K}_\text{b}$          & Set of UEs associated to $\text{gNB}_\text{b}$  \\ \hline
$\mathcal{C}_{i}$        & Set of CCs assigned to UE $i$  \\ \hline
$\mathcal{N}_\text{RB}^{i,j}$ & Set of RBs assigned to UE $i$ in CC $j$  \\ \hline
$\text{b}(i)$                   & Serving gNB for UE $i$     
 \\ \hline
$h^{\text{b}(i),i}_{j,k}$       & Path loss between UE $i$ and $\text{gNB}$ ${\text{b}(i)}$ \vspace{-1mm}  \\ 
& over RB $k$ in CC $j$ \\ \hline
$p^{(i,j,k)}$            & Allocated UL Tx power level on RB $k$  \vspace{-1mm} \\ 
& of CC $j$ in UE $i$ \\ \hline
$R^{(i,j,k)}$            & Achieved UL rate on RB $k$ of CC $j$ in UE $i$ \\ \hline
$p^i$                    & Total allocated UL Tx power level for UE $i$ \\ \hline
$R^i$                    & Total achieved UL rate of UE $i$ \\ \hline
$L^i_{\mathrm{c}}$                    & Coupling loss between the Tx and Rx of UE $i$ \\ \hline
$p^i_\text{SI}$                    & Self interference power at the input of UE $i$ Rx           \\ \hline
$D^i$                    & Delay of UE $i$ \\ \hline
$D_\text{QoS}$           & Delay requirement for QoS   \\ \hline
$\hat{q}^i$              & Average number of bits per burst of data for UE $i$  \\ \hline
$p_\text{max}$           & Maximum transmit power level of UEs \\ \hline
$\alpha^i_j \in \{0,1\}$  & CC activation indicator for CC $j$ in UE $i$ \\ \hline
$\beta^i_{j,k} \in \{0,1\}$          &  RB allocation indicator for RB $k$ of CC $j$ \vspace{-1mm} \\ 
& in UE $i$ \\ \hline
\end{tabular}
\vspace{0.3cm}
\caption{Table of network parameters and notations.}
\label{notations}
\vspace{-8mm}
\end{table}

\subsection{Network scenario and notations}
\label{subsec:net_scen}
The network consists of $N_\text{g}$ next-generation node Bs (gNBs), denoted by the set $\mathcal{B}=\{\text{gNB}_1,\cdots, \text{gNB}_{N_\text{g}}\}$. We consider that CA is allowed on the UL transmission at all the gNBs. The set of non-overlapping and orthogonal CCs allotted to each gNB is denoted by $\mathcal{M} = \left\{1, \cdots, M \right\}$. A resource block (RB) is the smallest unit of time-frequency resource that can be allocated to a UE. Each CC consists of several RBs. Let $N_j$ denote the number of RBs in CC $j$. The set of RBs in CC $j$ is denoted by $\mathcal{N}_j$, where $\mathcal{N}_j = \left\{1, \cdots, N_j\right\}$. Let $\mathcal{C}_i$ denote the set of CCs allocated to UE $i$. $\mathcal{C}_i$ consists of both PCC and SCCs for UE $i$. The PCC is the main carrier assigned to the UE, and it will always be activated for all the UEs in the network. Meanwhile, an SCC can be activated/deactivated at any time to boost the achievable data rate. The UEs associated with different gNBs may reuse the same RBs when they are allocated the same CCs, and hence they would interfere with each other through the assigned RBs.  Let $K$ denote the number of UEs distributed in the network area and $\mathcal{K} = \left\{1,\cdots,K\right\}$ denote the set of these ${K}$ UEs. The serving gNB for UE $i$ is denoted by $\text{b}(i)$. Now, let $K_\text{b}$ denote the number of UEs in the coverage area of $\text{gNB}_\text{b}$ and $\mathcal{K}_\text{b} = \left\{1,\cdots, K_\text{b}\right\}$ denote the set of these $K_\text{b}$ UEs. Let $h^{\text{b}(i),i}_{j,k}$ denote the path loss between UE $i$ and $\text{gNB}$ ${\text{b}(i)}$ over RB $k$ in CC $j$. Let $\alpha^i_j \in \{0,1\}$ denote the activation indicator for CC $j$ in UE $i$ (a `1' indicates that the CC is activated and a `0' indicates that the CC is not activated). Likewise, let $\beta^i_{j,k} \in \{0,1\}$ denote the allotment indicator for RB $k$ of CC $j$ in UE $i$ (a `1' indicates RB $k$ of CC $j$ is allocated to UE $i$). The variable $p^{(i,j,k)}$ denotes the allocated UL transmit power on RB $k$ of CC $j$ in UE $i$ and $R^{(i,j,k)}$ denotes the achieved rate on RB $k$ of CC $j$ in UE $i$. The variables $p^i$ and $R^i$ denote the total UL transmit power level and total achieved rate, respectively, of UE $i$, and $p_{\mathrm{max}}$ denotes the maximum transmit power level of all the UEs. The parameter $L^i_\text{c}$ denotes the coupling loss between the transmitter and receiver of UE $i$. The variable $p^i_\text{SI}$ denotes the power level of the self interference at the input of the UE $i$, where the SI is caused due to the UE transmit PA non-linearity. The SI model is presented in Sec. II \ref{subsec_SI_model}. {\em The problem of interest here is to obtain the CC and RB allocation vectors \big($\alpha^i_j$s and $\beta^i_{j,k}$s\big) and transmit powers \big($p^{(i,j,k)}$s\big) of the UEs that maximize the sum throughput \big(sum of $R^i$s\big), under maximum transmit power \big($p_\text{max}$\big) and self interference \big($p^i_\text{SI}$s\big) constraints}. The delay variable $D^i$ (delay of UE $i$) and $D_{\mathrm{QoS}}$ for all UEs (delay QoS requirement) are used to define the state space in the RL framework\footnote{Latency/delay requirements for information delivery are common in communication networks (e.g., 150 msec latency for packet voice). The more the throughput/rate, the lower can be the delay. In order for the RL framework to maximize the throughput, we use meeting (or otherwise) a delay QoS as an indicator to define the system state, based on which the RL agent takes action to allocate resources towards maximizing the throughput.},
as will be discussed later in Sec. \ref{sec:sec4}. For quick and easy reference, the above notations are summarized in Table \ref{notations}. Figure \ref{network} shows an illustration of the considered network environment.

\subsection{UE throughput model}
As mentioned earlier, the objective is to maximize the sum throughput across the UEs. To enable this, we introduce the model for the UE throughput in this subsection. Following the notations introduced in the previous subsection, the total transmit power for UE $i$ is given by
\begin{equation}
p^i = \sum_{j\in\mathcal{M}} \alpha^i_j \sum_{k \in \mathcal{N}_j} \beta^{i}_{j,k} \;p^{(i,j,k)},
\label{powermodel}
\end{equation}
where $\alpha^i_j \in \{0, 1\}$
is the activation indicator for CC $j$ in UE $i$, 
$\beta^i_{j,k} \in \{0, 1\}$ is the allocation indicator for RB $k$ of CC $j \:\in \:\mathcal{M}$ in UE $i$, and $p^{(i,j,k)}$ is the allocated power on RB $k$ of CC $j$ in UE $i$. The total UL transmit power for UE $i$ is bounded by 
\begin{equation}
p^i \leq p_\text{max},
\label{maxpower}
\end{equation}
where $p_\text{max}$ is the maximum allowed UL transmit power for each UE. Now,
let $\gamma^i_{j,k}$ denote the signal-to-interference plus noise (SINR) ratio on RB $k$ of CC $j$ in UE $i$. Then, $\gamma^i_{j,k}$ is given by 
\begin{equation}
\gamma^i_{j,k} = \frac{p^{(i,j,k)}{h^{\text{b}(i),i}_{j,k}}}{\underset{i \in \mathcal{K}, i^{\prime} \neq i}{\sum} \beta^{i'}_{j,k} p^{(i',j,k)}{h^{\text{b}(i),i'}_{j,k}}+\sigma^2},
\label{SINR}
\end{equation}
where $\sigma^2$ is the variance of additive white Gaussian noise (AWGN). By Shannon's capacity theorem, the achievable rate for UE $i$ on RB $k$ in CC $j$ is given by
\begin{equation}
R^{(i,j,k)} = B_{j,k} \text{log}_{2}(1+\gamma^i_{j,k}),
\label{capacity}
\end{equation} 
where $B_{j,k}$ is the bandwidth of RB $k$ in CC $j$. Using \eqref{capacity}, the total throughput for UE $i$ can be written as 
 \begin{equation}
R^i = \sum_{j\in\mathcal{M}} \alpha^i_j \sum_{k \in \mathcal{N}_j} \beta^{i}_{j,k} \;R^{(i,j,k)}.
\label{ratemodel}
\end{equation}

\subsection{Non-linearity induced SI model}
\label{subsec_SI_model}
\begin{figure}
\centering
\includegraphics[width=7cm,height=4cm]{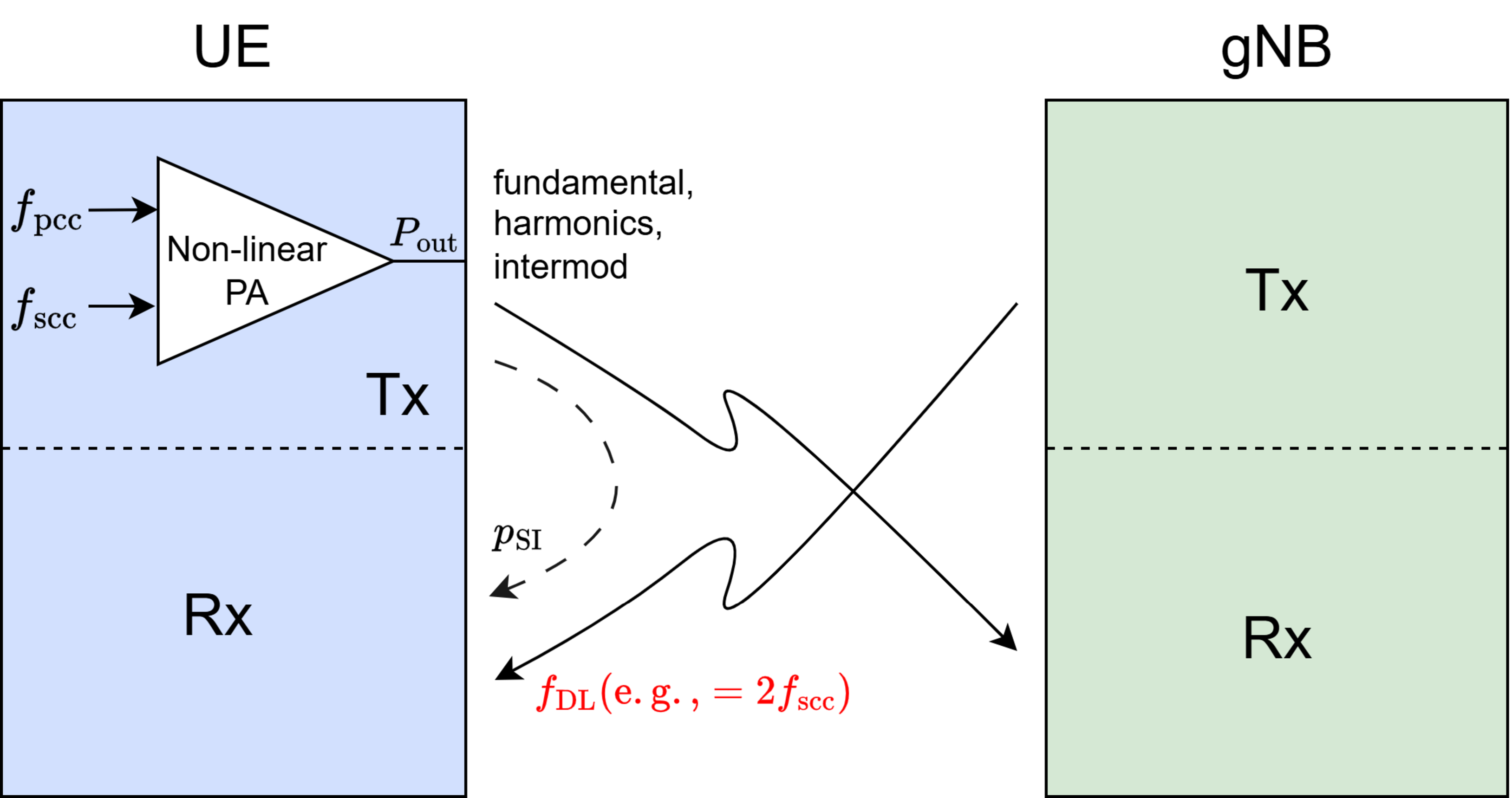}
\caption{SI due to PA non-linearity at UE.}
\label{Tx_Rx_SI}
\end{figure}

CA brings with it the issue of SI caused due to PA non-linearity. When a transmitter's non-linear PA is fed with multiple carriers at the input, the output will have undesired harmonic and intermod (IM) terms, which will couple with its own receiver. This will cause a loss in the sensitivity of the receiver if any of these harmonic/IM frequencies coincide with the receive frequency. This is illustrated in Fig. \ref{Tx_Rx_SI}, where the 2nd harmonic of one of the the UE's UL carrier (say, $f_\text{scc}$) coincides with the UE's receive frequency on the downlink (i.e., $f_\text{DL}=2f_\text{scc}$). A suitable model for the SI power at the receiver input is required in order to account for it in the optimization framework. Here, we present the SI model when the PA is fed with two carriers at the input.

Let the input signal $x(t)$ to the PA consist of two tones of frequencies ($\omega_1$,$\omega_2$) and amplitudes ($A_1$,$A_2$), respectively, i.e., $x(t)$ is given by
\begin{equation}
x(t) = A_1\cos\omega_1 t + A_2\cos\omega_2 t. 
\label{eqn:0}
\end{equation}
Considering up to third order non-linearity of the PA, the PA output $y(t)$ is given by \cite{book1},\cite{book2}
\begin{equation}
y(t)=c_1 x(t)+c_2 {x^2}{(t)}-c_3 {x^3}{(t)},
\label{eqn:1}
\end{equation}
where $c_1$, $c_2$, and $c_3$ are positive real coefficients in the non-linearity model. Substituting (\ref{eqn:0}) in (\ref{eqn:1}) and expanding yields the output frequencies and corresponding amplitudes as shown in Table \ref{tab2}. While the fundamental frequencies $\omega_1$ and $\omega_2$ are desired terms at the output, other harmonic frequencies ($2\omega_1$, $2\omega_2$, $3\omega_1$, $3\omega_2$) and IM frequencies ($\omega_1\pm \omega_2$, $2\omega_1\pm \omega_2$, $2\omega_2\pm \omega_1$) are undesired terms at the output. As an example, consider the 2nd harmonic output $2\omega_2$. The amplitude of this 2nd harmonic term is given by $\frac{1}{2}c_2A_2^2$, and the corresponding output power is given by $\frac{c_2^2A_2^4}{8}$. This implies that if the input power of fundamental $\omega_2$ is increased (or decreased) by 1 dB, the corresponding 2nd harmonic power at the output will increase (or decrease) by 2 dB. Likewise, it can be seen that the 3rd harmonic output will increase (or decrease) by 3 dB if the corresponding fundamental input power is increased (or decreased) by 1 dB. Let $G$ denote the linear power gain of the PA, which is defined as the ratio of the output power of the fundamental to the input power of the fundamental in the linear region, i.e., $c_1=\sqrt{G}$.

\begin{table}
\centering
\begin{tabular}{|c|l|}
\hline
\textbf{Frequency}  & \textbf{Amplitude} \\ \hline
\hline
DC                  & $\frac{1}{2}c_2(A_1^2+A_2^2)$ \\ \hline
$\omega_1$          & $c_1A_1 - \frac{3}{2}c_3\big(\frac{A_1^3}{2}+A_1A_2^2\big)$  
\\ \hline
$\omega_2$          & $c_1A_2 - \frac{3}{2}c_3\big(\frac{A_2^3}{2}+A_1^2A_2\big)$  
\\ \hline
$2\omega_{n=1,2}$   & $\frac{1}{2}c_2A_n^2 \ \ \text{for} \ n=1,2$ \\ \hline
$3\omega_{n=1,2}$   & $-\frac{1}{4}c_3A_n^3 \ \ \text{for} \ n=1,2$ \\ \hline
$\omega_1\pm \omega_2$   & $c_2A_1A_2$ \\ \hline
$2\omega_1\pm \omega_2$   & $-\frac{3}{4}c_3A_1^2A_2$ \\ \hline
$2\omega_2\pm \omega_1$   & $-\frac{3}{4}c_3A_1A_2^2$ \\ \hline
\end{tabular}
\vspace{0.3cm}
\caption{Non-linear PA output for two-tone input signal.}
\label{tab2}
\end{table}

In our considered setting described in the previous subsections, denoting the PA input amplitude corresponding to CC $j$ in UE $i$ as $A_{i,j}$, its PA input power \big(i.e., $A_{i,j}^2/2$\big) is related to the PA input power allocation per RB as 
\begin{equation}
P_\text{in}= \frac{A_{i,j}^2}{2}=\sum_{k=1}^{N_\text{RB}}p^{i,j,k},
\label{eqn}
\end{equation}
where $N_\text{RB}$ is the total number of RBs in the CC $j$, and the corresponding output power is given by
$P_\text{out} = GP_\text{in}$. We consider that the SI is due to 2nd harmonic. The power of the 2nd harmonic at the transmitter output is given by
\begin{equation}
(p_\text{2H})_\text{Tx,out} = \frac{c_2^2 A_{i,j}^4}{8}. 
\label{SI_Tx_out}
\end{equation}
Let $L_c$ denote the coupling loss between the transmitter and receiver. The SI power at the input of the receiver is then given by
\vspace{-2mm}
\begin{equation}
(p_\text{SI})_\text{Rx,in}  = \frac{(p_\text{2H})_\text{Tx,out}}{L_{\mathrm{c}}} = \frac{c_2^2 A_{i,j}^4}{8L_c}.
\label{SI}
\end{equation}

\begin{figure}
\centering
\includegraphics[width=7.5cm,height=5cm]{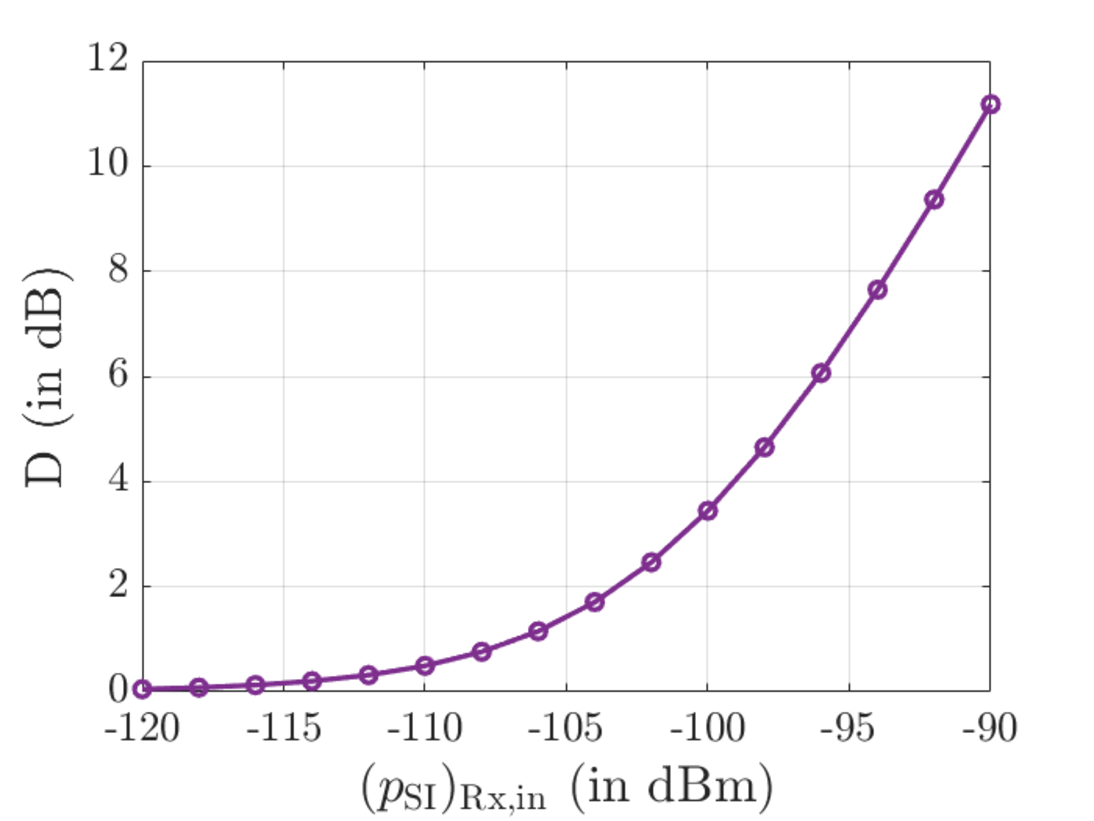}
\caption{Receiver sensitivity degradation as a function of SI power level $(p_\text{SI})_\text{Rx,in}$.}
\label{SID}
\vspace{-4mm}
\end{figure}

\subsubsection{Sensitivity degradation due to SI}
A consequence of the SI in (\ref{SI}) is that it results in a degradation in sensitivity of the UE receiver on the downlink. The receiver sensitivity degradation due to SI is given by
\begin{equation}
\text{D} \ (\text{in dB})=10 \text{log}_{10} \left(\frac{(p_\text{SI})_\text{Rx,in}+\sigma^2}{\sigma^2}\right),
\label{SI_Deg}
\end{equation}
where $\sigma^2=kTBF$ is the additive noise power, $k = 1.38\times 10^{-23}$ Joules/$^\circ$K is the Boltzmann's constant, $T$ is the operating temperature, $B$ is the bandwidth, and $F$ is the receiver noise figure. For $T=300^\circ K$, $B=10$ MHz, and $F=3$ dB, the noise power is $\sigma^2=-100$ dBm. For this noise power and an SI power of -105 dBm, the receiver sensitivity degrades by 1.19 dB. Figure \ref{SID} shows the variation of sensitivity degradation as a function SI power $(p_\text{SI})_\text{Rx,in}$.

Handling the SI in the optimization framework is a new contribution in this paper. In this regard, we note that while the allocated UE transmit power $p^{(i,j,k)}$ contributes to the SINR at the gNB receiver on the UL as per (\ref{SINR}), it also results in SI through the 2nd harmonic level at the UE receiver on the DL as per (\ref{SI}). The SINR contributes to the rate as per (\ref{capacity}) and (\ref{ratemodel}), which, in turn, contributes to the the sum rate maximization objective (described in the next section). On the other hand, the SI effect at the UE receiver is handled by proposing a novel penalty term (involving $p_\text{SI}$) in the reward function in the RL framework (as will be described in Sec. \ref{sec:sec4}).  

As seen in Fig. \ref{SID}, the amount of sensitivity degradation in the receiver is not a linear function of the SI power. Using a HA strategy (as discussed in Sec. \ref{sec:sec1}) leads to completely abandoning the UL and DL carrier combinations that cause SI. This is wasteful because, as Fig. \ref{SID} suggests, low SI powers may be tolerable. This implies that the UE transmit power allocation can suitably be adjusted to keep the degradation to acceptable levels. The proposed RL framework is tasked to jointly optimize the SCC activation pattern and transmit power allocation to the activated SCCs. During training of the RL agent, to take advantage of the tolerable SI power, we introduce the power allocation constraint in terms of SI, thereby making the RL agent aware of the SI induced receiver sensitivity degradation. This is achieved through a novel reward/penalty function introduced and is described in Sec. \ref{sec:sec3a}-\ref{subsec:reward_func}.
 
\vspace{-2mm}
\section{Problem formulation} 
\label{sec:sec3}
In this section, we formulate the problem of sum throughput maximization and discuss the rationale behind the choice of the RL approach to solve the problem. 

\vspace{-2mm}
\subsection{Sum throughput maximization}
\label{problem_formulation}
In this subsection, we formulate the problem of joint UL power, CCs, and RBs allocation to the UEs for sum throughput maximization. There is always a PCC that is activated for all the UEs. Hence, effectively, the CC allocation problem is only for the SCCs. Additionally, RB allocation on CCs also needs to be carried out. This work assumes that all gNBs use the same spectrum for their CCs. Therefore, the UEs transmitting through their activated CCs towards their associated gNBs can cause interference at other gNBs. Thus, the UL transmit power level at the UEs  should be adjusted carefully, while adhering to maximum transmit power and SI power constraints. The UL CA optimization problem can be formulated as follows.
     
\vspace{-5mm}
\begin{align*}
    &\textbf{Objective:} \text{ Maximize the sum throughput of the UEs} \nonumber \\
    &\quad \text{associated to } \text{gNB}_\text{b},\;\forall \text{b} \in \{1,2,\cdots, N_\text{g}\} \nonumber \\
    &\textbf{Variables:} \nonumber \\
    &\quad \text{1) Set of allocated CCs } \mathcal{C}_i  , \forall i \in \mathcal{K} \nonumber \\
    &\quad \text{2) Set of allocated RBs }\mathcal{N}^{i,j}_\text{RB}, \; \forall  i \in \mathcal{K} ,\; \forall j \in \mathcal{M} \nonumber \\
    &\quad \text{3) UL Transmit power levels } p^{(i,j,k)},\; \forall i \in \mathcal{K}, \nonumber \\
    &\quad \quad \forall j \in \mathcal{M},\;\forall k \in \mathcal{N}^{i,j}_\text{RB} \nonumber \\
    &\textbf{Constraints:}\\ & \quad
    \text{1) Maximum allowed UL Tx power }  p_\text{max}  \nonumber \\
    &\quad \quad \text{ and SI level } p_\text{SI} \text{ at the UEs} \nonumber \\ 
    & \quad \text{2) Maximum number of RBs allowed per CC, $N_\mathrm{max}$  } \nonumber
    \end{align*}
The above UL CA optimization problem can be formally written as
\begin{align}
&\hspace{-10mm}\underset{
    \substack{
        \{\alpha^i_j\},\{\beta^i_{j,k}\},\{p^{(i,j,k)}\}
    }
}{\text{maximize}} 
\quad \sum_{i\in\mathcal{K}}  R^i - f(p_{\mathrm{SI}}) \label{eq:opt_problem} \\
\text{s.t. } \quad
&  p^i \leq p^{\max}, \quad \forall i \in \mathcal{K} \nonumber\\
&  \sum_{i\in\mathcal{K}} \beta^i_{j,k} \leq 1, \quad \forall j \in \mathcal{M},\forall k=1, \cdots, N_{\max}\nonumber \\
&  \sum_{i\in\mathcal{K}} \sum_{k=1}^{N_{\max}} \beta^i_{j,k} \leq N_{\max}, \quad \forall j \in \mathcal{M}, \nonumber
\end{align}
\text{where}
\begin{align*}
R^i &= \sum_{j\in\mathcal{M}} \alpha^i_j \sum_{k \in \mathcal{N}_j} \beta^{i}_{j,k} B_{j,k} \log_2\left(1+\gamma^i_{j,k}\right), \\
f(p_{\mathrm{SI}}) &:\ \text{Penalty function of SI power level,} \\
\gamma^i_{j,k} &= \frac{p^{(i,j,k)}h^{\mathrm{b}(i),i}_{j,k}}{\sum\limits_{\substack{i' \in \mathcal{K} \\ i' \neq i}} \beta^{i'}_{j,k} p^{(i',j,k)}h^{\mathrm{b}(i),i'}_{j,k} + \sigma^2}, \\
p^i &= \sum_{j\in\mathcal{M}} \alpha^i_j \sum_{k \in \mathcal{N}_j} \beta^{i}_{j,k} p^{(i,j,k)}.
\end{align*}

\subsection{Why RL approach}
\label{sec:yrl}
As we have seen in the above formulation, the optimization problem of maximizing sum throughput involves three variables of optimization, which include two discrete variables (number of CCs, RBs) and one continuous variable (transmit power). Solving such a mixed-variable optimization problem in dynamic environments using conventional optimization frameworks can be tedious and hard to develop. The sum throughput maximization problem can be solved as a sequential decision-making problem. Sequential decision-making problems refer to a class of decision-making problems in which steps are taken in a sequence, and each step influences not only the immediate reward, but also the long-term reward. These problems involve making decisions over time under uncertainty, with the objective typically being to maximize some cumulative measure. Sequential decision-making problems can be solved in multiple ways, including RL, dynamic programming, and Monte Carlo approximation methods. 

\begin{figure}
\centering
\includegraphics[width=7.5cm,height=4.8cm]{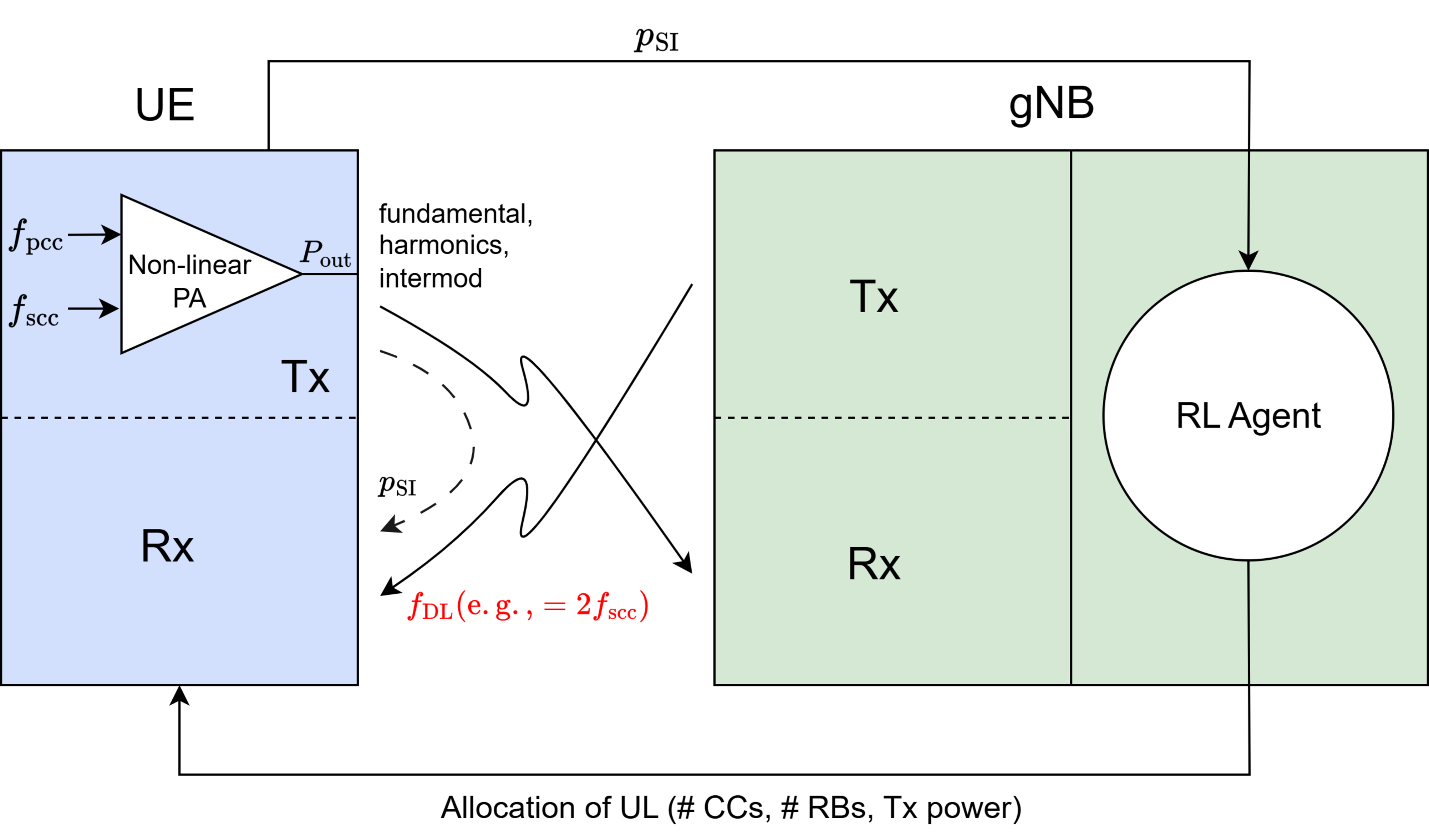}
\caption{RL approach with one agent per gNB.}
\label{Tx_Rx_SI_RL}
\vspace{-4mm}
\end{figure}

RL is a machine learning approach that uses agents to solve problems by exploration and exploitation. Deep RL (DRL) refers to the combination of RL with deep learning. The primary components of RL are the agent and the environment \cite{stat_2}. The environment represents the world in which the agent operates and interacts. During each interaction step, the agent receives an observation, which may be partial, of the current state of the environment. Based on this observation, the agent decides on an action to execute. The environment then undergoes a change as a result of the action of the agent. In addition, the agent receives a reward signal from the environment that quantifies the desirability of the action taken in the current state with a numerical value. The agent's objective is to maximize its cumulative reward. RL methods provide strategies for the agent to learn optimal behaviors to achieve this goal. Fundamentally, RL is the study of agents and their learning processes through trial and error. It formalizes the concept that providing rewards or punishments for an agent's actions influences the likelihood of those actions being repeated or avoided in the future. RL techniques do not require any initial data to start with. As training progresses, the networks learn from previous actions taken by the RL agent, making RL approach a good choice for dynamic and uncertain environments. So, we choose the RL approach to solve our problem. Also, as there are multiple gNBs, we provide each gNB with an agent, resulting in a multi-agent RL architecture that can scale well. Figure \ref{Tx_Rx_SI_RL} illustrates the proposed approach in which each gNB has an RL agent that learns and makes decisions on the allocation of resources to all its associated UEs. These RL agents at different gNBs communicate with each other in the backhaul to share the inter-gNB parameters needed in their respective optimizations.    

Different types of RL agents can be used to solve an optimization problem. The types of agents include model-dependent agents, model-free agents, value-based agents, policy-based agents, and actor-critic agents. In our considered system, with optimization over continuous and discrete variables, obtaining transition probabilities and policies in closed-form is difficult. Hence, we use model-free type of RL agents. Deep deterministic policy gradient (DDPG) algorithm \cite{ref4},\cite{ddpg2} is a model-free type of algorithm that does not need a model for the policies and transition probabilities. DDPG algorithm can optimize the policy function by maximizing the value function using actor-critic networks. However, DDPG can effectively handle only continuous action spaces \cite{ref4}. To handle discrete action spaces and to approximate the Q-function value at the critic, deep Q-networks (DQN)\cite{UL_CA8} are generally used. Hence, these traditional RL agents can work only with either purely continuous action spaces or purely discrete action spaces. To solve the problem of mixed variable optimization, we adapt CA2C algorithm which combines DDPG and DQN in an actor-critic framework to solve the optimization problem.

As mentioned earlier, the authors in \cite{ref1} consider the problem of CA enabled by activation/deactivation of CC with power and delay constraints. To do this, they propose a multi-agent Double Deep Q-Network (DDQN) based CC management algorithm. The DDQN is designed to minimize the average delay and average consumed power, with the constraints on number of active CCs and QoS requirements. In \cite{stat_1}, the authors simultaneously minimize the delay and power consumption for the UEs. For this, they propose a Joint Power-Sharing and Carrier Aggregation (JPSCA) algorithm, which is a type of compound-action actor-critic (CA2C) method \cite{stat_2}. However, as described in Sec. \ref{sec:sec2}-\ref{subsec_SI_model}, activating CCs without considering the downlink frequency leads to severe sensitivity degradation of the user's downlink receiver (see Fig. \ref{SID}), leading to poor performance on the downlink \cite{book2}. Considering its practical significance, we develop a non-linear model for SI to enable its integration into the proposed RL-based framework. This distinguishes the proposed work from that in \cite{stat_1}.

\vspace{-2mm}
\section{RL and CA2C algorithm}
\label{sec:sec3a}
As mentioned above, we model our resource allocation problem as a sequential decision-making problem and take an RL approach to solve it. A Markov decision process (MDP) provides a mathematical framework for modeling sequential decision-making problems where the outcomes are partly random and partly under the control of a decision-maker. An MDP is a 5-tuple, $\langle \mathcal{S}, \mathcal{A}, P, R, \pi_0 \rangle$, where $\mathcal{S}$ is the state space, $\mathcal{A}$ is the action space, $P$ is the transition probability function with $P(\mathbf{s}'| \mathbf{s},\mathbf{a})$ being the probability of transitioning into state $\mathbf{s}'$ starting from state $\mathbf{s}$ and
taking action $\mathbf{a}$, $R$ is the reward function, and $\pi_0$ is the starting state distribution.

\subsection{Value functions and Bellman equations}
Value functions are functions that estimate the expected cumulative future reward from a state or state-action pair under a particular policy. 
There are two primary types of value functions, which include 
\begin{itemize}
\item \textbf{State value function \(V^\pi(\mathbf{s})\)}: This function gives the expected cumulative reward starting from state \(\mathbf{s}\) and following policy \(\pi\).
\item \textbf{Action value function \(Q^\pi(\mathbf{s}, \mathbf{a})\)}: This function gives the expected cumulative reward starting from state \(\mathbf{s}\), taking action \(\mathbf{a}\), and subsequently following policy \(\pi\).
\end{itemize}
The Bellman equations are pair of equations that provide the mathematical expressions for the above value functions. The Bellman equations are given by
\begin{itemize}
    \item Bellman equation for \(V^\pi(\mathbf{s})\):
    \begin{equation}
    \hspace{-7.5mm}
    V^\pi(\mathbf{s}) = \mathbb{E}_{\mathbf{a} \sim \pi(\mathbf{s})} \Big[ R(\mathbf{s}, \mathbf{a}) + \gamma \sum_{\mathbf{s}'}\hspace{-0.5mm} P(\mathbf{s}' | \mathbf{s}, \mathbf{a}) V^\pi(\mathbf{s}') \Big],
    \end{equation}
    \item Bellman equation for \(Q^\pi(\mathbf{s}, \mathbf{a})\):
    \begin{eqnarray}
    \hspace{-12mm}
    Q^\pi(\mathbf{s}, \mathbf{a}) & \hspace{-2mm} = & \hspace{-2mm} R(\mathbf{s}, \mathbf{a}) \nonumber \\ 
    & \hspace{-2mm} & \hspace{-2mm} + \gamma \sum_{\mathbf{s}'} P(\mathbf{s}' | \mathbf{s}, \mathbf{a}) \mathbb{E}_{\mathbf{a}' \sim \pi(\mathbf{s}')} \left[ Q^\pi(\mathbf{s}', \mathbf{a}') \right],
    \end{eqnarray}
\end{itemize}
where $\gamma$ is the discount factor. The discount factor $\gamma \: \in\: [0,1]$ is a weight multiplied with the instant reward to get the cumulative reward. The cumulative reward for an agent is given by $\sum^{T}_{t=1}\gamma^t \eta(t)$, where $\eta = R(\mathbf{s}, \mathbf{a})$, $t$ is the time instance, $T$ is the total number of time steps that the agent takes.
 
The goal in RL is to find an optimal policy \(\pi^*\) that maximizes the expected cumulative reward from any state \(\mathbf{s}\). The value functions under the optimal policy, denoted as \(V^*(\mathbf{s})\) and \(Q^*(\mathbf{s}, \mathbf{a})\), are given by
\begin{itemize}
\item optimal state value function: 
\begin{equation}
V^*(\mathbf{s}) = \max_\pi V^\pi(\mathbf{s}),
\end{equation}
\item optimal action value function:
\begin{equation}
Q^*(\mathbf{s}, \mathbf{a}) = \max_\pi Q^\pi(\mathbf{s}, \mathbf{a}).
\end{equation}
\end{itemize}
The Bellman optimality equations describe these optimal value functions as follows:
\begin{itemize}
\item Bellman optimality equation for \(V^*(\mathbf{s})\):
\begin{equation}
\hspace{-8mm}
V^*(\mathbf{s}) = \max_{\mathbf{a}} \Big[ R(\mathbf{s}, \mathbf{a}) + \gamma \sum_{\mathbf{s}'} P(\mathbf{s}' | \mathbf{s}, \mathbf{a}) V^*(\mathbf{s}') \Big],
\end{equation}  
\item Bellman optimality equation for \(Q^*(\mathbf{s}, \mathbf{a})\):
\begin{equation}
\hspace{-8mm}
Q^*(\mathbf{s}, \mathbf{a}) = R(\mathbf{s}, \mathbf{a}) + \gamma \sum_{\mathbf{s}'} P(\mathbf{s}' | \mathbf{s}, \mathbf{a}) \max_{\mathbf{a}'} Q^*(\mathbf{s}', \mathbf{a}').
\end{equation}
\end{itemize}

\subsection{CA2C algorithm}
\label{CA2Calgo}
CA2C algorithm is an actor-critic, model-free RL algorithm developed to handle environments with both continuous and discrete action spaces \cite{stat_2}. It combines both DDPG and DQN together into an actor-critic framework. The actor network uses a deterministic policy, meaning it always outputs the same action for a given state. This contrasts with stochastic policies, which would output a probability distribution over actions. Using DDPG, the actor learns a deterministic policy, while the critic evaluates the policy by learning the Q-value function using DQN, both guided by the Bellman equations.

The actor network in the RL architecture represents the policy function of the environment. Here, let $\boldsymbol{\theta}$ be the parameters of the actor DNN. Then, $\pi(\mathbf{s}\mid \boldsymbol{\theta})$ parameterizes the policy function that takes state \(\mathbf{s}\) and outputs a deterministic action \(\mathbf{a}\). The objective of the actor is to maximize the expected cumulative reward $J(\boldsymbol{\theta}) = \mathbb{E}_{\mathbf{s} \sim \mathcal{D}} \left[ Q(\mathbf{s}, \pi(\mathbf{s} | \boldsymbol{\theta})) \right]$, and the actor is updated using the policy gradient \cite{stat_2}
\begin{equation}
\nabla_{\boldsymbol{\theta}} J \approx \mathbb{E}_{\mathbf{s} \sim \mathcal{D}} \big[ \nabla_\mathbf{a} Q(\mathbf{s}, \mathbf{a} | \mathbf{w}) \nabla_{\boldsymbol{\theta}} \pi(\mathbf{s} | \boldsymbol{\theta}) \big],
\end{equation}
where $\mathcal{D}$ is the replay buffer that stores experiences of the time steps that have been completed till the current time instance, and $\mathbf{w}$ denotes the parameters of the critic DNN. The critic network \(Q(\mathbf{s},\mathbf{a} | \mathbf{w})\) approximates the action value function. It takes a state \(\mathbf{s}\) and action \(\mathbf{a}\) and outputs the expected long-term cumulative reward. The critic is updated by minimizing the loss function $L(\mathbf{w})$ \cite{stat_2}, given by
\begin{equation}
L(\mathbf{w}) = \mathbb{E}_{(\mathbf{s}, \mathbf{a}, R, \mathbf{s}') \sim \mathcal{D}} \big[ \left( y - Q(\mathbf{s}, \mathbf{a}|\mathbf{w}) \right)^2 \big],
\label{lossfunction}
\end{equation}
where $y = R + \gamma Q(\mathbf{s}', \pi(\mathbf{s}'|\hat{\boldsymbol{\theta}})|\hat{\mathbf{w}})$ is the target Q-value, and $\hat{\boldsymbol{\theta}}$ and $\hat{\mathbf{w}}$ are the parameters of the target actor and critic DNNs, respectively.

While training, CA2C uses target networks for both the actor and critic. These target networks are slowly updated to track the learned networks, reducing the risk of divergent updates and improving the convergence properties. The training of the target networks is carried out as follows:
\begin{equation}
\hat{\boldsymbol{\theta}} \leftarrow \tau \boldsymbol{\theta} + (1 - \tau) \hat{\boldsymbol{\theta}}, 
\end{equation}
\begin{equation}
\hat{\mathbf{w}} \leftarrow \tau \mathbf{w} + (1 - \tau) \hat{\mathbf{w}},
\end{equation}
where \(\tau \leq 1\) is a constant and is termed as soft update parameter.
    
\section{Proposed RL architecture} 
\label{sec:sec4}
In this section, we describe the proposed RL architecture in detail. We present the set of agents, state space, action space, rewards, and RL architecture and algorithms that we use for our optimization problem. 

\subsection{Set of agents} Each gNB is provided with an RL agent. The set of gNBs, $\mathcal{B}=\left\{\text{gNB}_1,\cdots, \text{gNB}_{N_\text{g}}\right\}$, is taken to be the set of agents. Hence, this is a multi-agent model-free RL problem.

\subsection{State space}  In our multi-agent RL system, all agents have the same state space which is given in terms of the QoS delay requirement for the UEs \cite{stat_1}. The state vector for our RL system is a binary vector with 0's and 1's as elements. The element value 1 in index $i$ means that UE $i$ has met the delay QoS requirements, and 0 means that UE $i$ has not met the delay QoS requirements. The state space $\mathcal{S}$ for the system is defined as
\begin{equation}
\mathcal{S}=\left\{\mathbf{s} \mid \mathbf{s} = \left[s^{i}\right]_{i\in \mathcal{K}}, \text{ where}\; s^i \in \left\{0,1\right\} \right\}.
\label{statespace}
\end{equation}
The state vector gets updated after every time step based on the condition whether the delay of a UE is more than the QoS delay requirement or not, i.e.,      
\begin{align}
s^i = \begin{cases}
1, \quad\textrm{if } D^i \leq D_\text{QoS}\\
0 , \quad\textrm{otherwise,}
\end{cases}
\end{align}
where $D^i = \frac{\hat{q}^{i}}{R^i}$ is delay of UE $i$, $\hat{q}^{i}$ is average number of bits per burst of data for UE $i$, ${R^i}$ is total throughput for UE $i$ as defined in (\ref{ratemodel}), and $ D_\text{QoS}$ is the delay QoS requirement.

Each gNB can collect this information about its associated UEs and broadcast it to all the other gNBs. The overhead of broadcasting the state vector of zeros and ones among the gNBs would be low. 

\subsection{Action space}
In the multi-agent RL system, for a given $\text{gNB}_\text{b}$ (agent b), the action corresponds to adjusting the UL transmit power of UEs assigned to $\text{gNB}_\text{b}$ and activating/deactivating SCCs of those UEs. Let $\mathcal{A}_\text{b}$ denote the action space for $\text{gNB}_\text{b}$. It is defined as
\begin{equation}
\hspace{-1mm}
\mathcal{A}_\text{b} = \{\mathbf{a}_\text{b} = (\mathbf{p}_\text{b},\boldsymbol{\alpha}_{\text{b}}) \mid \mathbf{p}_\text{b} = [p^i]_{i\in {\mathcal{K}_\text{b}}} , \boldsymbol{\alpha}_{\text{b}} = [\alpha^i_j]_{i\in {\mathcal{K}_\text{b}}, j\in{\mathcal{M}}} \}.
\label{actionspace}
\end{equation}
In (\ref{actionspace}), $p^i$ is the UL transmit power allocated to UE $i$ by $\text{gNB}_\text{b}$ and $\alpha^i_j$ indicates the status of CC $j$ associated with UE $i$. Since $\mathbf{p}_\text{b}$ is a continuous action and $\boldsymbol{\alpha}_{\text{b}}$ is a discrete action, the action space $\mathcal{A}_\text{b}$ 
is a compound-action space.

\subsection{Policy}
A policy is a function map that maps states to actions that an agent should take. The policy profile for an agent b is given by 
\begin{equation}
\pi_\text{b} \colon \mathbf{s} \rightarrow \mathbf{a}_\text{b}.
\label{agent}
\end{equation}
In DRL, we deal with parameterized policies, i.e., the policies whose outputs are computable functions that depend on a set of parameters (e.g., the weights and biases of a neural network), which we can adjust to change the behavior via some optimization algorithm.

\begin{figure}
\centering
\includegraphics[width=8.2cm,height=2.45cm]{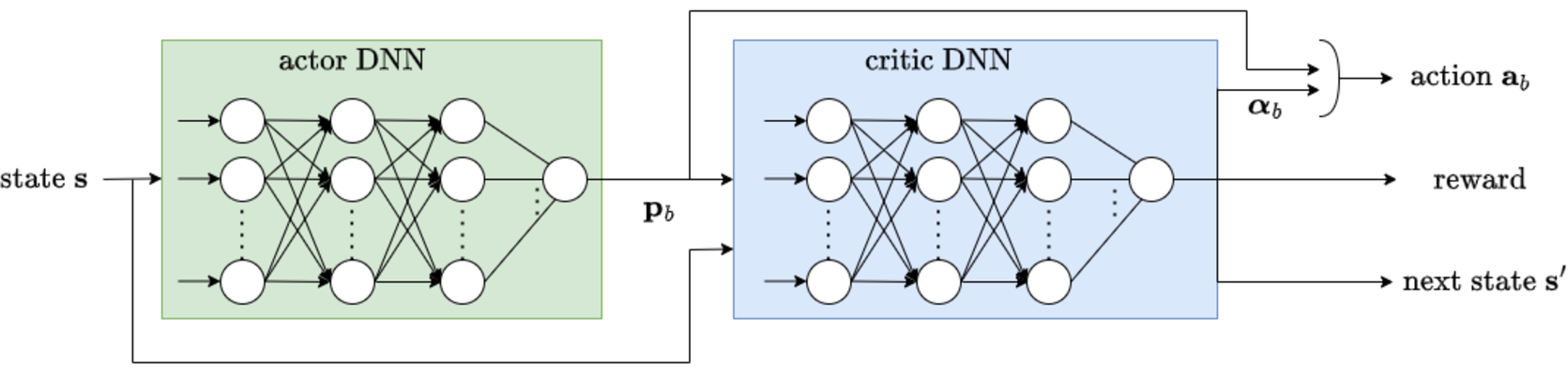}
\caption{RL architecture using actor-critic network.}
\label{RLarch}
\end{figure}

\begin{table}
\centering
\begin{tabular}{|l|c|c|}
\hline
{\textbf{Parameters}} & { \textbf{Actor DNN}} & { \textbf{Critic DNN}} \\ \hline \hline
\# hidden layers   & 3  & 3     \\ \hline
\# neurons in hidden layer 1 & 128   & 128  \\ \hline
\# neurons in hidden layer 2 & 512  & 512 \\ \hline
\# neurons in hidden layer 3 & 1024 & 1024  \\ \hline
Activation function & ReLU  & ReLU  \\ \hline
Optimization   & Adam  & Adam  \\ \hline
\end{tabular}
\vspace{4mm}
\caption{Table of parameters for the actor DNN and critic DNN.}
\label{RLarchtable}
\vspace{-4mm}
\end{table}

\subsection{DNN architecture of actor-critic networks:}
We use fully connected deep neural networks (DNN) for actor network and critic network as shown in Fig. \ref{RLarch}. The online networks and target networks have identical architecture. The parameters of the actor and critic DNNs are listed in the Table \ref{RLarchtable}.

\subsection{Reward function}
\label{subsec:reward_func}
Since the objective is to maximize the sum throughput, the primary component in the reward function is the sum of UE throughputs. In addition, we include another component in the reward function in order to account for the SI effect. Specifically, we include a penalty component in the reward function, where the penalty is a function of the SI power level. More the SI power level, more is the penalty. This encourages the RL framework to make allocations in such a way that the sensitivity degradation due to SI is within acceptable levels and heavily discourages (penalizes) high SI power levels. Accordingly, we propose the following reward function for our problem: 
\begin{equation}
R_\text{b} = \sum_{i \in \mathcal{K}_\text{b}}^{}R^i - \underbrace{\sum_{i \in \mathcal{K}_\text{b}}^{}{\frac{R^i_\text{p}}{d_{i}^{'}}}}_\text{penalty component},
\label{rew_fn}
\end{equation}
where $R^i$ is total throughput for UE $i$ as defined in (\ref{ratemodel}), $R^i_\text{p}$ is the proposed penalty function for UE $i$, defined as
\begin{equation}
R^i_\text{p} =
\begin{cases}
0, & \text{for $p^i_\text{SI}\leq\theta^i_{1}$} \\
\frac{\Omega^i(p^i_\text{SI} - \theta^i_{1})}{\theta^i_{2} - \theta^i_{1}}, & \text{for $\theta^i_{1}<p^i_\text{SI}<\theta^i_{2}$}\\
\Omega^i, & \text{for $p^i_\text{SI}\geq\theta^i_{2}$}, 
\end{cases}
\label{R_b}
\end{equation}
and $d_i^{'}$ is is defined as
\begin{equation}
d_{i}^{'} = \frac{d_{i}^{2}}{d_{\mathrm{max}}^{2}},
\label{d_norm}
\end{equation}
where $d_\text{max}$ is the maximum distance between  UE and gNB (i.e., cell radius). The normalization of UEs distance from gNB ensures fairness among UEs located at different distances from gNB. In (\ref{R_b}), $\theta^i_{1}$ corresponds to the $p^i_\text{SI}$ value below which sensitivity degradation is acceptable and $\theta^i_{2}$ corresponds to the $p^i_\text{SI}$ value above which the degradation of sensitivity is not acceptable. For example, suppose a sensitivity degradation in the range 3 dB to 6 dB is acceptable, then, from (\ref{SI_Deg}), $\theta_1^i$ and $\theta_2^i$ can be chosen as -100 dBm and -95 dBm, respectively. Also, the constant $\Omega^i$ in (\ref{R_b}) is chosen to be a large value that heavily penalizes/discourages unacceptable SI power levels. 

\textit{Remark:} We note that custom penalty functions have been proposed before in the literature \cite{sefati2022hybrid, makantasis2019deep}. The motivation behind penalty function design is to enable the RL agent to explore the adverse space. In other words, instead of having a binary penalty function that prevents the RL agent to completely avoid an undesired space, it is desired that penalty functions dynamically penalize the RL agent based on how ``deep'' the RL agent has forayed into the undesired space. This allows the RL agent to learn to avoid the undesired space with some tolerance, which in most practical cases might be acceptable. Along these lines, we propose a loss function that allows the RL agent to explore the space with low penalty for lower levels of SI, but gets heavily penalized if it makes the system unstable. The design of the reward function for the described non-linearity induced SI model has not been attempted before, which contributes to the novelty. Also, the proposed penalty function allows the RL framework to achieve better sum throughput than naive schemes such as the hard avoidance scheme (see Fig. \ref{Single_UE_sumrate}).

\begin{figure}
\centering
\includegraphics[width=8.5cm,height=5cm]{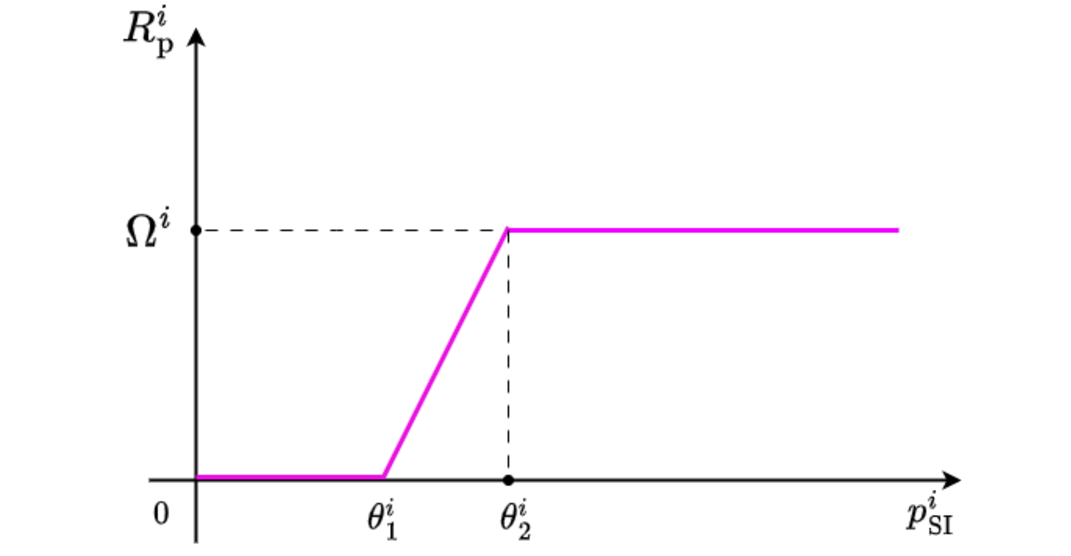}
\caption{Proposed penalty function.}
\label{reward_fn}
\vspace{-2mm}
\end{figure}

\subsection{Algorithms for allocation}
In this subsection, we present the algorithms that we use to get the joint allocation of UL power, number of CCs, and number of RBs for the UEs. First, we present the complete algorithm on how the optimization problem is solved and training is performed. Then, we present the algorithms of CA2C, CC selection (CCS), and RB allocation. 

\begin{figure*}
\hspace{3mm}
\includegraphics[width=18cm,height=7.5cm]{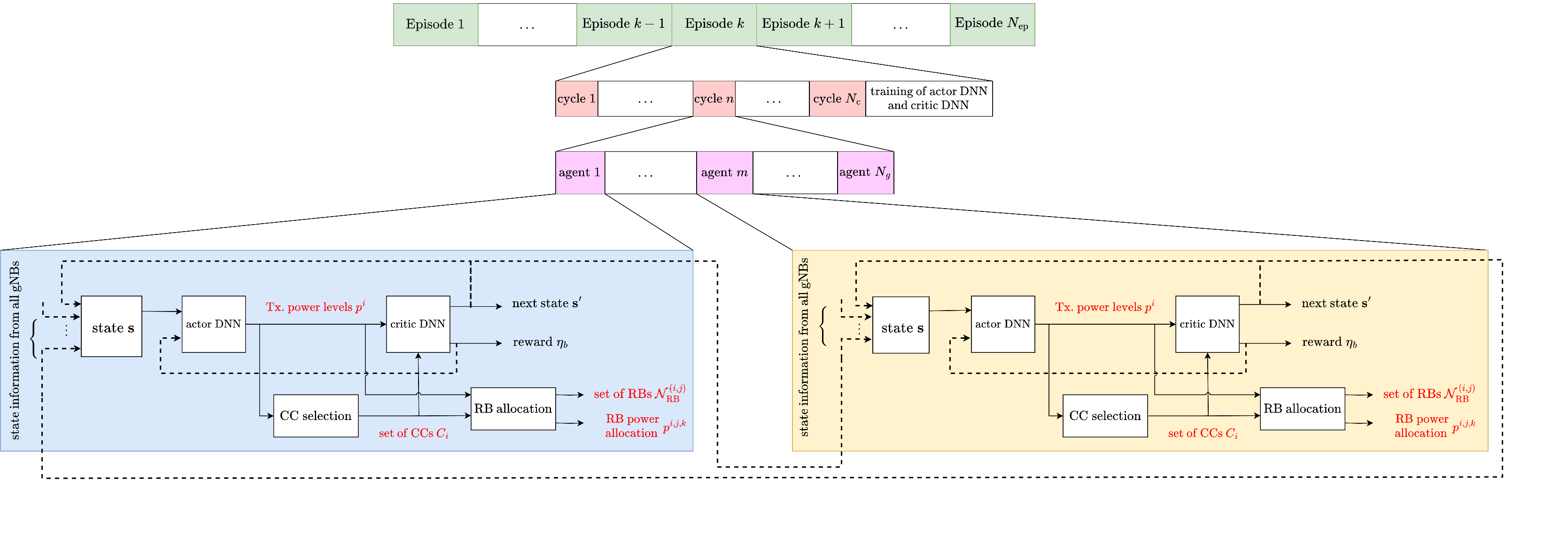}
\vspace{-13mm}
\caption{RL algorithm for resource allocation in UL CA.}
\label{RLalgo}
\vspace{-3mm}
\end{figure*}

We propose a learning approach to derive the optimal policy to adjust the continuous action of UL transmit power levels of each UE. Let $\pi=\left(\pi_{1}, \cdots, \pi_{N_\text{g}}\right)$ be the policy profile for the gNBs. Specifically, for a given $\mathrm{gNB}_\text{b}, \pi_\text{b}$ is a function from a given state $\mathbf{s}$ to action $\mathbf{a}_\text{b}$, i.e., $\pi_\text{b}: \mathcal{S} \mapsto \mathcal{A}_\text{b}$. So, the policy profile $\pi$ corresponds to the patterns of behavior for the gNBs at different states of the environment, and thus the policy for them should be optimized to maximize the reward for the $\mathrm{gNBs}$. Let $\pi^{*}=\big(\pi_{1}^{*},\cdots, \pi_{N_\text{g}}^{*}\big)$ be the optimal policy profile for the gNBs. To find the optimal policy for a given $\mathrm{gNB}_\text{b}$, both the current and future rewards for the agent should be taken into account. We use the notation $Q_\text{b}\left(\mathbf{s}, \mathbf{a}_\text{b}\right)$ for the Q-function of $\mathrm{gNB}_\text{b}$. Specifically, $Q_\text{b}\left(\mathbf{s}, \mathbf{a}_\text{b}\right)$ is defined in terms of the expectation of the weighted sum of the short-term reward for the agent \cite{ref3}. Based on what is discussed in \cite{stat_2} and \cite{UL_CA10}, the optimal policy for an agent $\mathrm{gNB}_\text{b}$ is the policy under which the $\mathrm{Q}$-function for that agent is maximized, i.e., 
\begin{equation}
\mathbf{a}_\text{b}^{*}=\pi_\text{b}^{*}(\mathbf{s})=\arg \max _{\mathbf{a}_\text{b} \in \mathcal{A}_\text{b}} Q_\text{b}^{*}\left(\mathbf{s}, \mathbf{a}_\text{b}\right), \ \forall \mathbf{s} \in \mathcal{S}.
\label{optimalpolicy}
\end{equation}
For $\text{gNB}_\text{b}$, the state $\mathbf{s}$, reward $R_\text{b}$, and $\boldsymbol{\alpha}_\text{b} =[\alpha^i_{j}]_{i\in\mathcal{K}_{b},j\in\mathcal{M}}$, where $\boldsymbol{\alpha}_\text{b}$ indicates the set of activated CCs for the UEs in $\mathcal{K}_{b}$, are given as the input to the actor DNN. Using the CA2C algorithm presented in the previous section, we derive $\mathbf{Algorithm \;
\ref{alg:sum_throughput_maximization}}$
(listed in next page) for sum throughput maximization by obtaining UL transmit powers, set of CCs, and set of RBs assigned to the UEs (see Fig. \ref{RLalgo}). For given state $\mathbf{s}$ and discrete action $\boldsymbol{\alpha}_{\text{b}}$, each agent (gNB) b $\in \mathcal{B}$ uses actor DNN to update the continuous action corresponding to UL transmit power levels $\mathbf{p}_\text{b}$. Similarly, given the state $\mathbf{s}$ and continuous action $\mathbf{p}_\text{b}$, each agent b $\in \mathcal{B}$ uses critic DNN to obtain discrete action $\boldsymbol{\alpha}_{\text{b}}$ corresponding to CCs activation. Using this continuous action $\mathbf{p}_\text{b}$, the CC selection (CCS) 
algorithm (listed in $\mathbf{Algorithm\; 
\ref{alg:cc_selection}}$) finds the optimum CC allocation that needs to be done for the UEs in order to maximize the Q-values of the critic network. A round-robin (RR) algorithm (listed in $\mathbf{Algorithm \; 
\ref{alg:rb_allocation}}$) is used to assign RBs to the UEs sharing a common CC. The experiences are stored in replay buffer. At the end of every $N_\text{c}$ cycles in each episode, the actor and critic DNNs are trained using the samples from replay buffer and the parameters are updated for the DNNs.

\subsubsection{Training}
To train the actor and critic DNNs, we combine the DDPG and DQN training algorithms as in \cite{ref4} and \cite{UL_CA8}. In the training process, to prevent over-optimism and instability, the concept of the target network and online network are employed. Specifically, the actor and critic target networks have parameters $\hat{\boldsymbol{\theta}}_\text{b}$ and $\hat{\mathbf{w}}_\text{b}$, respectively. Additionally, the parameters for the actor and critic online networks are $\boldsymbol{\theta}_\text{b}$ and $\mathbf{w}_\text{b}$, respectively. At each step, we use the soft update method for updating the parameters for the target networks, given by $\hat{\boldsymbol{\theta}}_\text{b}=\tau \boldsymbol{\theta}_\text{b}+(1-\tau) \hat{\boldsymbol{\theta}}_\text{b}$ and $\hat{\mathbf{w}}_\text{b}=\tau \mathbf{w}_\text{b}+(1-\tau) \hat{\mathbf{w}}_\text{b}$, where $\tau$ is the fixed soft update parameter. The replay buffer is used to train the online actor and critic DNNs.
Specifically, for a given agent $\text{b}$, at a given time $t$, let us denote the tuple of current space, current action, next state, and reward as the experience $\mathbf{e}_{b}^{t}=\left[\mathbf{s}(t), \mathbf{a}_\text{b}(t), \mathbf{s}^{\prime}, R_\text{b}(t)\right]$. 
The replay buffer for the agent $\text{b}$ is denoted by $\mathcal{D}_\text{b}=\left\{\mathbf{e}_\text{b}^{t}\right\}$. At each step, a mini-batch of the transmissions is uniformly chosen and used to train the actor and critic DNNs. In addition, we denote the functions evaluated by the actor and critic networks by $J_\text{b}\left(\boldsymbol{\theta}_\text{b}\right)$ and $L_\text{b}\left(\mathbf{w}_\text{b}\right)$, respectively. The CA2C algorithm training process is explained in Sec. IV \ref{CA2Calgo}. The function $J_\text{b}\left(\boldsymbol{\theta}_\text{b}\right)$ is given in terms of the average $\mathrm{Q}$-function for agent $\text{b}$, and $L_\text{b}\left(\mathbf{w}_\text{b}\right)$ is given in terms of the average difference between the $\mathrm{Q}$-function and target value for agent $\text{b}$ as in \eqref{lossfunction}. 
The updating function for $\boldsymbol{\theta}_\text{b}$ and $\mathbf{w}_\text{b}$ is \cite{stat_2}
\begin{align}
\label{actortraining}
\boldsymbol{\theta}_\text{b} & =\boldsymbol{\theta}_\text{b}+\kappa \nabla_{\boldsymbol{\theta}_\text{b}} J_\text{b}\left(\boldsymbol{\theta}_\text{b}\right), \\ \label{critictraining}
\mathbf{w}_\text{b} & =\mathbf{w}_\text{b}-\kappa \nabla_{\mathbf{w}} L_\text{b}\left(\mathbf{w}_\text{b}\right),
\end{align}
where $\kappa$ is the learning rate.

\begin{algorithm}[t]
\caption{Sum throughput maximization algorithm}
\label{alg:sum_throughput_maximization}
\begin{algorithmic}
\STATE \textbf{Input:} Number of training episodes $N_\text{ep}$, number of cycles $N_\text{c}$, learning rate $\kappa$, batch size, action space $\mathcal{A}_\text{b}, \forall \: \text{b} \: \in \: \mathcal{B}$, state space $\mathcal{S}$, maximum allowed transmit power $p_\text{max}$
\STATE \textbf{Output:} UL power allocation for the UEs $p^{(i,j,k)}$, set of CCs assigned to the UEs $\mathcal{C}_i $, and set of RBs assigned in each CC $\mathcal{N}_\text{RB}^{i,j}$, $\forall i \: \in \mathcal{K},\;\forall j\:\in\:\mathcal{M},\;\forall k\:\in\:\mathcal{N}_\text{RB}^{i,j}$  
\STATE \textbf{Initialization} : Experience memory $D_\text{b}$, $\forall \: \text{b} \:\in\: \mathcal{B}$, actor
online network parameters $\boldsymbol{\theta}_\text{b}$, $\forall \text{b} \:\in \: \mathcal{B}$, actor target
network parameters $\hat{\boldsymbol{\theta}}_\text{b}=\boldsymbol{\theta}_\text{b}$, $\forall \text{b} \:\in\: \mathcal{B}$, critic online
network parameters $\mathbf{w}_\text{b}$, $\forall \text{b} \:\in\: \mathcal{B}$, critic target network
parameters $\hat{\mathbf{w}}_\text{b}=\mathbf{w}_\text{b}$, $\forall \text{b} \:\in \:\mathcal{B}$, network state $\mathbf{s}$, UE locations
\FOR{episode $= 1 \;$to$\; N_\text{ep}$}
\FOR{cycle $= 1 \;$to$\; N_\text{c}$}
\FOR{$\text{b} = 1\;$to$\;N_\text{g}$}
    \STATE \textbf{Step 1:}
    \FOR{$\text{b}^{\prime} = 1\;$to$\;N_\text{g}$}
        \STATE Collect and update $\mathbf{s}_{\mathcal{K}_{\text{b}^{'}}}$ via backhaul network
    \ENDFOR
    \STATE \textbf{Step 2:} Given state $\mathbf{s}$, previous cycle CC activation indicators $\alpha_j^{i}$, and previous cycle reward $R_\text{b}$ as input to actor DNN, obtain $p^i$ as the output
    \STATE \textbf{Step 3:} Given the UL transmit power levels of the UEs as the input to  $\mathbf{Algorithm\;2}$, get the CC activation vector $\boldsymbol{\alpha}_\text{b}$ as the output    
    \STATE \textbf{Step 4:} Given the transmit powers $p^i$, state $\mathbf{s}$, and CC activation vector $\boldsymbol{\alpha}_\text{b}$ as the input to critic DNN, get reward $R_\text{b}$ and next state $\mathbf{s}^{\prime}$ as the output 
    \STATE \textbf{Step 5:} Use 
    $\mathbf{Algorithm\;3}$ to get the RBs allocation for the CCs in UEs 
    \STATE \textbf{Step 6:} Store the experiences $\mathbf{e}_\text{b}$ in the replay buffer $\mathcal{D}_\text{b}$ for training of the DNNs 
\ENDFOR
\ENDFOR
\FOR{$\text{b} = 1\;$to$\;N_\text{g}$}
\STATE \textbf{Step 1:} Train actor networks by taking batch size number of training experiences  from $\mathcal{D}_\text{b}$ and performing backpropagation on the networks  using \eqref{actortraining}
\STATE \textbf{Step 2:} Train critic networks by taking batch size number of training experiences  from $\mathcal{D}_\text{b}$ and performing backpropagation on the networks using \eqref{critictraining}
\ENDFOR
\ENDFOR
\end{algorithmic}
\end{algorithm}

\begin{algorithm}[t]
\caption{CC selection algorithm}
\label{alg:cc_selection}
\begin{algorithmic}
\STATE \textbf{Input:} UL transmit power levels $p^i$ of the UEs
\STATE \textbf{Output:} Optimum CC selection vector
$\boldsymbol{\alpha}_{\text{b},\text{opt}}$
\STATE \textbf{Initialization:} CC selection vector set  $\mathcal{V_{\text{b}}}$, which consists of all the $2^{MK_\text{b}}$ binary vectors each of length ${MK_\text{b}}$, best reward value $R_{\text{b},\text{opt}}$ which is initialized as zero 
\FOR{$i= 1 \;$to$\; 2^{MK_\text{b}}$}
    \STATE \textbf{Step 1:} Let 
    $\boldsymbol{\alpha}_{\text{b}}$ be the $i$th element in the CC selection vector set $\mathcal{V_\text{b}}$ and compute the reward function $R_\text{b}$ for this CC selection configuration

\IF{$R_{\text{b},\text{opt}}\leq R_\text{b}$ AND $\alpha_1^{i}=1,\;\forall i\:\in\:\mathcal{K_\text{b}}$}
    \STATE \textbf{Step 2:} Update $R_{\text{b},\text{opt}}\leftarrow R_\text{b}$ and
    $\boldsymbol{\alpha}_{\text{b},\text{opt}} \leftarrow \boldsymbol{\alpha}_{\text{b}}$
\ENDIF
\ENDFOR
\end{algorithmic}
\end{algorithm}
\vspace{-1mm}
\begin{algorithm}[t]
\caption{RB allocation algorithm}
\label{alg:rb_allocation}
\begin{algorithmic}
\STATE \textbf{Input:} UL transmit power levels $p^i$ of the UEs, CC activation vector 
$\boldsymbol{\alpha}_{\text{b}}$
\STATE \textbf{Output:} $\mathcal{N}_\text{RB}^{i,j}$ which is set of RBs allocated to CC $j$ in UE $i$, and power allocation $p^{(i,j,k)}$
\FOR{$b= 1 \;$to$\; N_\text{g}$}
\FOR {$j=1\;\text{to}\;M$}
   \STATE \textbf{Step 1:} For each UE $i\:\in\: \mathcal{K}_\text{b}$ assign RBs for CC $j$ in a round robin way, and set $\beta^i_{j,k}=1$
\ENDFOR
\FOR{$i\: \in\:\mathcal{K}_\text{b}$}
\STATE \textbf{Step 2:} Assign $p^{(i,j,k)} = \frac{p^i}{\underset{j\in\mathcal{M},k\in\mathcal{N}_\text{RB}^{i,j}}{\sum}\beta^i_{j,k}}$ 
\ENDFOR
\ENDFOR
\end{algorithmic}
\end{algorithm}

\vspace{1mm}
\subsubsection{On RB allocation}
The total number of RBs in each CC is 50. All the 50 RBs in the PCC (CC1) are equally shared among the UEs in a gNB. The RBs in the SCC (CC2) are allocated based on {\bf Algorithm \ref{alg:rb_allocation}}. We consider three different resolutions in the allocation of RBs in CC2. These resolutions include resolution-50 (denoted by `Res50'), resolution-25 (denoted by `Res25'), and resolution-10 (denoted by `Res10'). Res50 is the coarsest resolution that is possible, where either all 50 RBs are allotted or 0 RBs are allotted. For example, in a single-user scenario, the user is allotted all 50 RBs in CC1 and either 50 or 0 RBs in CC2. In a two-user scenario, the 50 RBs in CC1 are equally shared among the two UEs, i.e., both UEs get 25 RBs each in CC1. In CC2, either 50 or 0 RBs are allotted to the UEs in a round robin way. The Res50 results in a compact CC selection vector set $\mathcal{V_{\text{b}}}$ of size $2^{MK_\text{b}}$ (each bit in every $MK_\text{b}$-length binary vector $\boldsymbol{\alpha}_{\text{b}}$ in $\mathcal{V_{\text{b}}}$ indicates whether a CC is activated or not for the UEs). However, this coarse resolution of RB allocation does not give high throughputs. This is because if the RL framework finds that allocation of all 50 RBs in CC2 (and associated increased transmit power) is not favorable due to corresponding increase in SI power, the only option it has is to reduce the RB allocation to zero which essentially amounts to not using the SCC. This leads to inefficient use of RBs and loss in throughput. This can be alleviated using finer resolution of RB allocation which allows a better distribution of RBs among users.

Fine resolution of RB allocation (e.g., Res25, Res10) is realized by increasing the number of bits that represent the state of CC2 in the CC selection vector. In Res50, one bit defines the status of CC2. Instead, say, $n_m$ bits are used for CC2 in the CC selection vector. Let $\boldsymbol{\alpha}_{\text{b},\text{ext}}$ be the $(M+n_\text{m}-1)K_\text{b}$-length extended binary CC selection vector and $\mathcal{V_{\text{b},\text{ext}}}$ be the corresponding vector set of size $2^{(M+n_\text{m}-1)K_\text{b}}$.  In the extended CC selection vector, CC1 is represented by 1 bit and CC2 is represented by $n_\text{m}$ bits. If all the $n_m$ bits are set to $0$, then the number of RBs allocated in CC2 is $0$. Instead, if one or more of the $n_\text{m}$ bits are set to 1, then the number of RBs allocated in CC2 for each bit set as 1 is $n_{a}$, where
\begin{equation}
\hspace{-2mm}
n_a = 
\begin{cases}
\frac{N_\text{max}}{Kn_\text{m}}, & \hspace{-4mm} \text{for $N_\text{max}\hspace{-2mm}\mod{Kn_\text{m}}=0$} \\
\text{share RBs among } \\ 
\text{set $n_\text{m}$ bits in $\boldsymbol{\alpha}_{\text{b},\text{ext}}$} \\
\text{using round robin,}
& \hspace{-4mm} \text{for $N_\text{max}\hspace{-3mm}\mod{Kn_\text{m}} \neq 0$},
\end{cases}
\label{finer_RB}
\end{equation}
where $N_\text{max}$ is the total number of RBs in a CC. The total number of RBs allocated in CC2 is $n_1n_a$, where $n_1$ is the number of bits set to 1 of the $n_\text{m}$ bits. For Res25, $n_\text{m}$ = 2, and for Res10, $n_{m}=5$. 

\begin{table}
\centering
\begin{tabular}{|l|c|}
\hline
$\mathbf{Parameter} $ & $\mathbf{Value}$  \\ \hline \hline
Max. UL Tx. power at UEs & 27 dBm \\ \hline
Total number of CCs & 2 \\ \hline
Carrier frequencies & mid-band (3.5 GHz) \\ \hline
Bandwidth per CC & 10 MHz  \\ \hline
Total number of RBs in a CC & 50 \\ \hline
Cell coverage radius & 50 m  \\ \hline
Number of cycles per episode & 100  \\ \hline
Learning rate  & 0.01   \\ \hline
Replay buffer size  & 500  \\ \hline
Discount factor ($\gamma$) & 0.99 \\ \hline  Exploration rate ($\epsilon$) & 0.9 \\ \hline
\end{tabular}
\vspace{4mm}
\caption{System parameters used in the simulations.}
\label{system_Param}
\vspace{-8mm}
\end{table}

\section{Results and discussions} 
\label{sec:sec5}
In this section, we present the simulation results for the sum throughput maximization problem that we solved using the RL framework described in the previous sections. We first present results for a single cell scenario ($N_\text{g}=1$) with single user and multiple users ($K=1,2$). We illustrate how the network adapts to dynamic changes in the environment (e.g., an existing UE in a multiuser network exits). We also present results for a multi-cell scenario ($N_\text{g}=2)$. The coverage radius of a cell is taken as 50 m. We consider that there are two CCs. The first CC (CC1) is the PCC and the second CC (CC2) is the SCC. The CCs operate in the mid-band frequency range (3.5 GHz). We consider that the SI is caused by the second harmonic of CC2 (SCC). The maximum transmit power at the UEs is 27 dBm (0.5 W). The power gain of the PA at the UEs is taken to be $G=30$ dB and $c_2$ in the non-linearity model is taken to be 0.1417. 
Noise power $\sigma^2$ is -70 dBm, average number of bits per burst of data $\hat{q}^{i}, \forall i$ is taken as 1000, and delay QoS requirement   $D_\text{QoS}$ is 0.15 sec\cite{stat_1}. 
Tables \ref{RLarchtable} and \ref{system_Param} list the system parameters used in the simulations. 

\subsection{Single user scenario}
\begin{figure}[t]
\centering
\includegraphics[width=9.1cm,height=7.5cm]{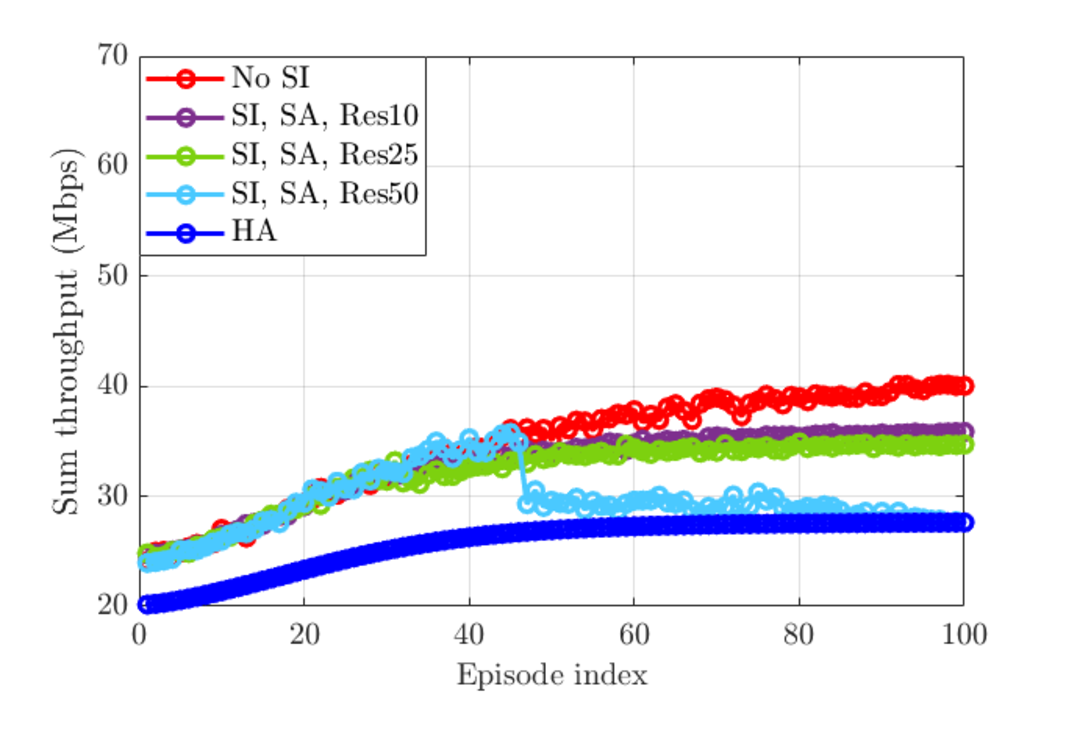}
\vspace{-8mm}
\caption{Sum throughput performance with single gNB and single UE.}
\label{Single_UE_sumrate}
\vspace{-4mm}
\end{figure}

Here, we consider a network environment which consists of one gNB and one UE associated with it. 
The UE is located at a distance of 25 m from the gNB. The RB bandwidth is 180 KHz. The values of the parameters in the penalty function defined in (\ref{R_b}) are taken to be ($\theta^1_{1}$, $\theta^1_{2}, \Omega^1$) = (-100 dBm, -95 dBm, $10^7$). The self coupling loss between the UE transmitter and UE receiver is taken to be $L_\text{c}^1=35$ dB. In Fig. \ref{Single_UE_sumrate}, we show the evolution of the achieved sum throughput as a function of episode index for the following five different cases:
\begin{itemize}
\item {\bf `HA'} indicates hard avoidance, where CC2 is effectively unused (i.e., no RBs are allocated on CC2), because it would cause SI if used. 
\item {\bf `No SI'} 
indicates the scenario where CC2 is used and it does not cause SI. Since $p_\text{SI}$ is zero in this case, the second term (penalty component) in the reward function in (\ref{rew_fn}) vanishes and the reward function value is only due to the first term (sum rate) in (\ref{rew_fn}). The resolution of RB allocation is 50, i.e., RBs are allocated in units of 50 (implying that the possible number of RBs allocated is 50 or 0).
\item {\bf `SI, SA, Res50'} indicates the scenario where CC2 is used and it causes SI. Allocation is made based on the proposed soft avoidance scheme using the reward function in (\ref{rew_fn}) which includes the SI-dependent penalty term given by (\ref{R_b}). The resolution of RB allocation is 50. 
\item {\bf `SI, SA, Res25'} scenario is same as that of the previous case, except that the RB allocation is done in units of 25 (i.e., possible number of RBs allocated: 50/25/0).  
\item {\bf `SI, SA, Res10'} scenario is same as those of the previous two cases, except that the RB allocation is done is units of 10 (i.e., possible number of RBs allocated: 50/40/30/20/10/0).  
\end{itemize}

\begin{figure*}[htbp]
\subfloat[No. of CCs allocated to UE vs episode index]{\includegraphics[width=4cm,height=4cm]{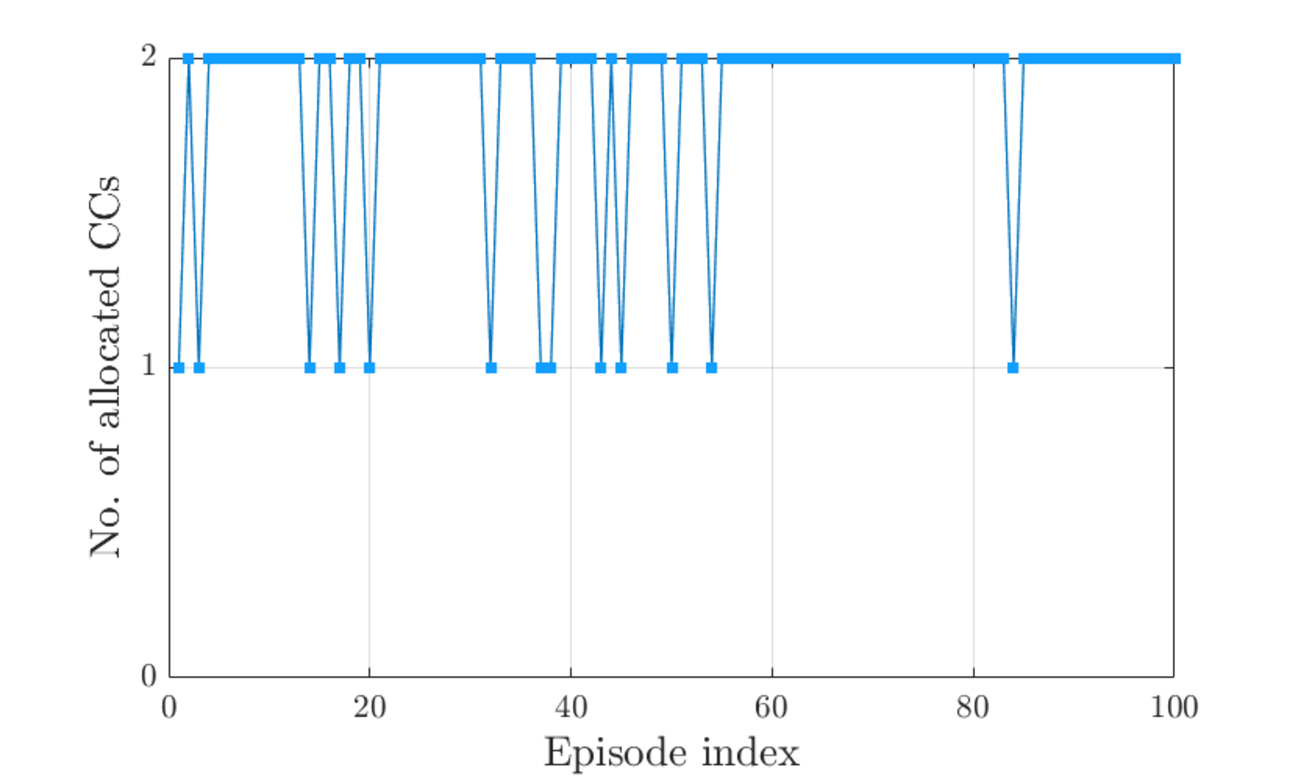}}\hfill
\subfloat[No. of allocated RBs in CC1 vs episode index]{\includegraphics[width=4cm,height=4cm]{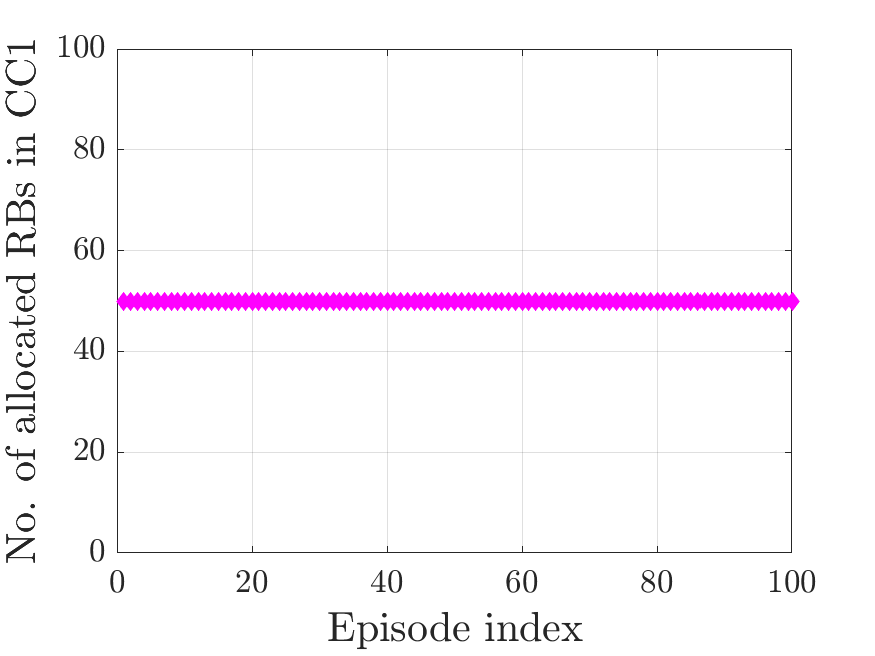}}\hfill
\subfloat[No. of RBs allocated in CC2 vs episode index]{
\label{subfig:SU_d}
\includegraphics[width=4cm,height=4cm]{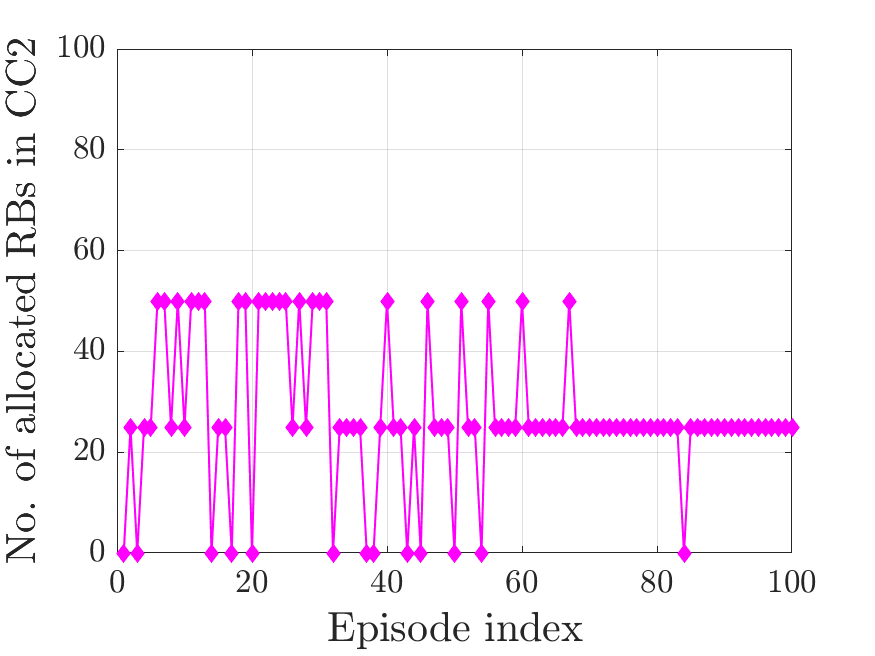}}\hfill
\subfloat[Total no. of RBs allocated to UE vs Episode index]{\includegraphics[width=4cm,height=4cm]{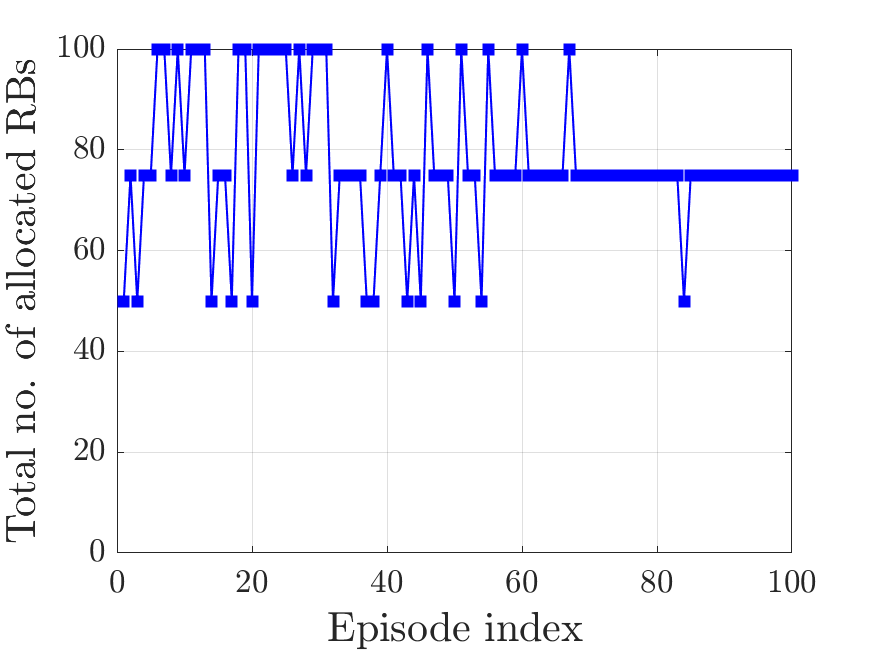}}\vfill
\subfloat[Total Tx. power at UE vs episode index]{\includegraphics[width=4cm,height=4cm]{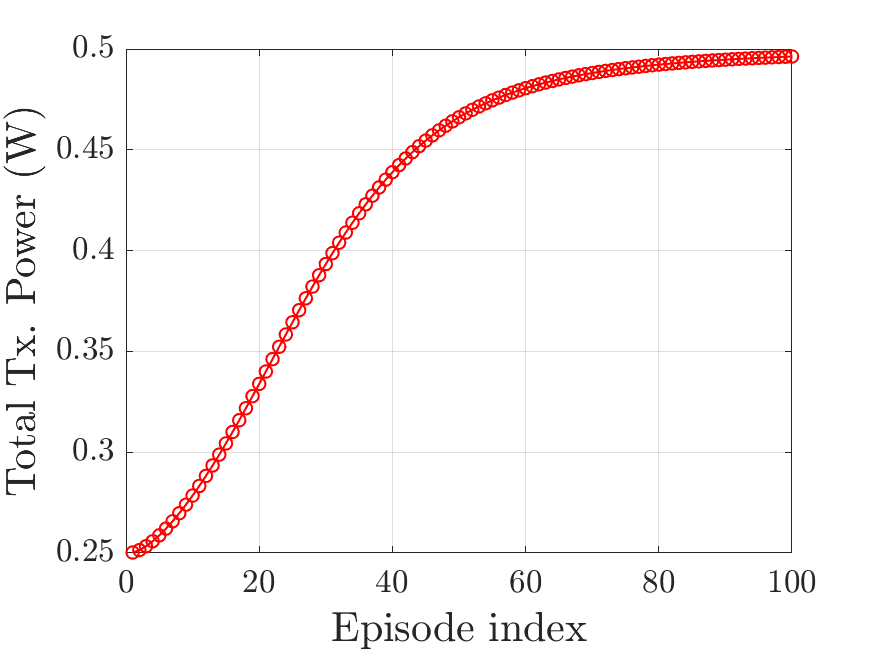}}\hfill 
\subfloat[Tx. power on CC1 vs episode index]{\includegraphics[width=4cm,height=4cm]{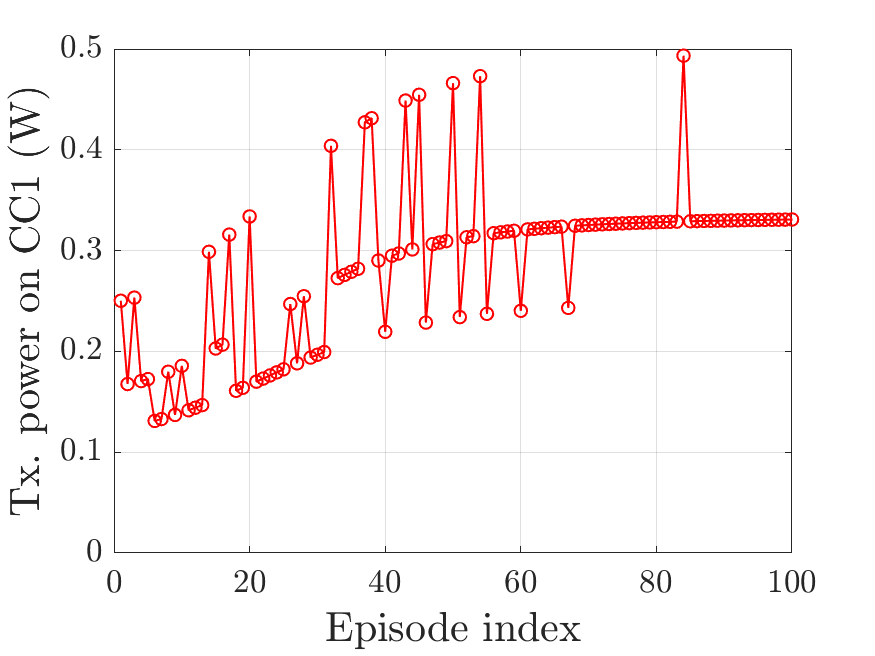}}\hfill
\subfloat[Tx. power on CC2 vs episode index]{\includegraphics[width=4cm,height=4cm]{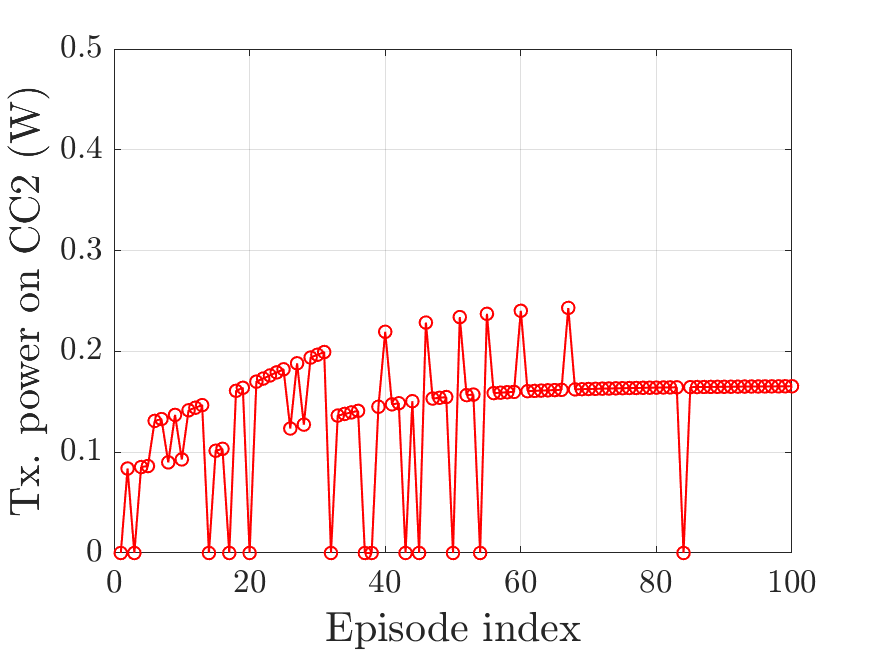}}\hfill
\subfloat[$p_\mathrm{SI}$ due to CC2 vs episode index]{\includegraphics[width=4cm,height=4cm]{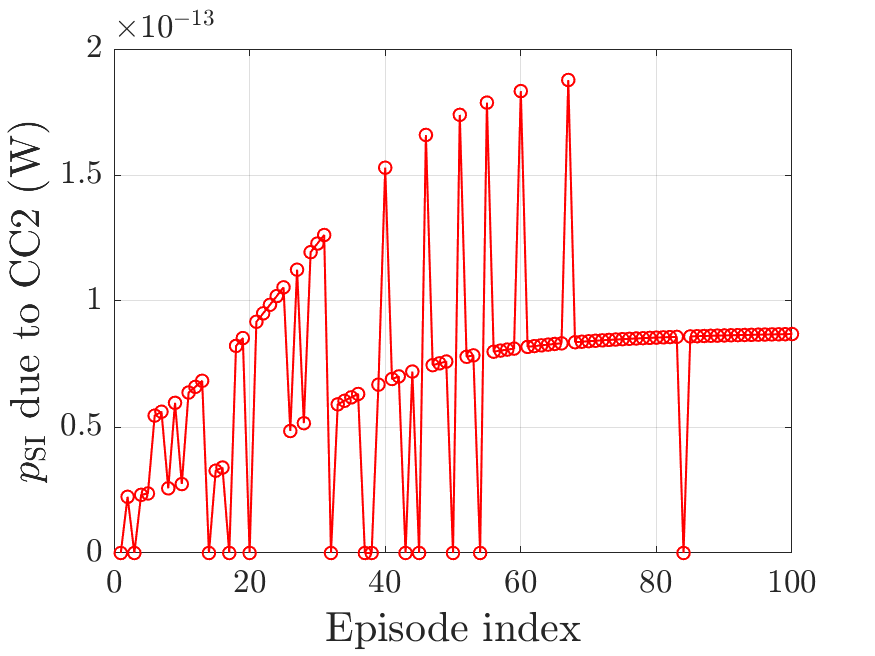}}\vfill
\caption{Traces of evolution of the variables in the optimization for the case of `SI, SA, Res25' with single gNB and single UE.}
\label{Single_UE_traces}
\vspace{-4mm}
\end{figure*}

In Fig. \ref{Single_UE_sumrate}, it can be seen that `HA' gives the least throughput and `No SI' gives the maximum throughput. This is because `HA' avoids the use of SCC that would cause SI. This leaves the UE to use only the PCC, and hence the throughput contribution is only due to PCC. On the other hand, in the case of `No SI', both PCC and SCC are active and there is no SI. This leads the throughput to accrue from both PCC and SCC. While the `No SI' case can serve to provide an upper bound on the achieved throughput, the system has to work with allocation of carriers that will encounter the SI issue. The proposed SA scheme addresses this requirement. 

For the case of `SI, SA, Res50', we can observe in Fig. \ref{Single_UE_sumrate} that the network initially allocates RBs on both CCs and hence the sum throughput grows imitating that of the 'No SI' case. As more RBs are allocated, the transmit power increases correspondingly, which in turn increases SI. The reward function penalizes this increased SI and the agent responds by not allocating any RBs on the CC causing the SI (because RB allocation is done on an all or none basis in Res50). This results in a dip in the overall sum throughput and the performance now starts imitating that of `HA'. This behavior is alleviated through fine resolution of RBs, as can be seen from the throughput behavior for the cases of `SI, SA, Res25' and `SI, SA, Res10'. In these Res25 and Res10 cases, as in the case of Res50, the throughput initially grows through RBs assigned on both CCs, imitating that of the `No SI' case. But there is no dip in the throughput later as witnessed in the case of Res50. This is because, with fine resolution, instead of resorting to allocating no RBs altogether, few RBs can be allotted. This non-zero RB allocation is able to sustain the throughput without deteriorating to the level of HA throughput. Also, Res10 is found be achieve better throughput compared to Res25, implying that finer the resolution better could be the throughput. Thus, the SA scheme with the proposed reward function and fine resolution of RB allocation is able to achieve better throughput performance compared to HA.

\begin{figure}
\centering
\includegraphics[width=9.1cm,height=7.5cm]{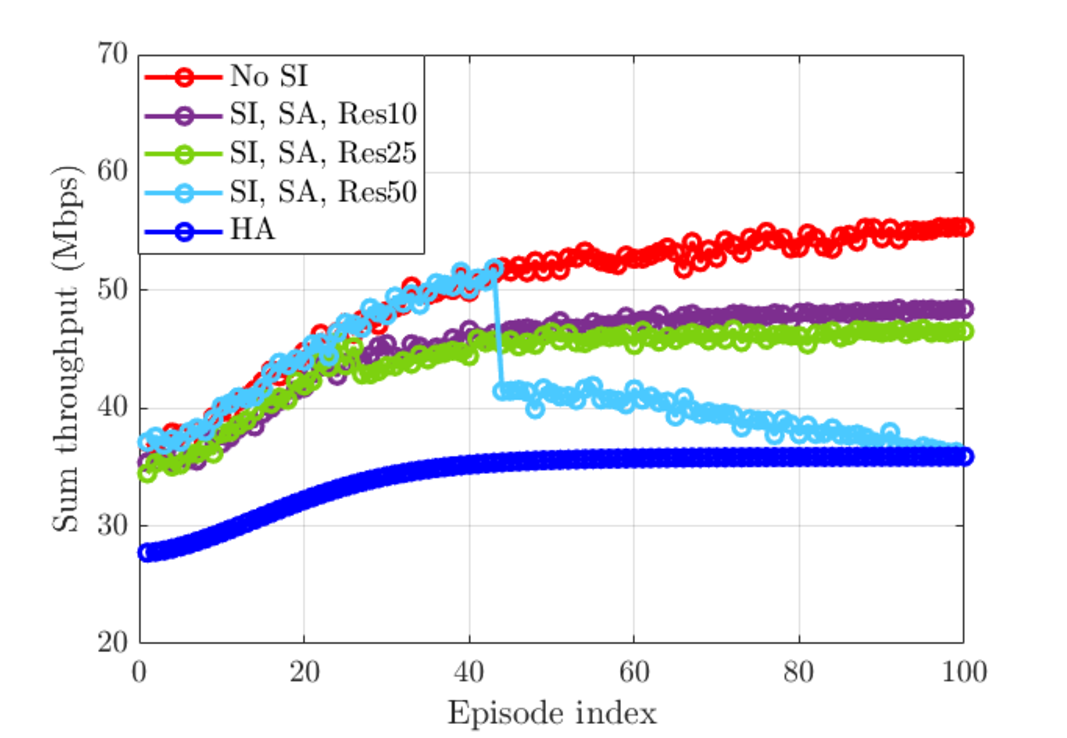}
\caption{Sum throughput performance with single gNB and two equidistant UEs.}
\label{Eq_multi_UE_sumrate}
\vspace{-2mm}
\end{figure}

\begin{figure*}[htbp]
\subfloat[No. of CCs allocated to UE1 vs episode index]{\includegraphics[width=4cm,height=4cm]{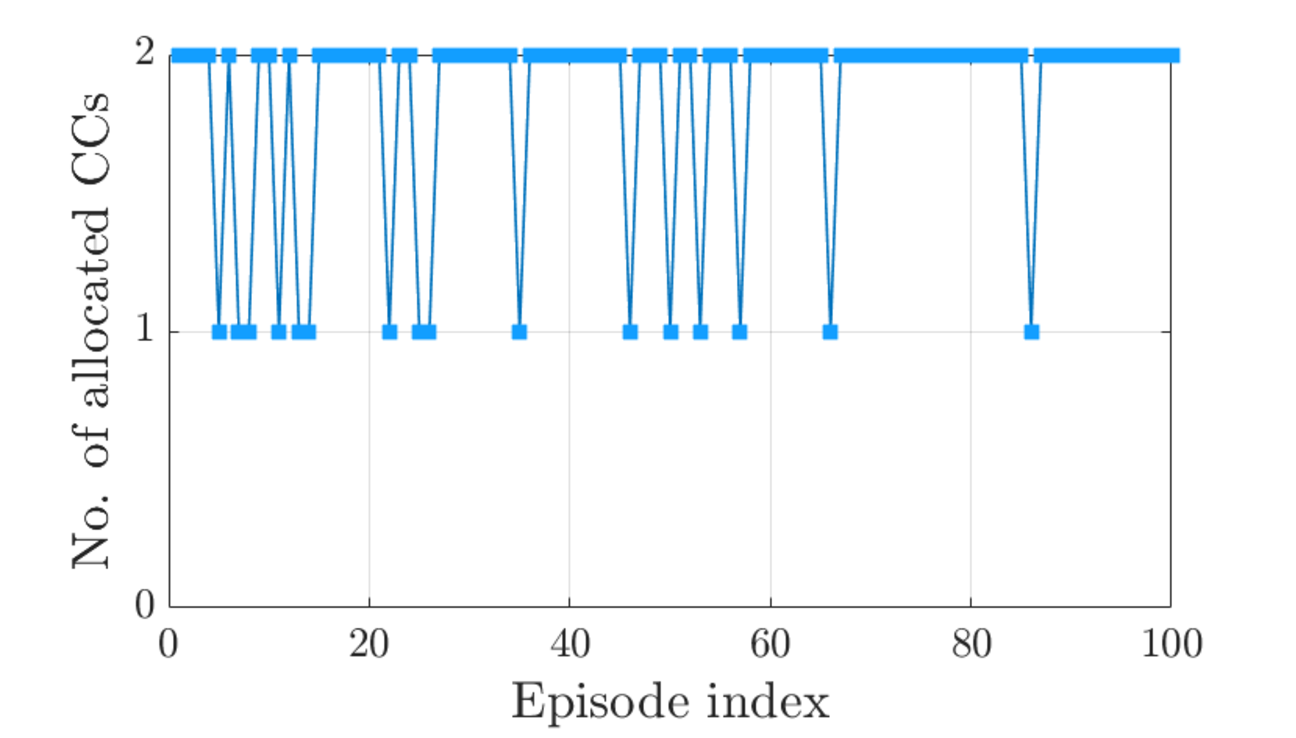}}\hfill
\subfloat[Allocated no. of RBs to UE1 in CC1 vs episode index]{\includegraphics[width=4cm,height=4cm]{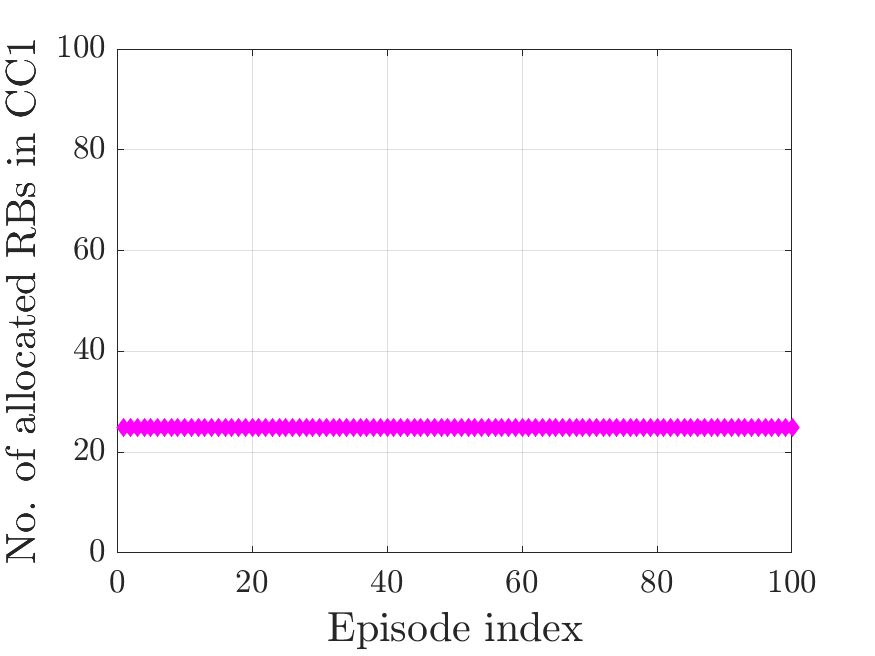}}\hfill
\subfloat[Allocated no. of RBs to UE1 in CC2 vs episode index]{\label{subfig:MU_eq_d}
\includegraphics[width=4cm,height=4cm]{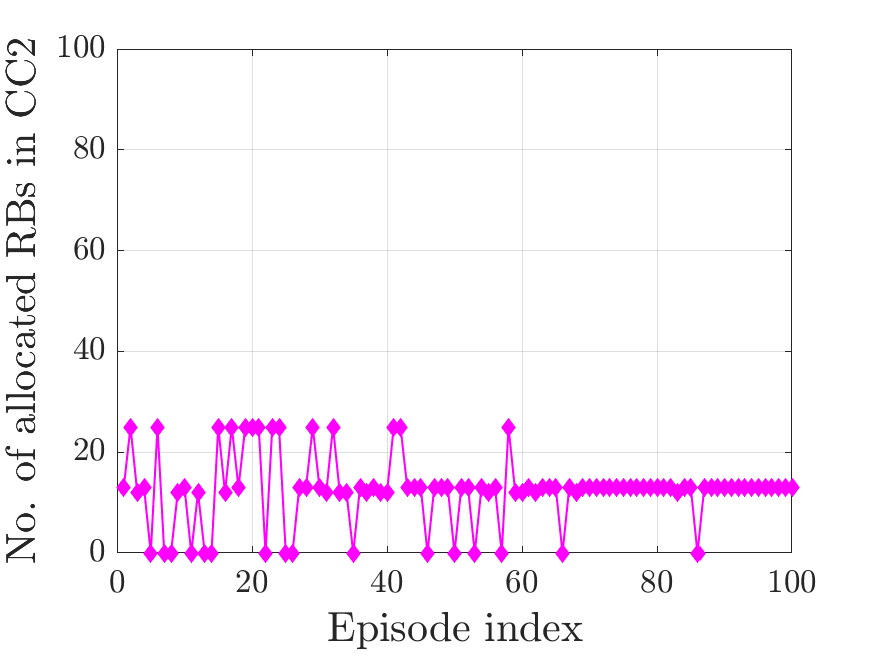}}\hfill
\subfloat[Total no. of RBs allocated to UE1 vs episode index]{\includegraphics[width=4cm,height=4cm]{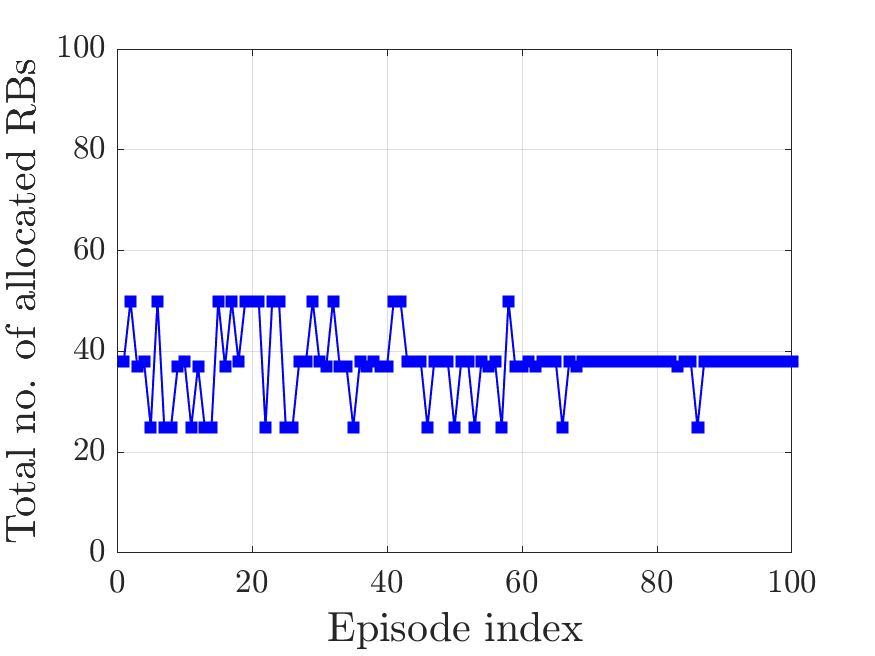}}\vfill
\subfloat[Total Tx. power at UE1 vs episode index]{\includegraphics[width=4cm,height=4cm]{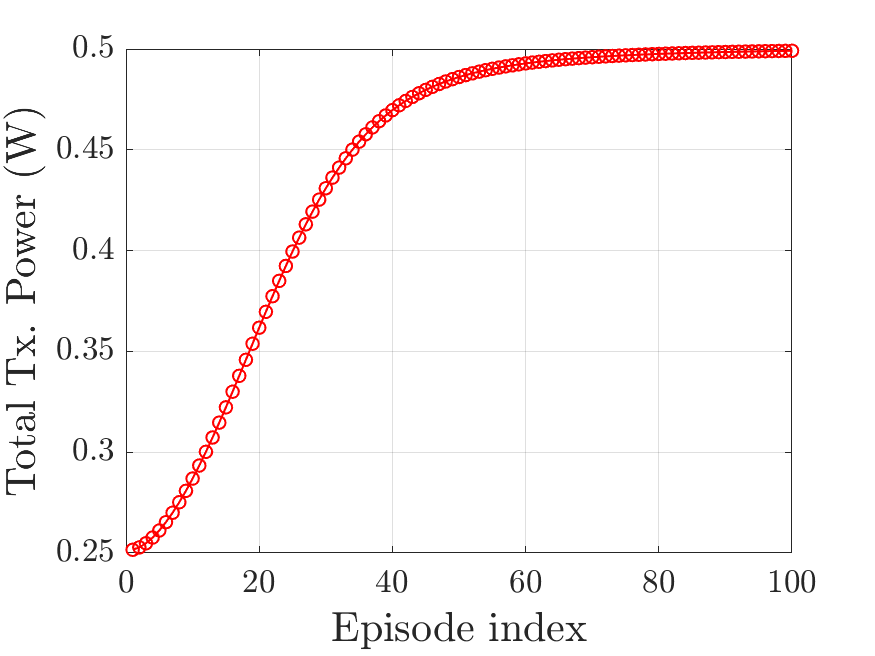}}\hfill
\subfloat[Tx. power on CC1 at UE1 vs episode index]{\includegraphics[width=4cm,height=4cm]{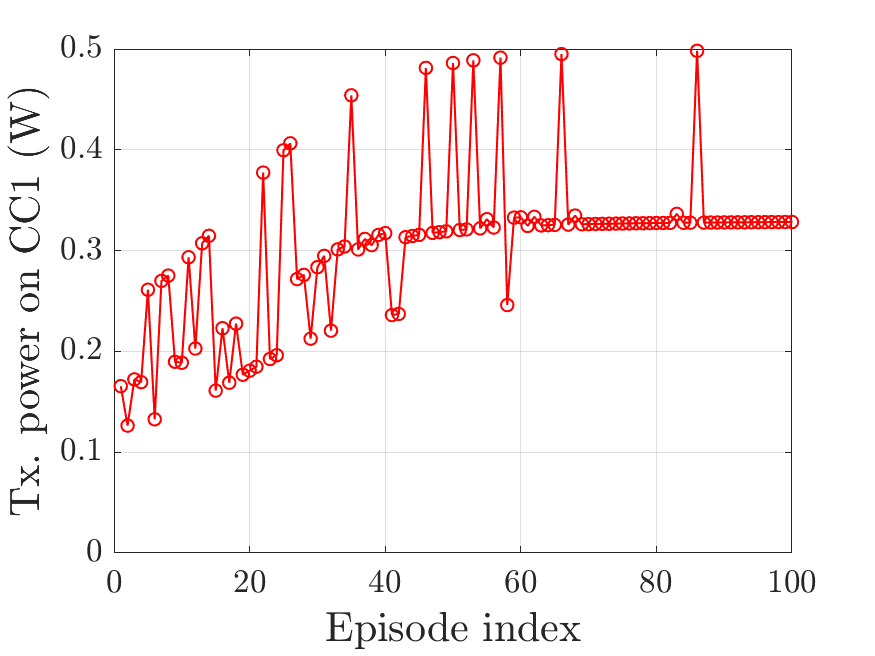}}\hfill
\subfloat[Tx. power on CC2 at UE1 vs episode index]{\includegraphics[width=4cm,height=4cm]{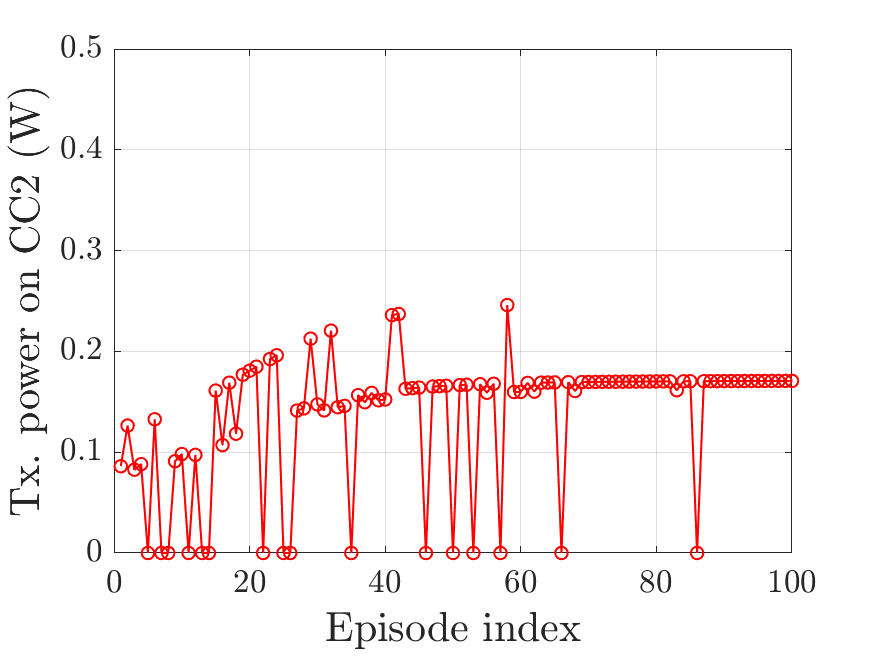}}\hfill
\subfloat[$p_\mathrm{SI}$ due to CC2 at UE1 vs episode index]{
\includegraphics[width=4cm,height=4cm]{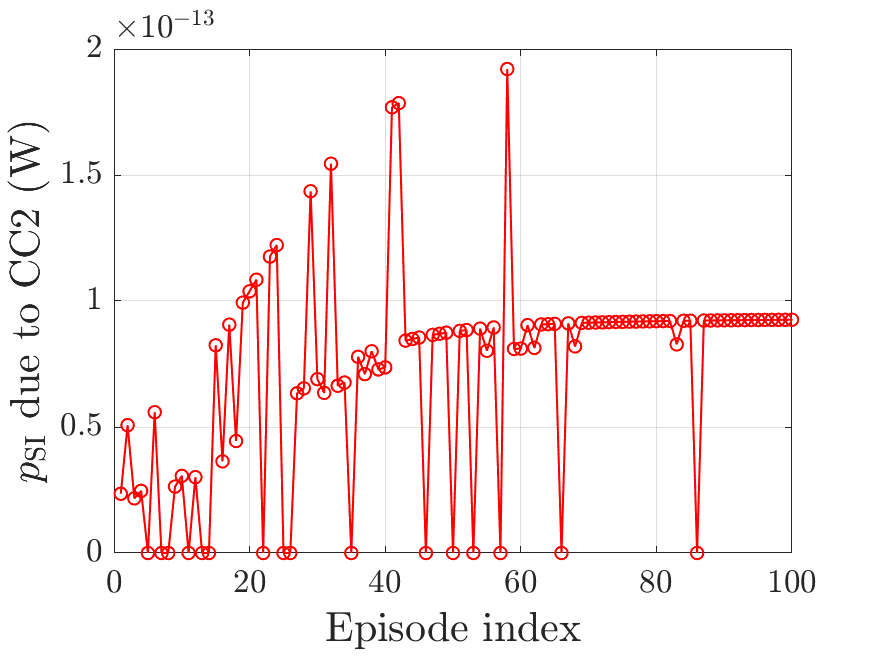}}
\caption{Traces of evolution of the variables in the optimization at UE1 for the case of `SI, SA, Res25' with single gNB and two equidistant UEs.}
\label{Eq_multi_UE1_traces}
\vspace{-4mm}
\end{figure*}

Figures \ref{Single_UE_traces} (a)-(h) illustrate the traces of evolution of the variables in the optimization as a function of episode index for the case of `SI, SA, Res25'. These variables include number of allocated CCs, number of RBs allocated in CC1 and CC2, total number of allocated RBs, total transmit power, transmit power allocated to CC1 and CC2, and the corresponding $p_\mathrm{SI}$ values due to CC2. 

\subsection{Multiuser scenario}
Here, we present the performance results for a system with a single gNB and multiple users associated with it. For clarity of exposition, we consider two UEs in the network. We consider two cases of UEs placement. In the first case, both UEs are equidistant from the gNB. In the second case, the UEs are at different distances from the gNB. 

\begin{figure}[t]
\centering
\includegraphics[width=9.1cm,height=7.4cm]{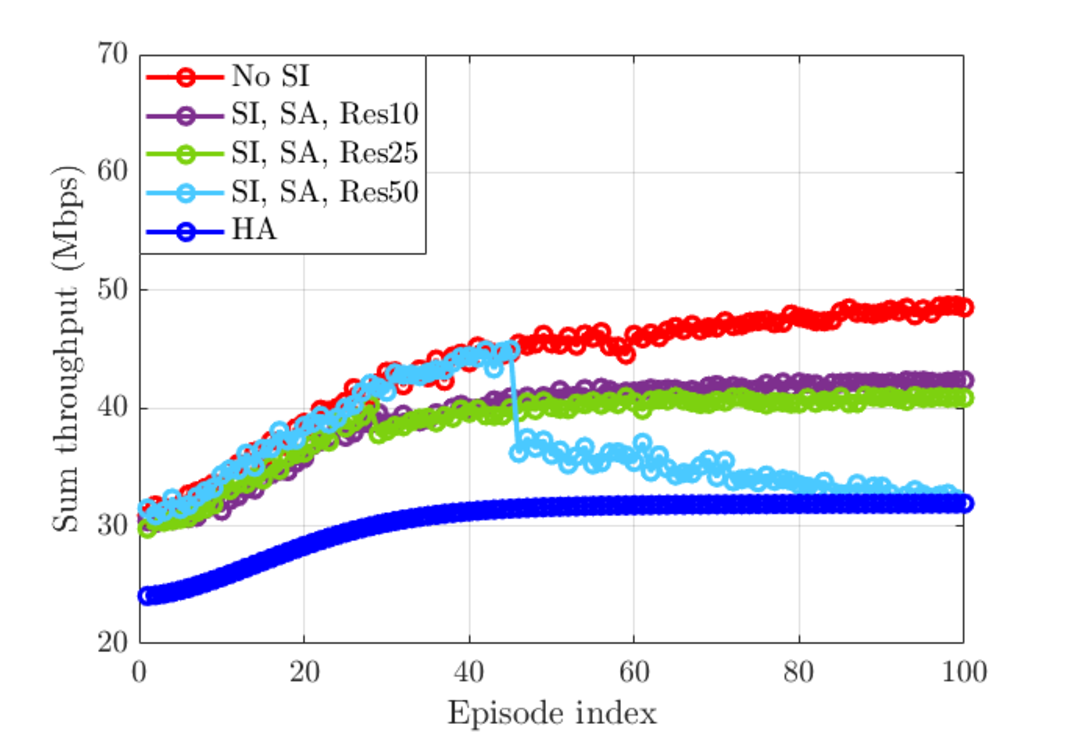}
\caption{Sum throughput performance with single gNB and two non-equidistant UEs.}
\label{Non_Eq_multi_UE_sumrate}
\vspace{-4mm}
\end{figure}

\subsubsection{Equidistant UEs}
In the equi distant UEs setup, both the UEs are located at 25 m from the gNB. The penalty function parameters are taken as ($\theta^i_{1}$, $\theta^i_{2}, \Omega^i$) = (-100 dBm, -95 dBm, $0.625\times 10^7$) for $i=1,2$. The self coupling losses at the UEs are taken as $L^i_\text{c}=35$ dB for $i=1,2$. Figure \ref{Eq_multi_UE_sumrate} shows  the sum throughput performance for the five scenarios considered earlier in Fig. \ref{Single_UE_sumrate}. As observed in the single UE results in Fig. \ref{Single_UE_sumrate}, here also we find that the proposed SA scheme with fine resolution of RB allocation achieves improved sum throughputs compared to that of HA. Also, comparing the results in Figs. \ref{Eq_multi_UE_sumrate} and \ref{Single_UE_sumrate}, it is interesting to note that the sum throughputs achieved in the two-user system are higher compared to those in the single-user system (e.g., sum throughputs of about 48 Mbps for two-user system versus about 36 Mbps for single-user system, both under `SI, SA, Res25' scenario). This can be explained as follows. In the single-user system, the number of allocated RBs per user is high (e.g., all 50 RBs in the PCC are allocated to the single user). 
However, the maximum available transmit power ($p_\text{max}$) is distributed among all the RBs, making the power per RB less. On the other hand, in the two-user system, the allocated RBs per user is less (e.g., each user gets only 25 RBs from PCC), whereas each user has its own $p_\text{max}$ power to distribute on these RBS, making the power per RB more. This results in higher SINRs and throughputs in favor of the two-user system. The above points can be observed in the traces of the evolution of the variables for UE1 shown in Fig. \ref{Eq_multi_UE1_traces} (a)-(h) for the case of `SI, SA, Res25' scenario.

\begin{figure*}[htbp]
\subfloat[No. of CCs allocated to UE1 vs episode index]{\includegraphics[width=4cm,height=4cm]{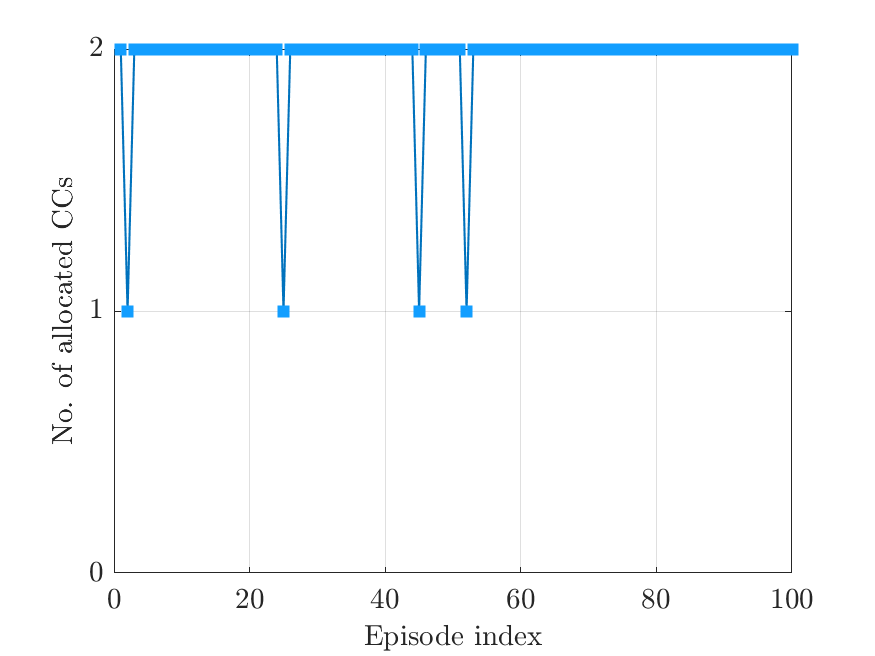}}\hfill
\subfloat[Allocated no. of RBs to UE1 in CC1 vs episode index]{\includegraphics[width=4cm,height=4cm]{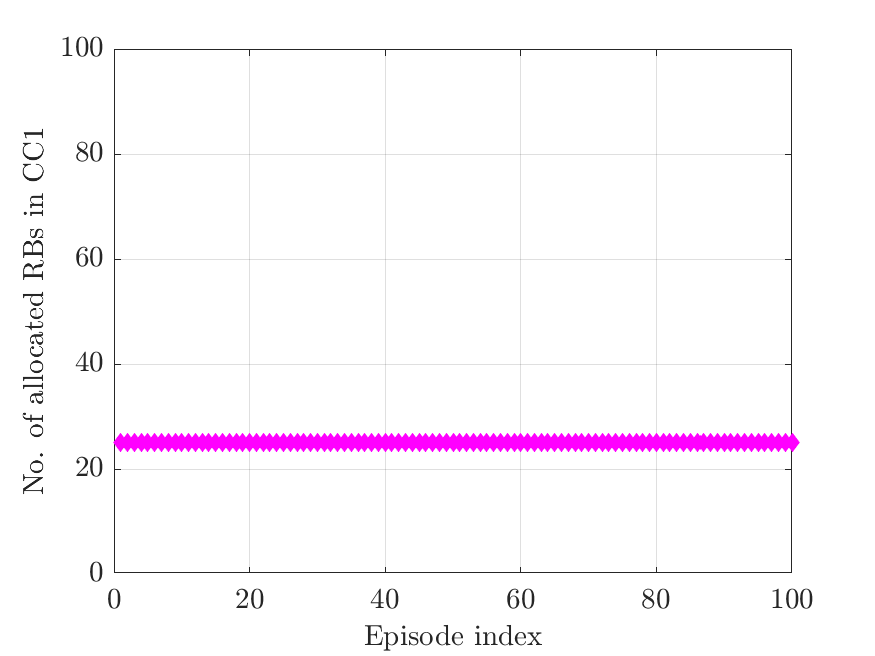}}\hfill
\subfloat[Allocated no. of RBs to UE1 in CC2 vs episode index]{\label{subfig:MU_neq_d}
\includegraphics[width=4cm,height=4cm]{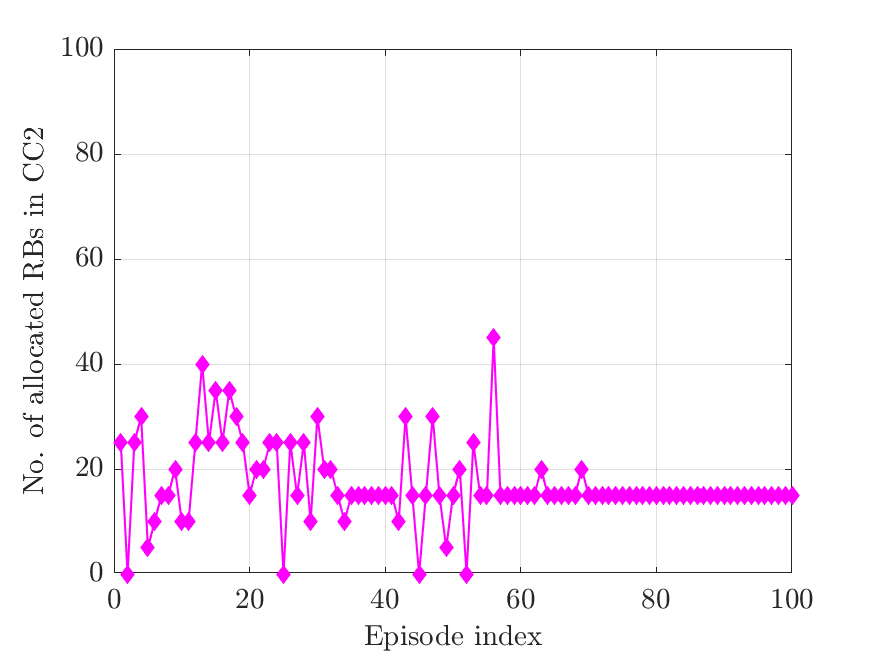}}\hfill
\subfloat[Total no. of RBs allocated to UE1 vs episode index]{\includegraphics[width=4cm,height=4cm]{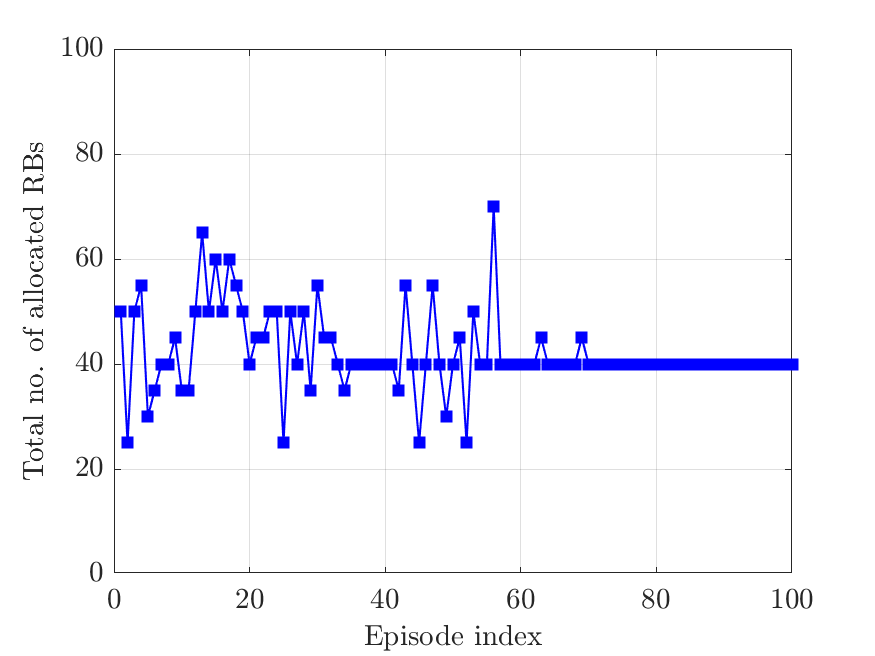}}\vfill
\subfloat[Total Tx. power at UE1 vs episode index]{\includegraphics[width=4cm,height=4cm]{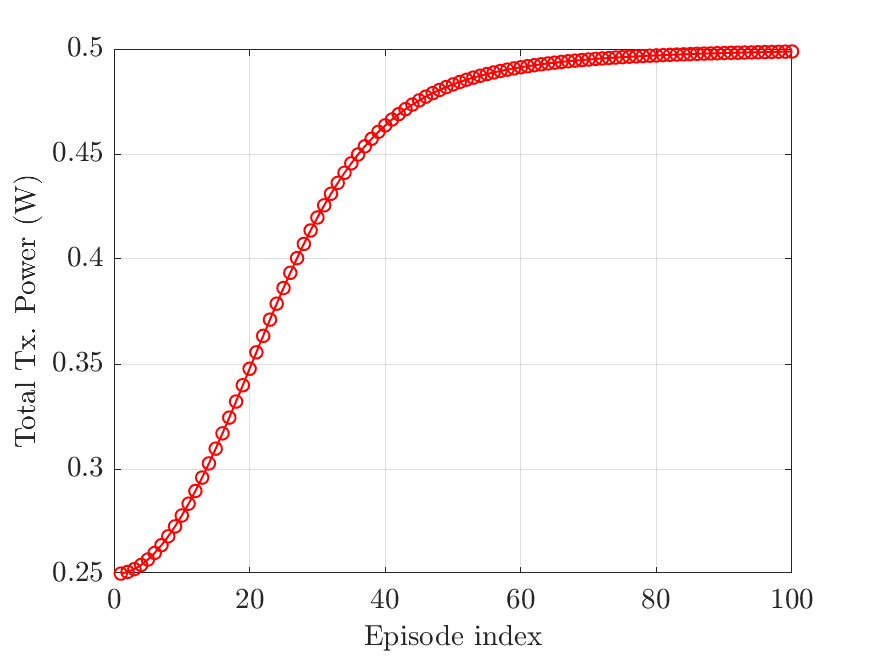}}\hfill
\subfloat[Tx. power on CC1 at UE1 vs episode index]{\includegraphics[width=4cm,height=4cm]{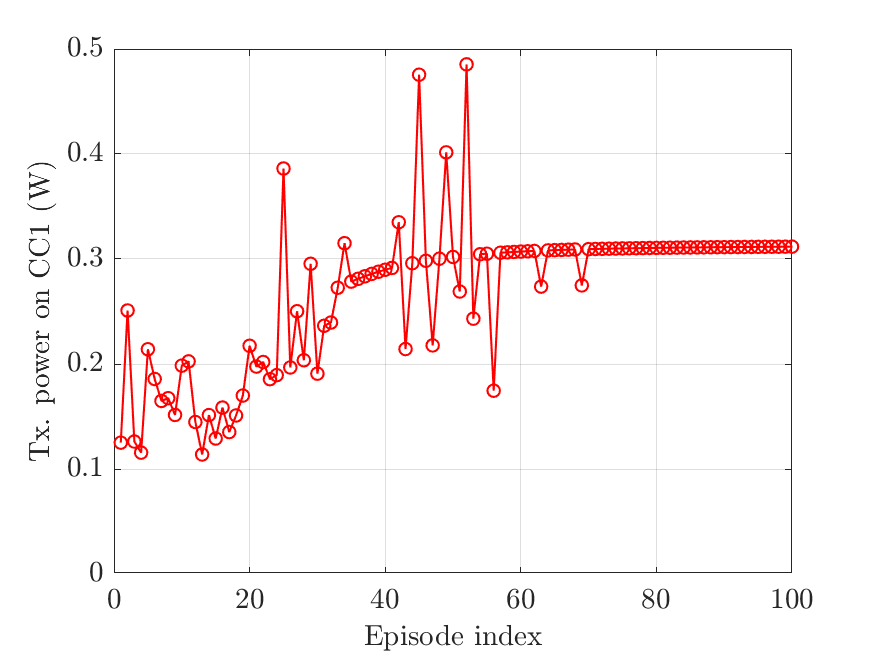}}\hfill
\subfloat[Tx. power on CC2 at UE1 vs episode index]{\includegraphics[width=4cm,height=4cm]{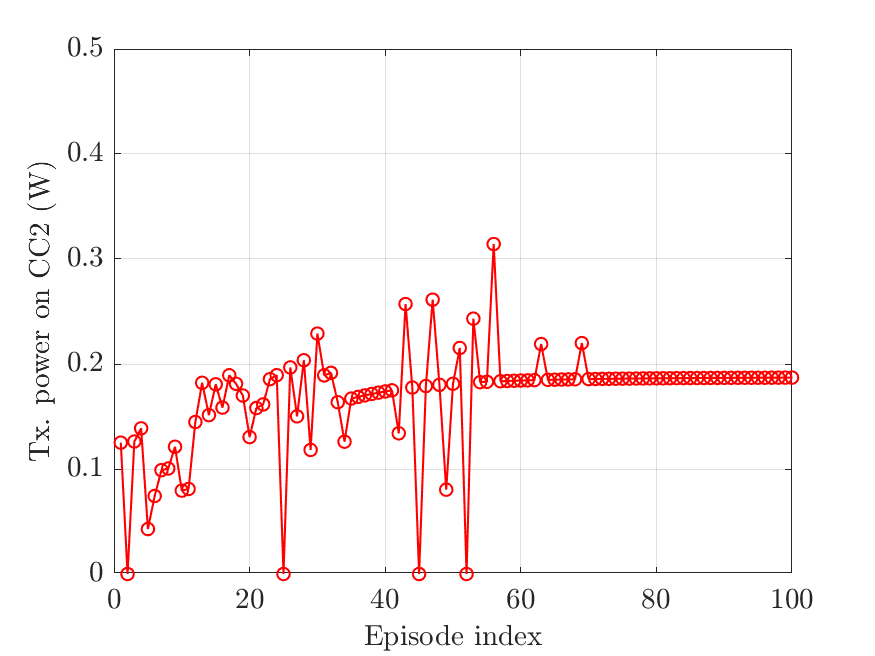}}\hfill
\subfloat[$p_\mathrm{SI}$ due to CC2 at UE1 vs episode index]{
\includegraphics[width=4cm,height=4cm]{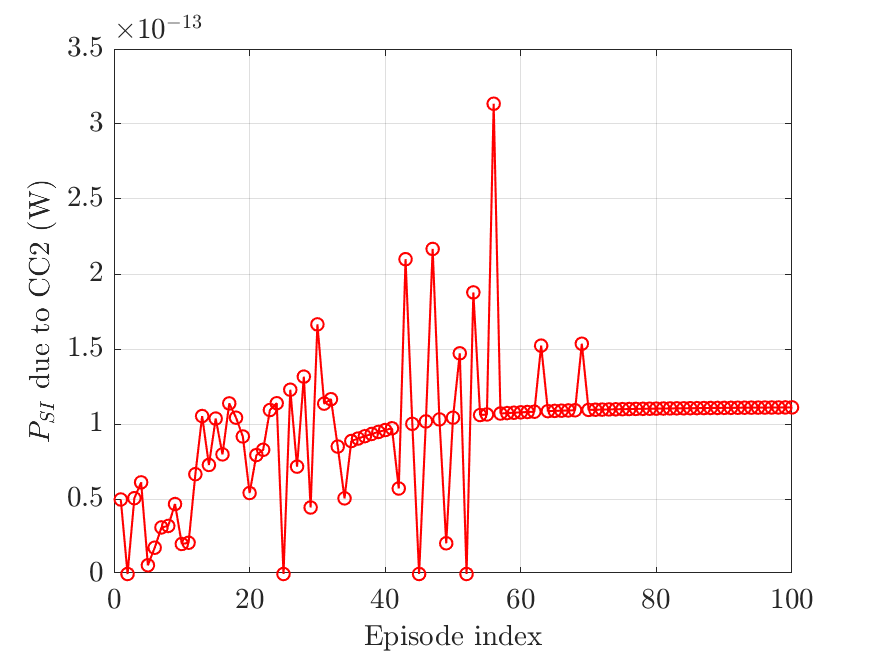}}\vfill
\subfloat[No. of CCs allocated to UE2 vs episode index]{\includegraphics[width=4cm,height=4cm]{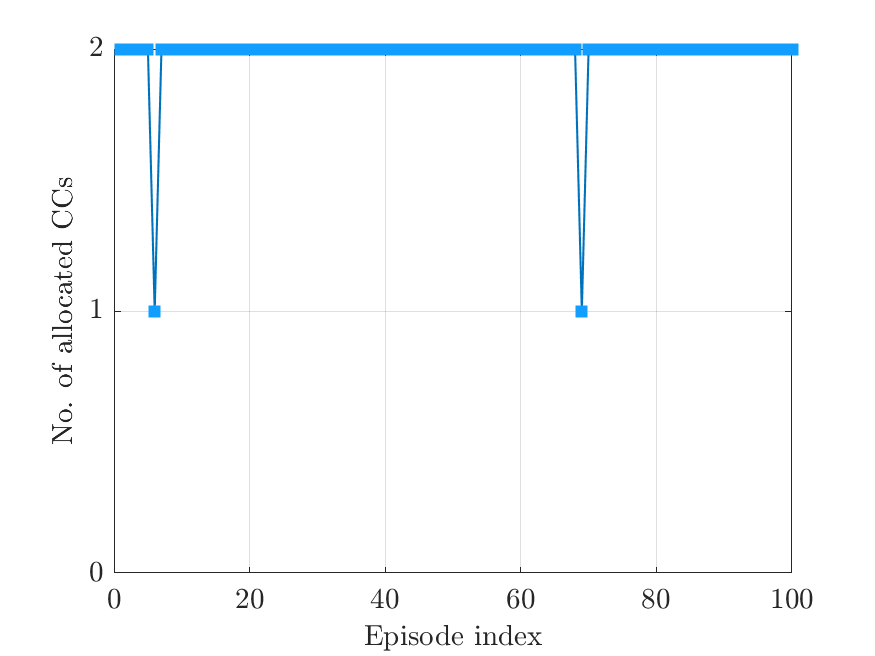}}\hfill
\subfloat[Allocated no. of RBs to UE2 in CC1 vs episode index]{\includegraphics[width=4cm,height=4cm]{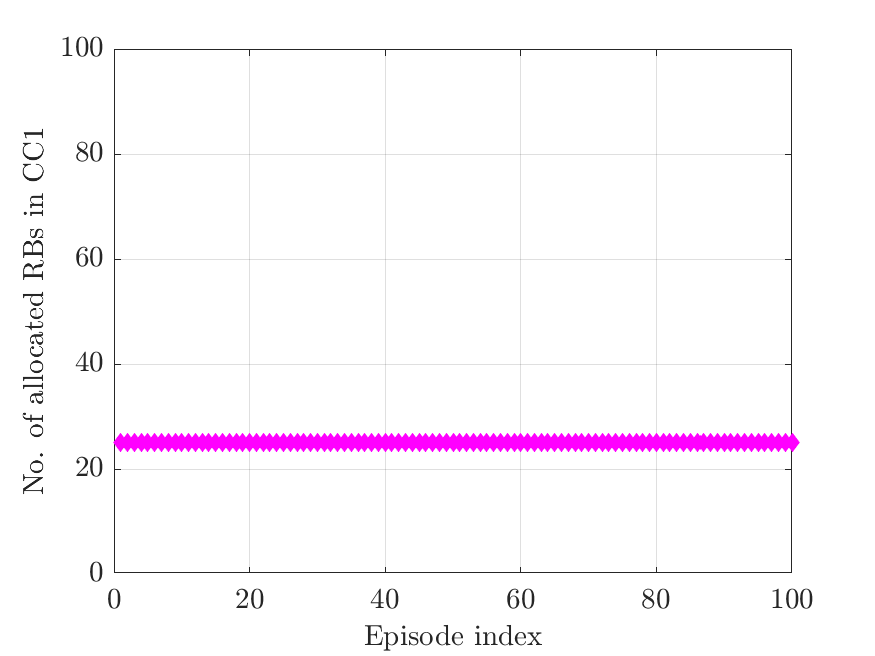}}\hfill
\subfloat[Allocated no. of RBs to UE2 in CC2 vs episode index]{\label{subfig:MU_neq_d}
\includegraphics[width=4cm,height=4cm]{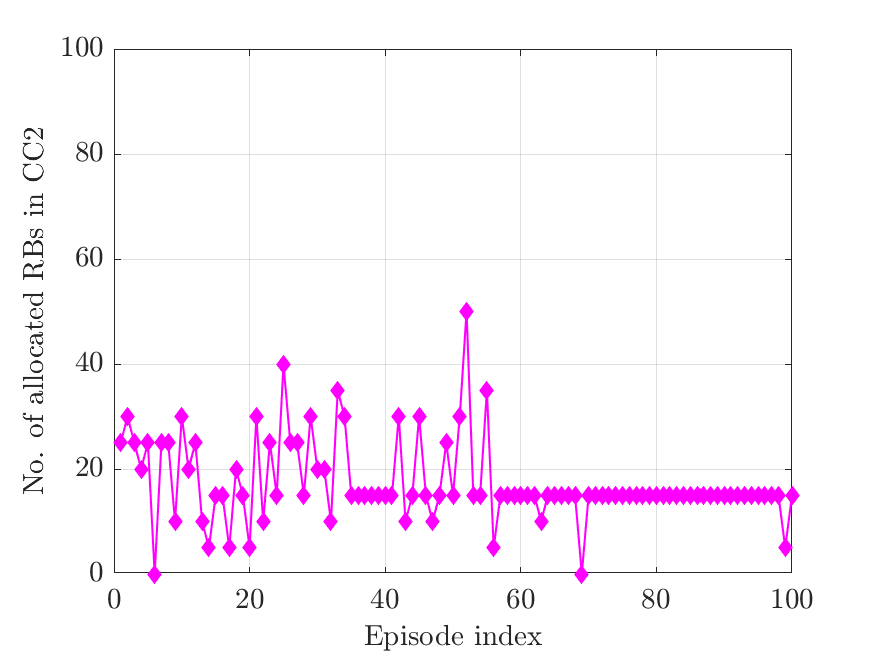}}\hfill
\subfloat[Total no. of RBs allocated to UE2 vs episode index]{\includegraphics[width=4cm,height=4cm]{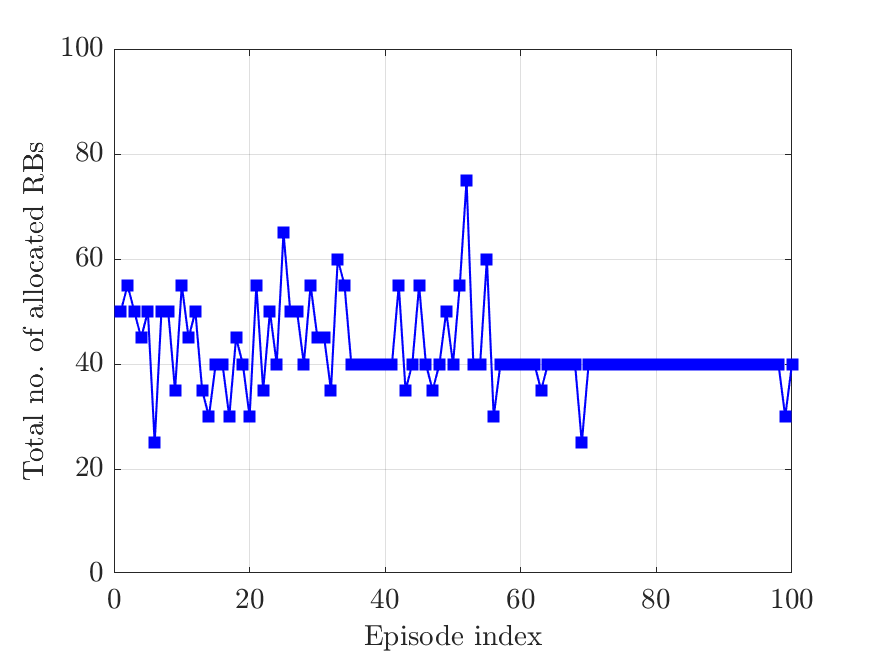}}\vfill
\subfloat[Total Tx. power at UE2 vs episode index]{\includegraphics[width=4cm,height=4cm]{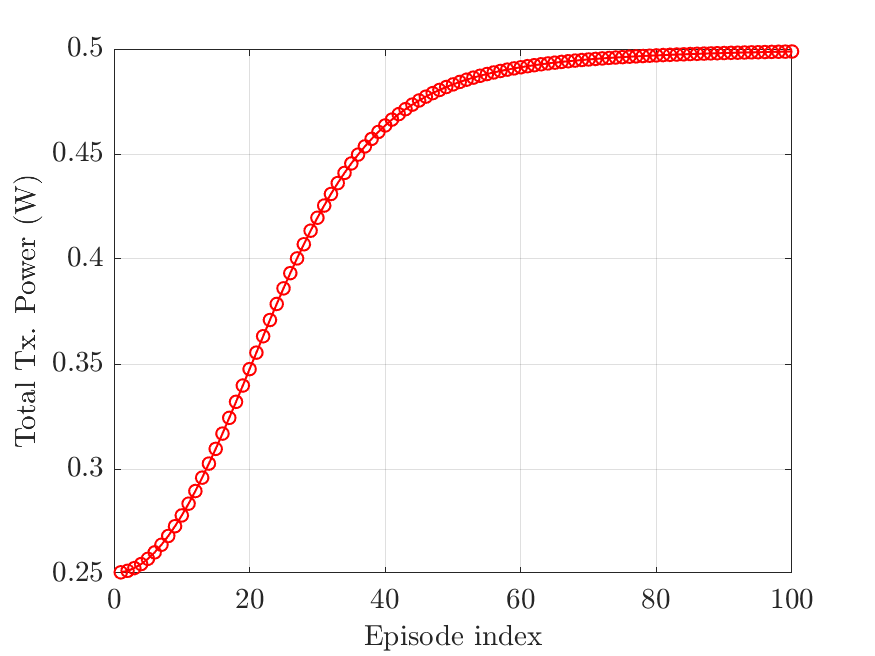}}\hfill
\subfloat[Tx. power on CC1 at UE2 vs episode index]{\includegraphics[width=4cm,height=4cm]{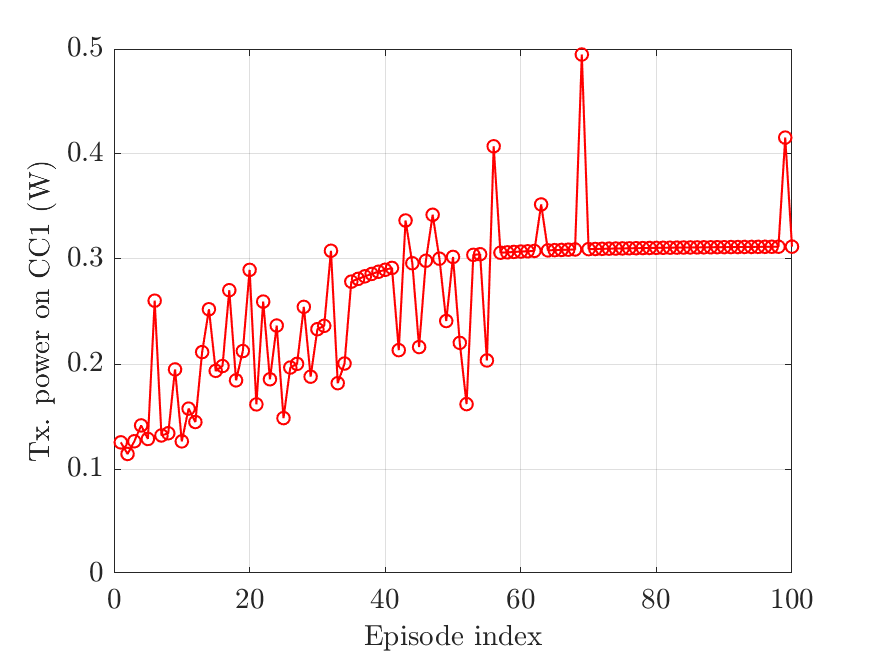}}\hfill
\subfloat[Tx. power on CC2 at UE2 vs episode index]{\includegraphics[width=4cm,height=4cm]{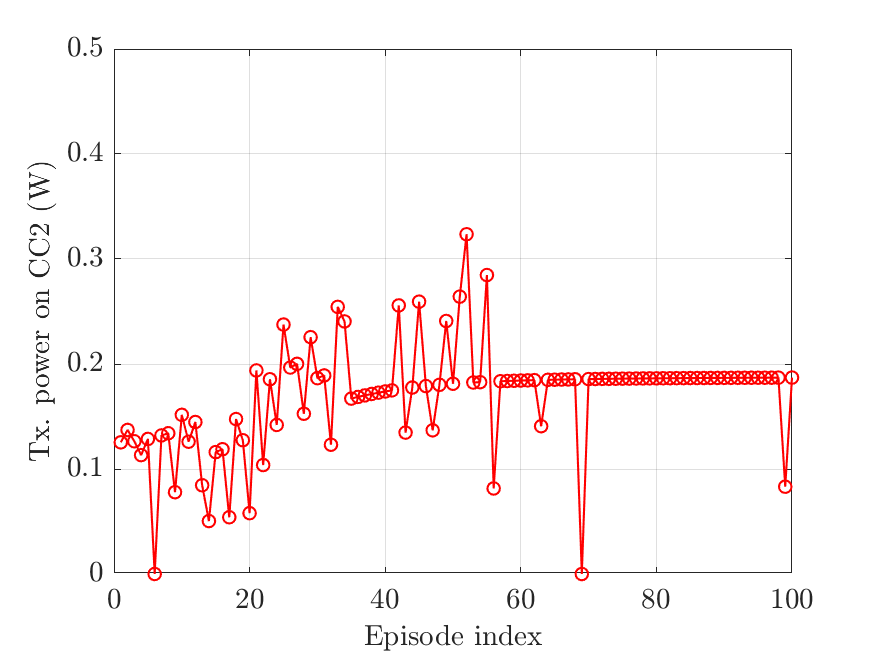}}\hfill
\subfloat[$p_\mathrm{SI}$ due to CC2 at UE2 vs episode index]{
\includegraphics[width=4cm,height=4cm]{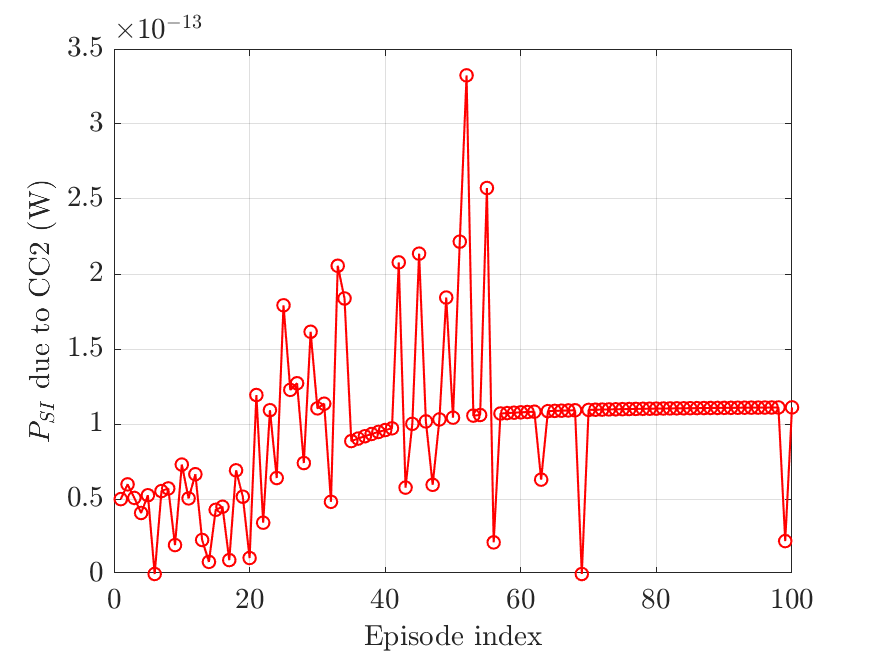}}
\caption{Traces of evolution of the variables in the optimization at UE1 and UE2 for the case of `SI, SA, Res10' with single gNB and two non-equidistant UEs.}
\label{Non_Eq_multi_UE1_UE2_traces}
\vspace{-4mm}
\end{figure*}

\begin{figure*}[h]
\subfloat[Sum throughput vs episode index]
{\label{subfig:Eq_Neq_multi_UEs_sumrate}\includegraphics[width=5.5cm,height=4cm]{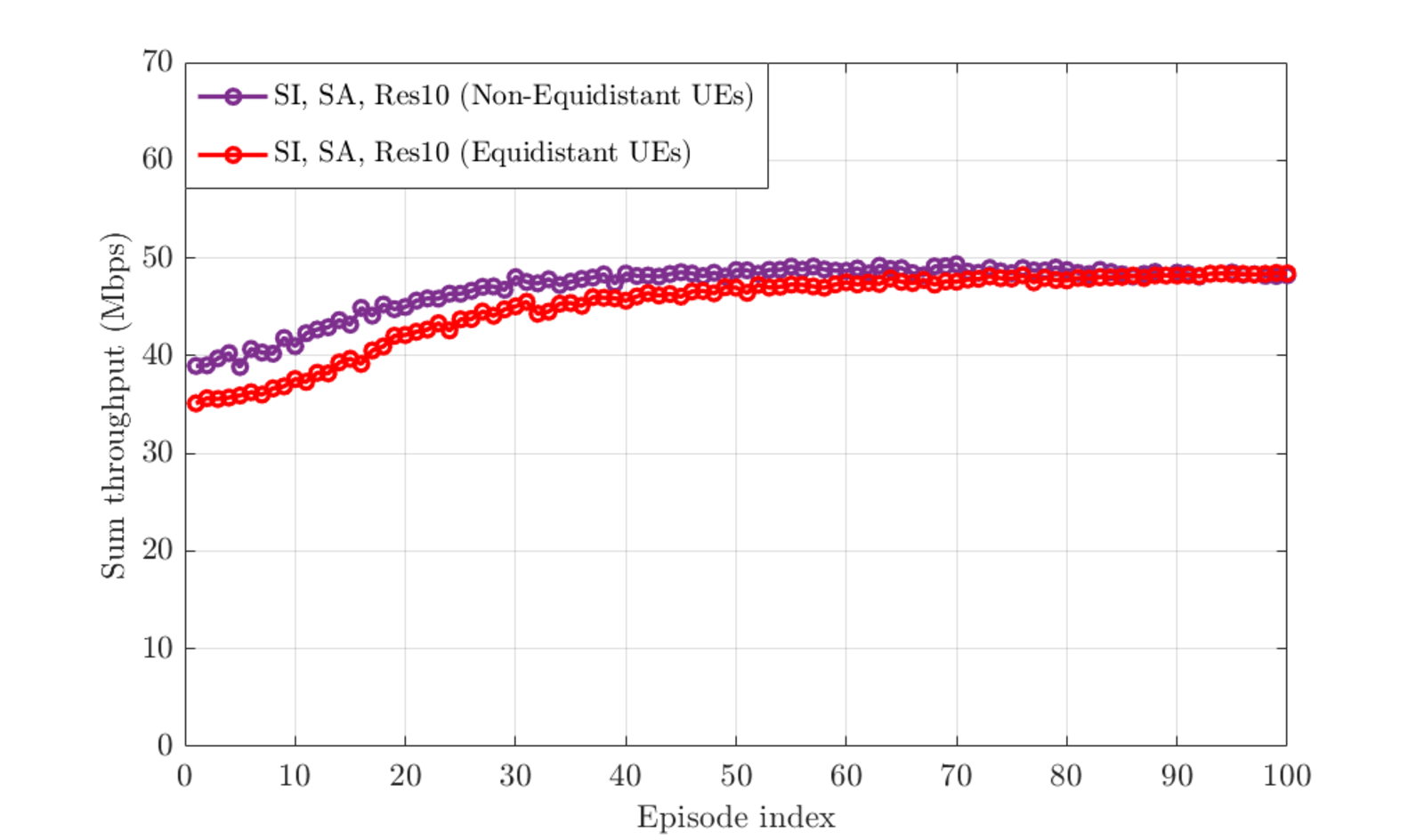}}\hfill
\subfloat[UE1 throughput vs episode index]{\label{subfig:Eq_Neq_rate_UE1}\includegraphics[width=5.5cm,height=4cm]{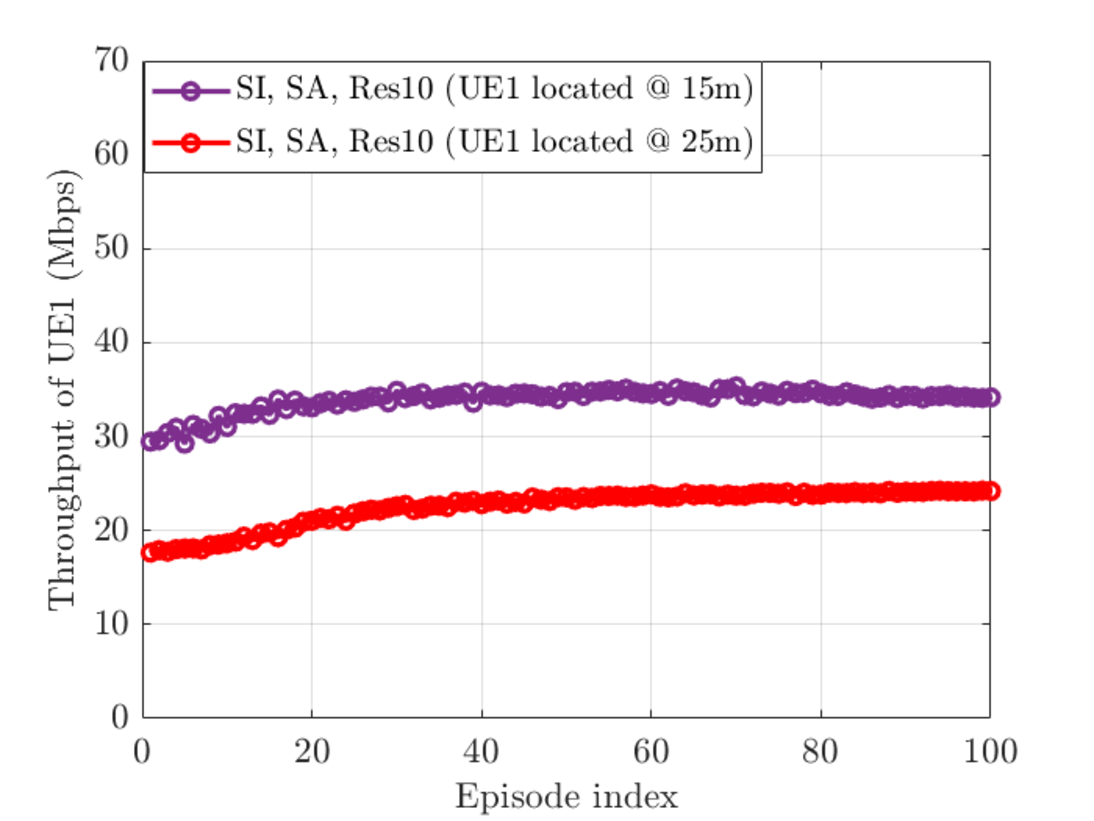}}\hfill
\subfloat[UE2 throughput vs episode index]{\label{subfig:Eq_Neq_rate_UE2}
\includegraphics[width=5.5cm,height=4cm]{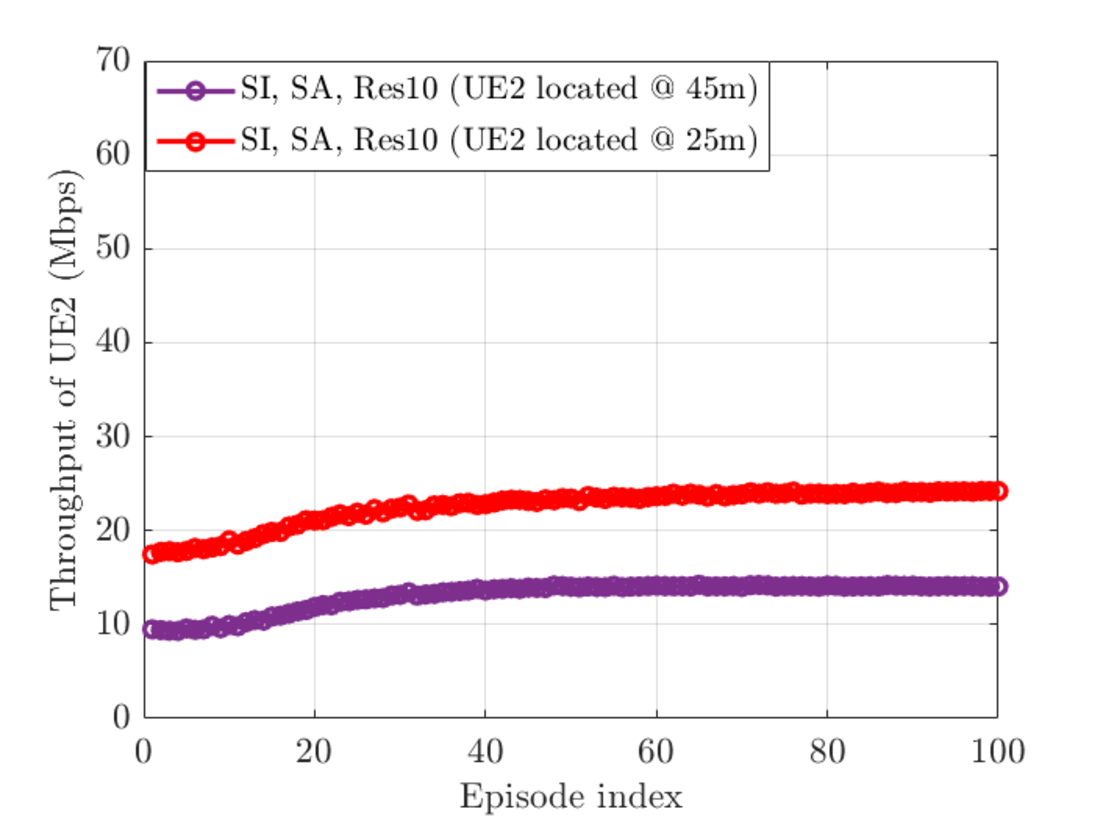}}
\vspace{2mm}
\caption{Illustration of throughput performance for two non-equidistant  and equidistant UEs for the case of `SI, SA, Res10' with single gNB.}
\label{Sumrate_UE1_UE2_single_cell}
\vspace{-2mm}
\end{figure*}

\subsubsection{Non-equidistant UEs}
\label{sec:nque}
For the non-equidistant UEs scenario with `SI, SA, Res25', UE1 is located at 25 m from gNB and UE2 is at 35 m from gNB. The penalty function parameters for UE1 are ($\theta^1_{1}$, $\theta^1_{2}, \Omega^1$) = (-100 dBm, -95 dBm, $0.5\times 10^7$) and for UE2 are ($\theta^2_{1}$, $\theta^2_{2}, \Omega^2$) = (-100 dBm, -95 dBm, $10^7$). The self coupling losses at the UEs are $L^i_\text{c} = 35$ dB for $i = 1,2$. Figure \ref{Non_Eq_multi_UE_sumrate} shows the sum throughput performance where the effectiveness of the proposed SA scheme with fine resolution of RB allocation can be observed. Comparing the performance plots in Figs. \ref{Non_Eq_multi_UE_sumrate} and \ref{Eq_multi_UE_sumrate}, we see that the system with non-equidistant users achieves less sum throughput compared to the system with equidistant users. This is because the far user (UE2 at 35 m distance) is disadvantaged in terms of reduced SINR at the gNB for a given transmit power, due to a larger path loss. This leads to a reduced throughput of about 17 Mbps at UE2. UE1, on the other hand, achieves a throughput of about 23 Mbps, resulting in a sum throughput of about 40 Mbps (which is higher than the 36 Mbps sum throughput achieved in the single user system in Fig. \ref{Single_UE_sumrate}).  

In Figs. \ref{Non_Eq_multi_UE1_UE2_traces} and \ref{Sumrate_UE1_UE2_single_cell}, we illustrate the performance results for another non-equidistant UEs scenario. Figures \ref{Non_Eq_multi_UE1_UE2_traces}(a)-(p) show the traces of evolution of variables of two non-equidistant UEs for the case of `SI, SA, Res10', with UE1 and UE2 located at 15 m and 45 m from gNB, respectively. In the case of ‘SI, SA, Res10’, RB allocation is done in units of 10 on CC2, shared by both UE1 and UE2. The following parameters are used: ($\theta^1_{1}$, $\theta^1_{2}, \Omega^1$) = (-100 dBm, -95 dBm, $0.5\times 10^7$), ($\theta^2_{1}$, $\theta^2_{2}, \Omega^2$) = (-100 dBm, -95 dBm, $10^7$), and $L^i_\text{c} = 35$ dB, $i = 1,2$. Figures \ref{Non_Eq_multi_UE1_UE2_traces}(a)-(h) are the traces for UE1 and Figs. \ref{Non_Eq_multi_UE1_UE2_traces}(i)-(p) are the traces for UE2. The corresponding throughput plots are shown in Fig. \ref{Sumrate_UE1_UE2_single_cell}. While Fig. \ref{Sumrate_UE1_UE2_single_cell}(a) shows the sum throughput, Figs. \ref{Sumrate_UE1_UE2_single_cell}(b) and (c) show the individual throughputs achieved by UE1 and UE2, respectively. The throughput plots for the case of two equidistant UEs (both UE1 and UE2 at 25 m from gNB) are also shown for comparison.
From Figs. \ref{Non_Eq_multi_UE1_UE2_traces} and \ref{Sumrate_UE1_UE2_single_cell}, it can be seen that the algorithm learns and allocates resources that maximize the sum throughput. It can be seen that, due to its smaller path loss (which is proportional to $\frac{1}{15^2}$), UE1 achieves a higher throughput compared to UE2 whose path loss is proportional to $\frac{1}{45^2}$. For the same reason, UE1's throughput in the considered non-equidistant scenario is higher than its throughput in the equidistant scenario where the path loss is proportional to $\frac{1}{25^2}$. Similarly, since UE2 is far from gNB in the non-equidistant case (45 m vs 25 m), its achieved throughput is less compared to that in the equidistance case.

\begin{figure*}[htbp]
\subfloat[Sum throughput vs episode index]
{\label{subfig:Online_rem_add_sum}\includegraphics[width=5.5cm,height=4cm]{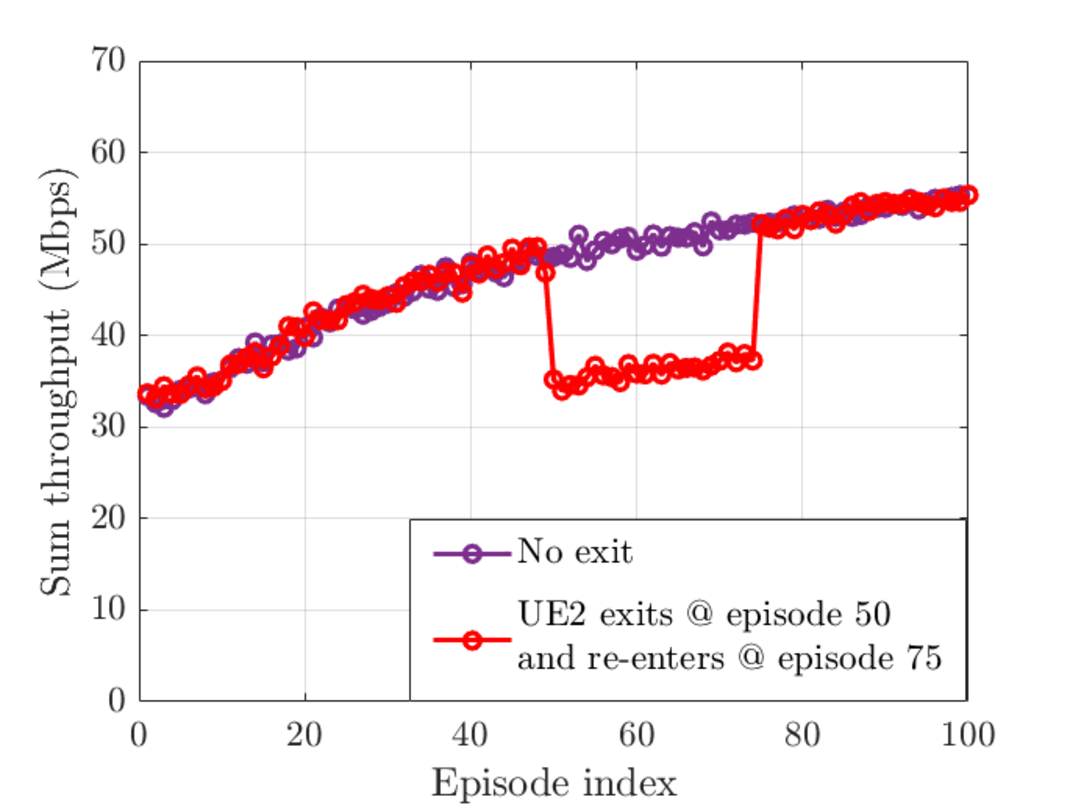}}\hfill
\subfloat[UE1 throughput vs episode index]{\label{subfig:Online_rem_add_UE1}\includegraphics[width=5.5cm,height=4cm]{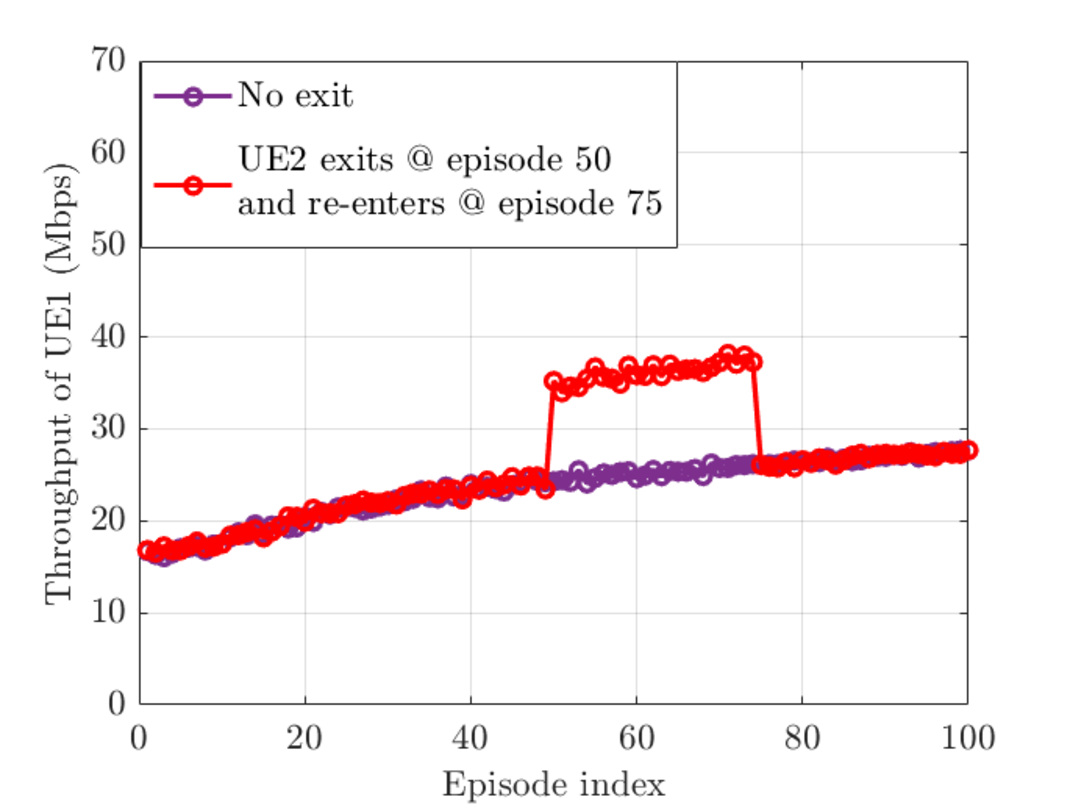}}\hfill
\subfloat[UE2 throughput vs episode index]{\label{subfig:Online_rem_add_UE2}
\includegraphics[width=5.5cm,height=4cm]{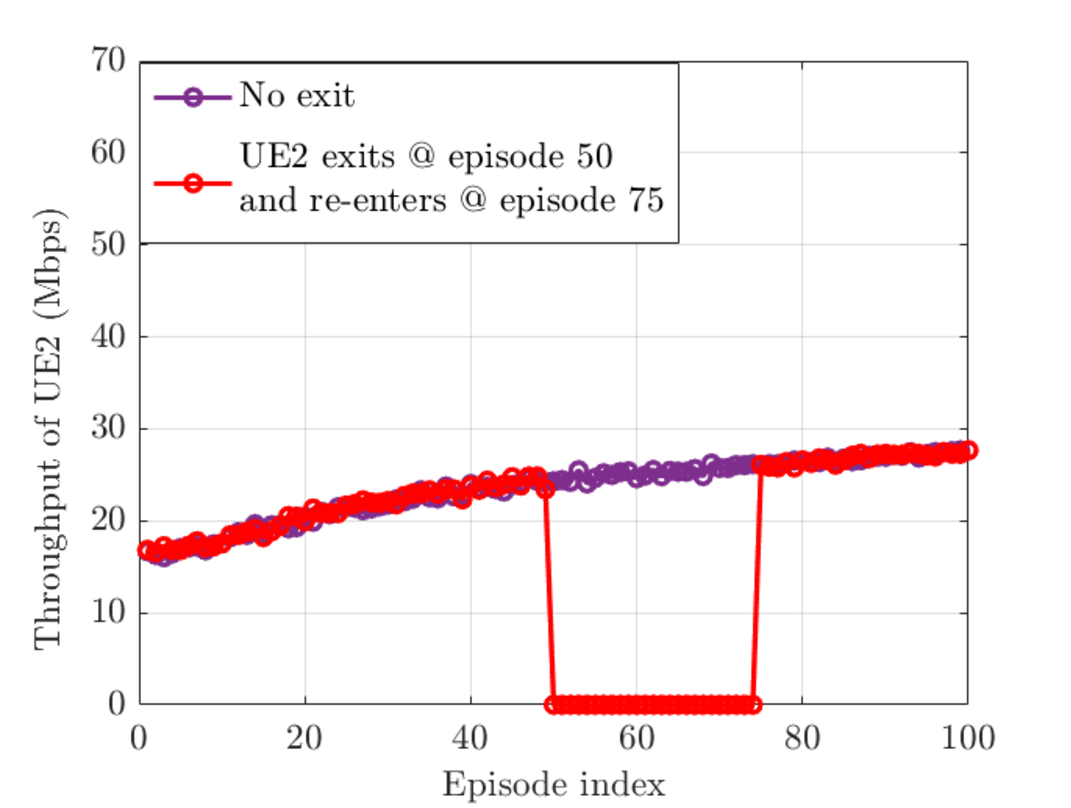}}
\vspace{2mm}
\caption{Illustration of online adaptation when a UE exits and re-enters the network.}
\label{Online_adapt_remove_add}
\vspace{-2mm}
\end{figure*}

\subsubsection{Online adaptation to change in number of users}
In Fig. \ref{Online_adapt_remove_add}, we provide an illustration of the ability of the proposed RL framework to adapt online to changes in the number of users in the network. For this, we consider the scenario in Fig. \ref{Eq_multi_UE_sumrate}, where there are two equidistant UEs in a single cell and there is no SI.  Both UEs are present in the network till the 50th episode. After 50 episodes, UE2 exits the network and rejoins the network for a new call after the 75th episode. The RL framework initially allocates CCs and RBs to both UEs and the sum throughput evolves in a way similar to that in Fig. \ref{Eq_multi_UE_sumrate} up to the 50th episode, as can be seen in Fig. \ref{subfig:Online_rem_add_sum}. When UE2 leaves the network after 50 episodes, its allocated RBs and transmit power are deallocated. As a result, there is a drop in the sum throughput (see Fig. \ref{subfig:Online_rem_add_sum}) and the deallocated RBs become available for reallocation to existing UEs in the network. In this case, they get allocated to UE1 and because of this there is a rise in the  UE1 throughput (see Fig. \ref{subfig:Online_rem_add_UE1}). The resulting sum throughput tends to maximize towards a value similar to that observed for the single-cell single-UE scenario in Fig. \ref{Single_UE_sumrate}. However, when UE2 enters the network again after 75 episodes, the resource allocation and throughput dynamics change. That is, the RBs are reallocated to both the UEs and the sum throughput evolution once again follows that in Fig. \ref{Eq_multi_UE_sumrate} (see \ref{subfig:Online_rem_add_sum}). Figures \ref{subfig:Online_rem_add_UE1} and \ref{subfig:Online_rem_add_UE2} illustrate the corresponding evolution of the individual throughputs of UE1 and UE2, respectively, illustrating how the RL framework adapts to the exit and re-entry of UE2.

\begin{figure}[t]
\centering
\includegraphics[width=9.1cm,height=7.4cm]{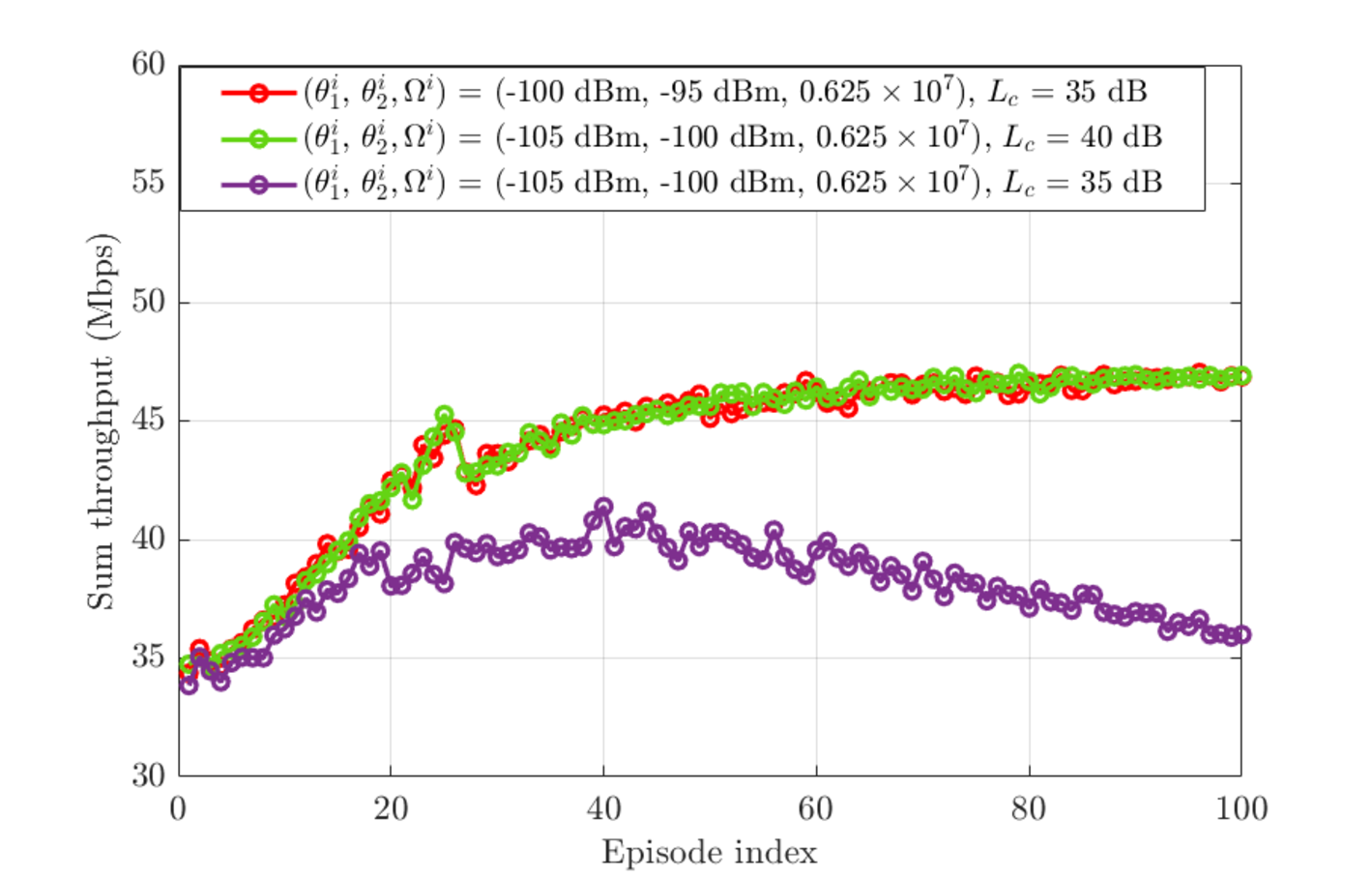}
\vspace{-5mm}
\caption{Sum throughput performance for different choices of penalty factors for the case of ‘SI, SA, Res25’ with single gNB and two equidistant UEs.}
\label{Penalty_factor}
\vspace{-5mm}
\end{figure}

\subsubsection{Choice of penalty factor}
\label{subsubsec:penalty_factor_section}
The choice of penalty factor in the reward function directly influences the agent’s learning priorities by adjusting the weight of the penalty relative to the sum throughput term. The proposed penalty function defined in (\ref{R_b}) is dependent on the parameters ($\theta^i_{1}$, $\theta^i_{2}, \Omega^i$). Previously, in the two equidistant UEs setup, the penalty function parameters were taken as ($\theta^i_{1}$, $\theta^i_{2}, \Omega^i$) = (-100 dBm, -95 dBm, $\text{0.625}\times {10^7}$) for $i=1,2$ and the self-coupling losses at the UEs were taken as $L^i_\text{c} = \text{35}$ dB for $i=1,2$. In Fig. \ref{Penalty_factor}, for a different choice of penalty function parameters, ($\theta^i_{1}$, $\theta^i_{2}, \Omega^i$) = (-105 dBm, -100 dBm, $\text{0.625}\times {10^7}$) and $L^i_\text{c} = \text{35}$ dB for $i=1,2$, the sum throughput in ‘SI, SA, Res25’ case initially grows imitating that of the ‘No SI' case, and then as more RBs are allocated, the transmit power increases with a corresponding increase in the SI power. This increased $p^i_\text{SI}$ lies beyond the chosen $\theta^i_2$ and as a result the reward function penalizes the increased SI to deteriorate the sum throughput to ‘HA' case. When $L_c$ is considered to be 40 dB instead of 35 dB, the corresponding $p^i_\text{SI}$ lies in between the chosen $\theta^i_1$ and $\theta^i_2$. Now the reward function penalizes in such a way that a few RBs are allocated, instead of no RBs being allocated. This resulting sum throughput performance is similar to that of the former choice of penalty factor.

\begin{figure*}[htbp]
\subfloat[Variation of learning rate]
{\label{subfig:Learning_rate}\includegraphics[width=5.5cm,height=4cm]{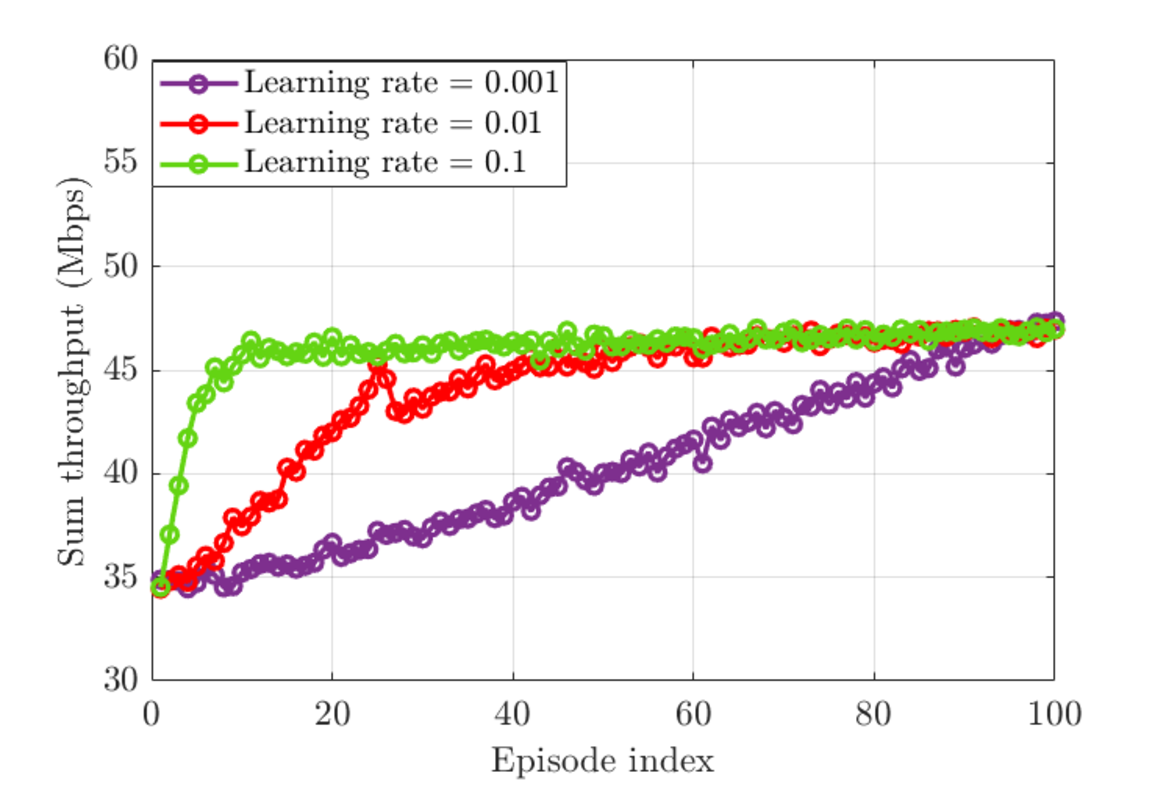}}\hfill
\subfloat[Variation of discount factor]{\label{subfig:Discount_factor}\includegraphics[width=5.5cm,height=4cm]{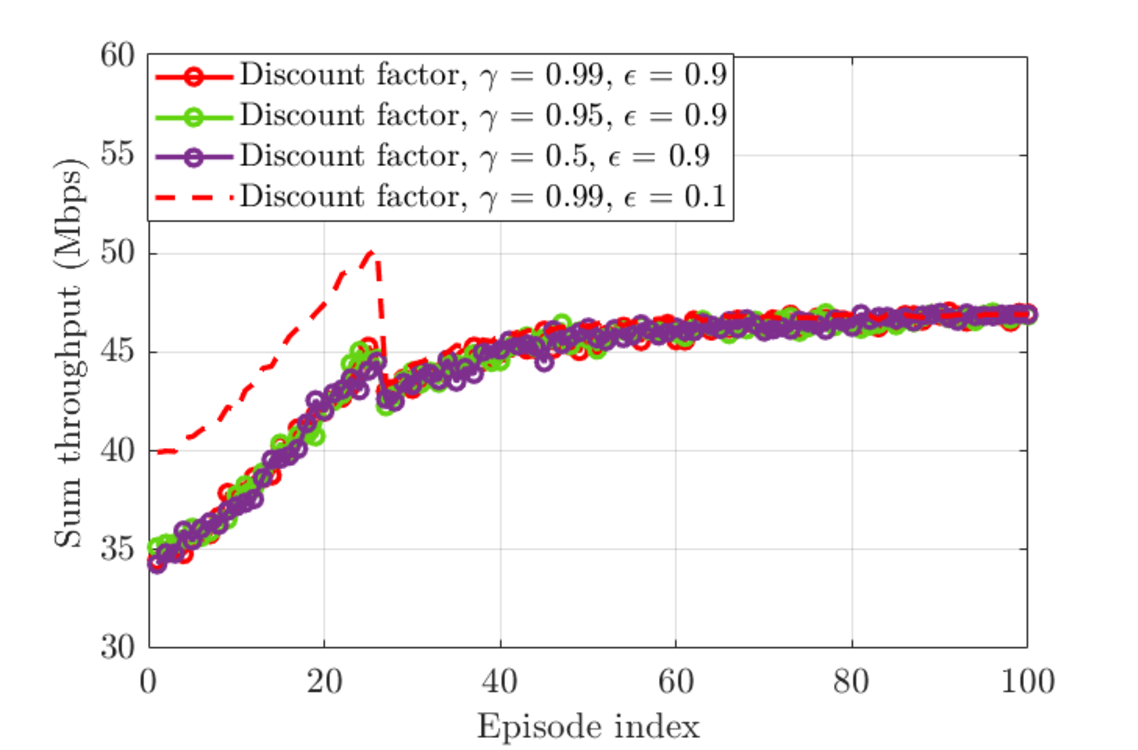}}\hfill
\subfloat[Variation of replay buffer size]{\label{subfig:Replay_buffer_size}
\includegraphics[width=5.5cm,height=4cm]{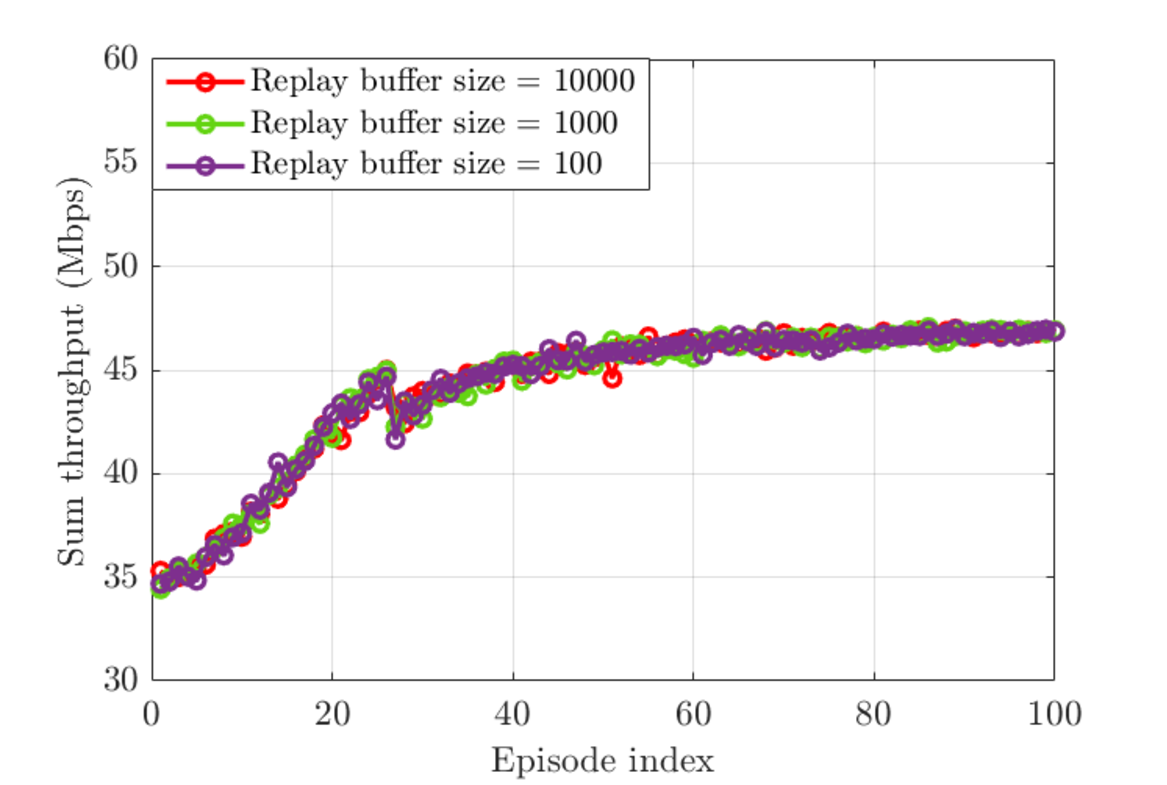}}
\vspace{2mm}
\caption{Performance of proposed approach to variations in key hyperparameters (learning rate, discount factor, replay buffer size) for the case of ‘SI, SA, Res25’ with single gNB and two equidistant UEs.}
\label{Hyperparameters_single_cell}
\vspace{-4mm}
\end{figure*}

\subsubsection{Choice of hyperparameters}
\label{subsubsec:hyperparameters_section}
In Fig. \ref{Hyperparameters_single_cell}, we provide the performance of the proposed approach as a function of hyperparameters such as learning rate, discount factor, and replay buffer size.
Figure \ref{subfig:Learning_rate} shows the sum throughput for different learning rates of 0.001, 0.01, 0.1. It can be seen that varying the learning rate affects the speed of convergence rather than the final sum throughput performance. Figure \ref{subfig:Discount_factor} shows the sum throughput performance for different discount factors. For the exploration rate of $\epsilon=0.9$, we do not see any difference in performance for varying discount factors of $\gamma=0.5, 0.95, 0.99$. This is because the high exploration rate dominates the agent's behavior and masks the effect of the discount factor. In Fig.\ref{subfig:Replay_buffer_size}, the observed insensitivity of performance to replay buffer size is attributed to the system configuration which has only 2 UEs and 2 CCs. In such a setting, the state and action spaces are small, and state-action transitions are quickly and repeatedly experienced during training. As a result, even a relatively small replay buffer is sufficient to capture various relevant experiences needed for effective learning. This is further supported by the deterministic nature of our channel model and the dense reward signal, which together ensure that the agent can converge to a robust policy without requiring a large memory of past experiences.

\subsubsection{Comparison with baseline algorithms}
\label{subsubsec:comparison}
\begin{figure}[t]
\centering
\includegraphics[width=9.1cm,height=7.4cm]{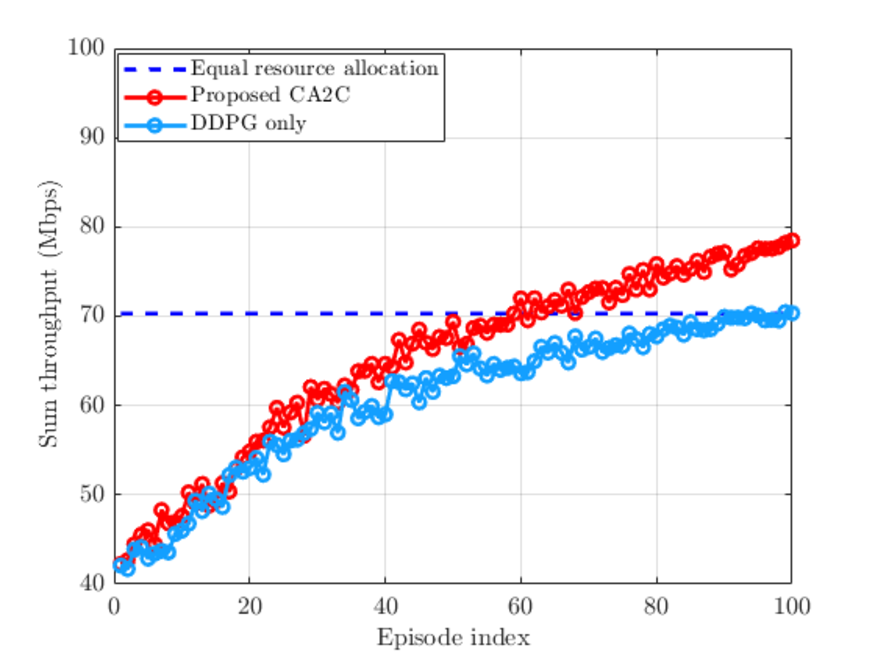}
\vspace{-6mm}
\caption{Sum throughput performance comparison between the proposed CA2C algorithm, the DDPG-only algorithm, and the ERA algorithm for a single gNB serving two non-equidistant UEs.}
\label{conventional_methods}
\vspace{-5mm}
\end{figure}

In Fig. \ref{conventional_methods}, we present a sum throughput performance comparison between the CA2C algorithm and two baseline algorithms, namely, 1) equal resource allocation (ERA) algorithm, a conventional resource allocation algorithm where all users are assigned equal power and resource blocks, independent of SI or delay conditions, and 2) DDPG-only algorithm, a model-free RL baseline using the standard DDPG algorithm which can only handle continuous action spaces. A single-cell system with two non-equidistant UEs placed at 15 m and 45 m, and 1 PCC and 2 SCCs is considered. In ERA and DDPG-only schemes, the discrete action of CC selection is done such that all the UEs are allocated on all the CCs available in the network. To compare with the baseline algorithms, the sum throughput performance under `No SI' is obtained. From Fig. \ref{conventional_methods}, it is seen that the DDPG-only algorithm adapts and reaches only the   ERA performance. Whereas, the CA2C algorithm achieves well beyond the ERA performance. This is because the ERA algorithm does not adapt and the DDPG-only algorithm adapts only the continuous variable (power). The CA2C algorithm, on the other hand, adapts both continuous and discrete variables, and thus achieves better sum throughput.

\subsection{Impact of imperfect CSI on performance}
\label{subsubsec:imperfect_CSI_section}
\begin{figure}
\centering
\includegraphics[width=9.1cm,height=7.4cm]{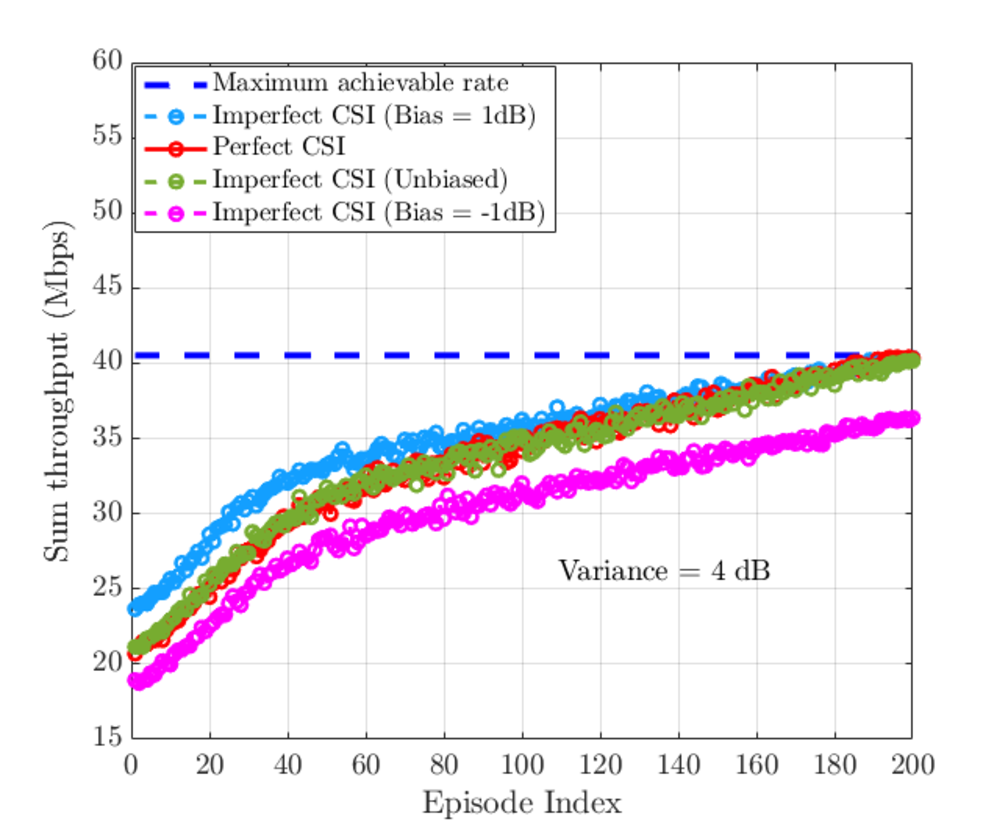}
\caption{Performance of the proposed approach in the presence of SINR estimation error for the case of ‘No SI’ with single gNB and single UE.}
\label{Imperfect_CSI_new}
\vspace{-4mm}
\end{figure}

In obtaining the simulation results, we have assumed perfect channel state information (CSI). For the path loss model considered, this amounts to assuming that the signal-to-noise-plus-interference ratios (SINR) at the gNBs are perfectly known, which is difficult to achieve in real-world scenarios. To study the impact of imperfect SINR knowledge on performance, we model the measured SINR as the true SINR corrupted by an SINR estimation error. The estimation error in dB is taken to be Gaussian \cite{sinr_est}.
Depending on the estimation method used, the estimate can be unbiased (mean = 0) or biased (mean $\neq$ 0) with a small or large error variance. We carry out simulations for biased/unbiased estimates (bias = 0 dB, 1 dB, -1 dB) with an error variance of 4 dB. Figure \ref{Imperfect_CSI_new} shows the sum throughput performance under imperfect CSI conditions described above for the case of `No SI' with single gNB and single UE (located at 25 m from gNB). The blue dashed line represents the maximum achievable rate under the perfect CSI assumption. The red curve is the baseline which represents the throughput performance under perfect CSI. The green curve is for the unbiased estimate, and it closely follows the perfect CSI performance. The blue curve (positive bias of 1 dB) slightly overshoots baseline performance during early episodes due to optimistic (high) SINR estimates and consequent aggressive resource allocation. The pink curve (negative bias of 1 dB) undershoots due to pessimistic (low) SINR estimates and consequent conservative resource allocation. This simulation highlights the impact of the SINR error statistics on the proposed algorithm performance.

\begin{figure*}[htbp]
\subfloat[State evolution]
{\label{subfig:state_dynamics}\includegraphics[width=5.8cm,height=4cm]{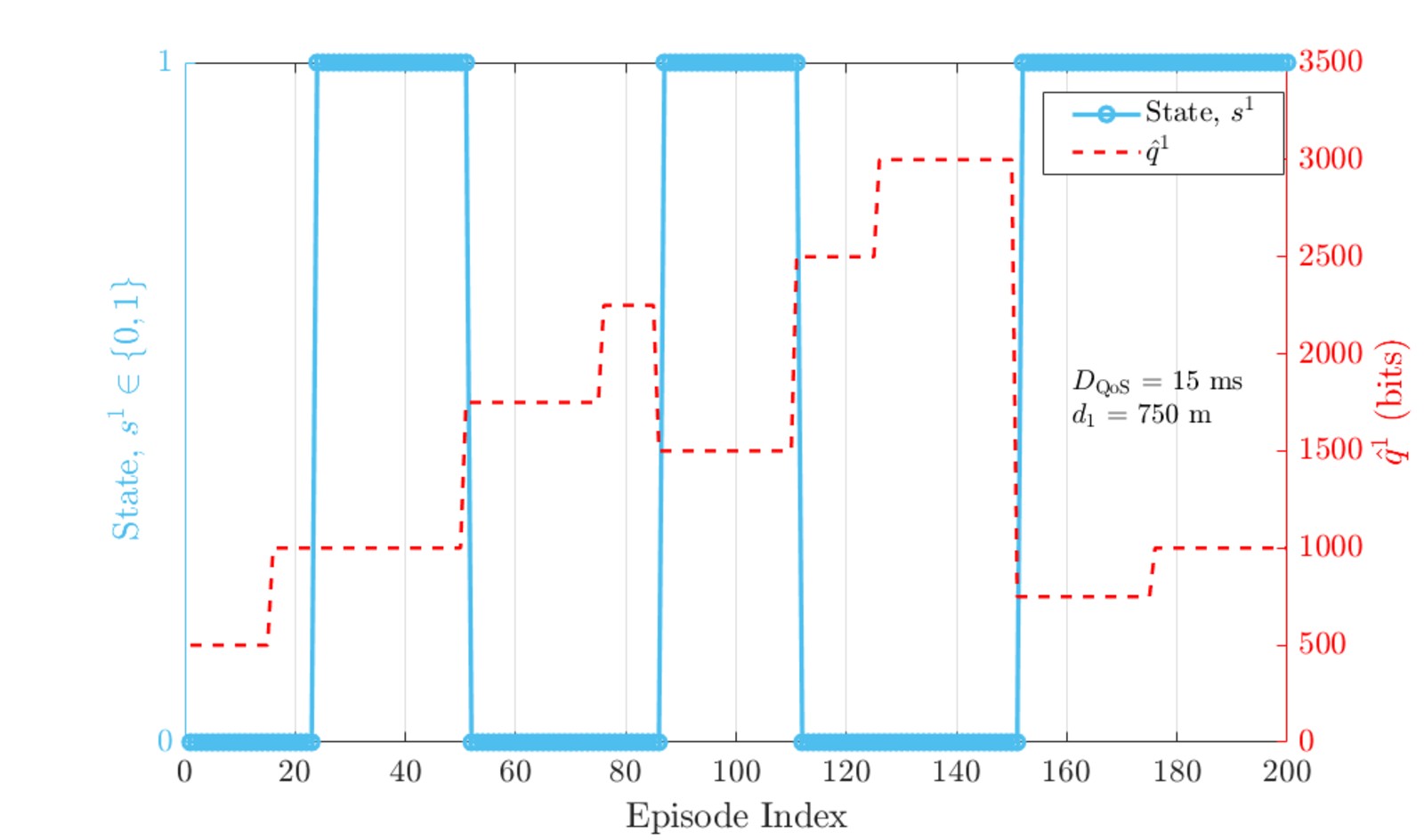}}\hfill
\subfloat[Power evolution]{\label{subfig:power_dynamics}\includegraphics[width=5.8cm,height=4cm]{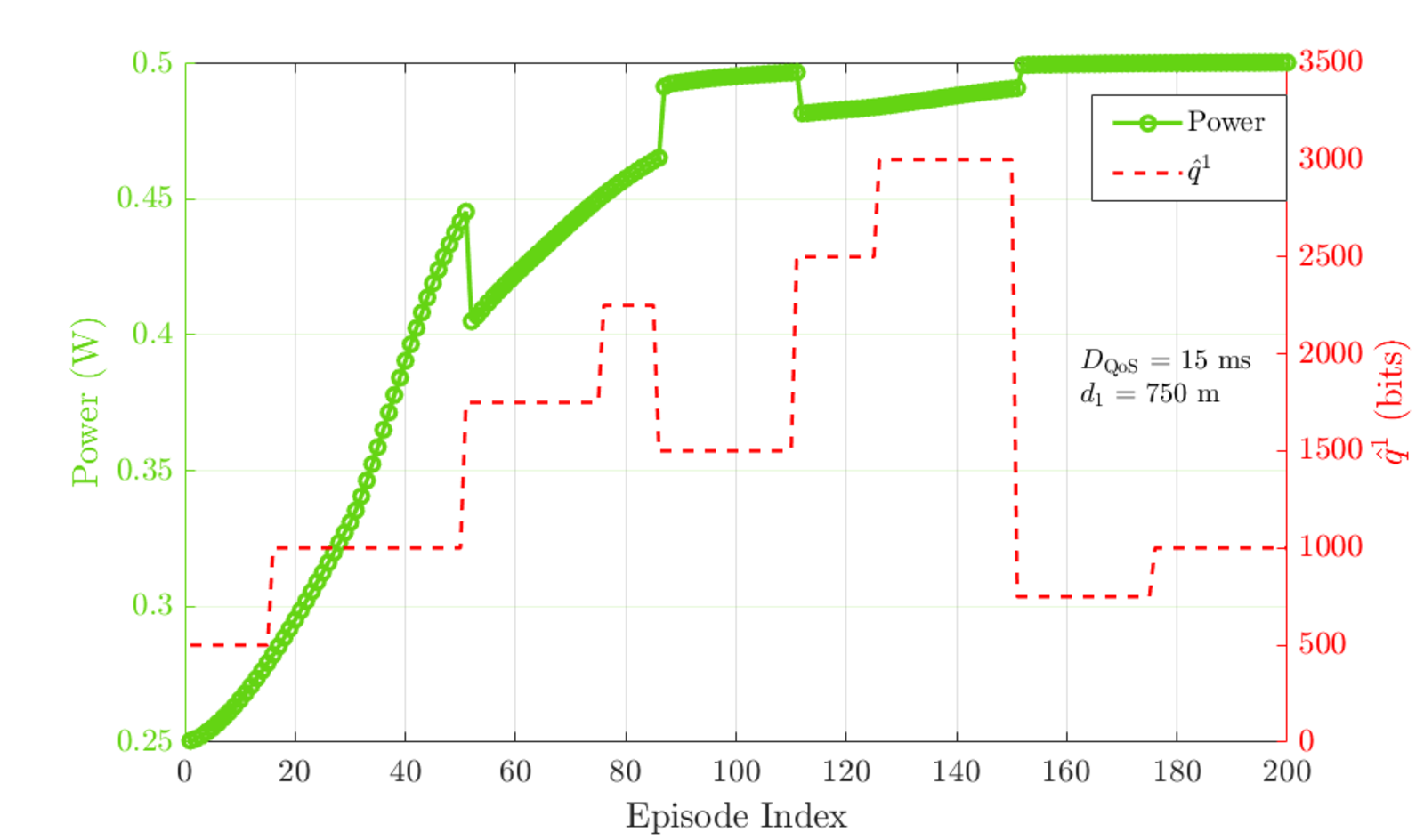}}\hfill
\subfloat[Sum throughput evolution]{\label{subfig:rate_dynamics}
\includegraphics[width=5.8cm,height=4cm]{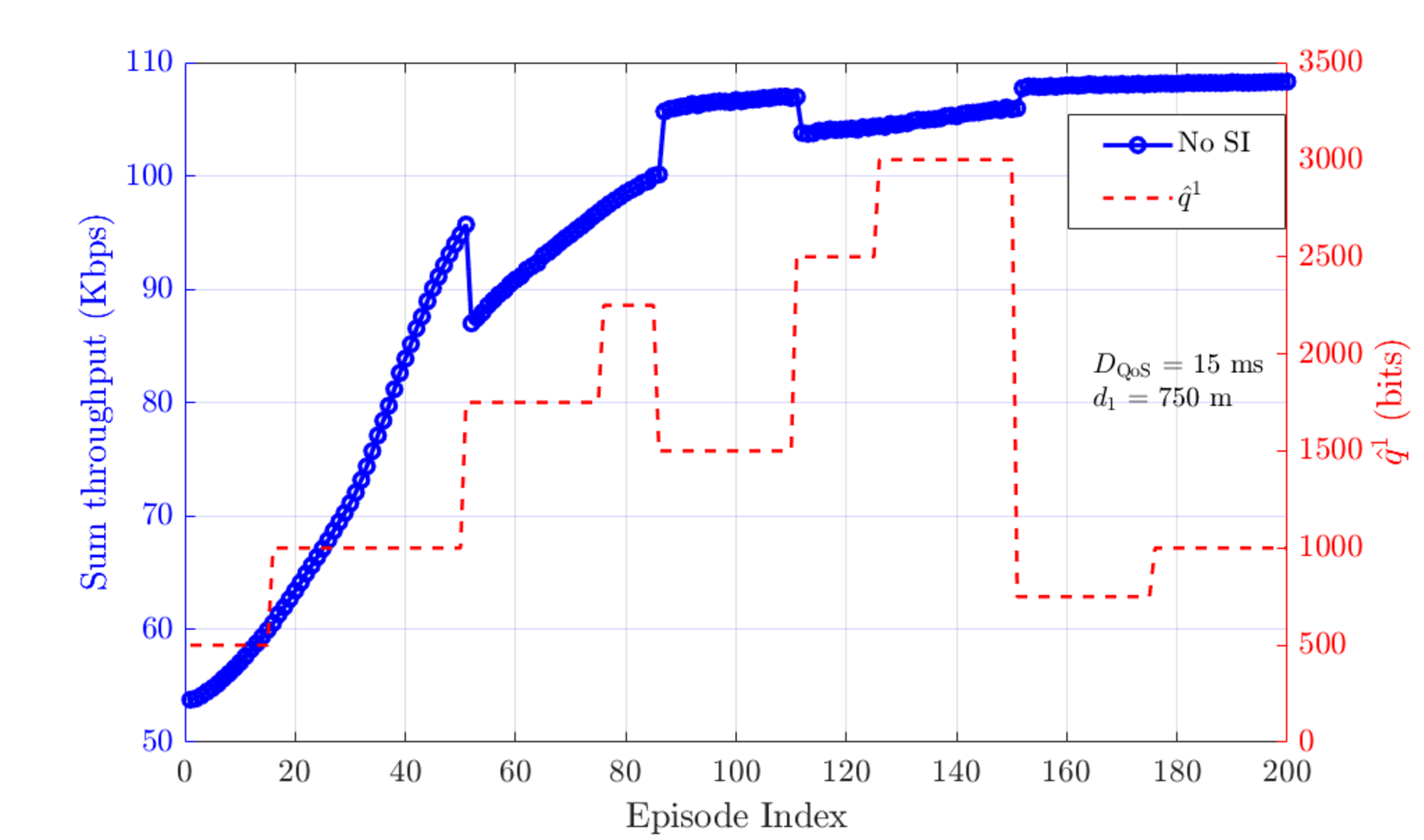}}
\vspace{2mm}
\caption{Illustration of the evolution of state, power, and sum throughput for time-varying traffic pattern for the case of ‘No SI’ with single gNB and single UE.}
\label{Traffic_dynamics}
\vspace{-2mm}
\end{figure*}

\subsection{Dynamic traffic modeling and resource allocation}
\label{subsubsec:traffic_model}
In this subsection, we present simulation results to illustrate how the RL network/algorithm learns to adapt to the variations in the UE traffic patterns. For this, we varied the average number of bits per burst of data (represented by the parameter $\hat{q}^i$) at different epochs in time to model different traffic load conditions and analyzed the evolution/traces of the variables in optimization, system state, and sum throughput. The results are captured in Figs. \ref{Traffic_dynamics}(a)-(c) for a single-gNB single-UE scenario with `No SI'. For illustration purposes, the following parameters are used in the simulations: $d_1$ = 750 m, $D_\text{QoS}$ = 15 ms. The red dashed plot in these figures shows the variation of $\hat{q}^1$ as a function of episode index. Figure \ref{Traffic_dynamics}(a) shows how the system state $s^1\in \{0,1\}$, 1 if $D_\mathrm{QoS}$ is met and 0 otherwise, changes in response to the variation in $\hat{q}^1$. Figures \ref{Traffic_dynamics}(b) and (c) depict the corresponding evolution of power and sum throughput, respectively. It can be seen that initially the system starts being in 0 state (not meeting $D_\mathrm{QoS}$ due to inadequate power) and transitions to 1 state (meeting $D_\mathrm{QoS}$) at 24th episode due to increasing power allocation by the algorithm. The system transitions back to the 0 state at 51st episode at which time $\hat{q}^1$ increases to 1750 bits, resulting in not meeting $D_\mathrm{QoS}$. However, the algorithm is able to recognize this transition and respond by momentarily reducing the power allocation, but recovers from this by increasing the power subsequently to get back to state 1 at 86th episode. Similar transitions to 0 state (111th episode) and subsequent recovery to 1 state (151th episode) can be observed in the evolution, illustrating the algorithm's ability to effectively learn and adapt to traffic variations. A similar evolution trend has been observed in SI case.

\subsection{Multi-cell scenario}
{Here, we consider a network which consists of two gNBs and two UEs. The first gNB ($\mathrm{gNB}_1$) is positioned at origin (0,0), while the second gNB ($\mathrm{gNB}_2$) is located at 100 m away at (100,0). UE1 is 25 m from $\mathrm{gNB}_1$ at (25,0), and UE2 is 40 m from  $\mathrm{gNB}_2$ at (60,0). Each UE is communicating with its own associated gNB. UE1's transmission to gNB$_1$ interferes of UE2's signal at gNB$_2$. Likewise, UE2's transmission to gNB$_2$ interferes with UE1's signal at gNB$_1$. The penalty function parameters for UE1 are ($\theta^1_{1}$, $\theta^1_{2}, \Omega^1$) = (-100 dBm, -95 dBm, $10^7$) and for UE2 are ($\theta^2_{1}$, $\theta^2_{2}, \Omega^2$) = (-100 dBm, -95 dBm, $10^7$). The self-coupling losses at both the UEs are $L_\text{c} = 35$ dB.

\begin{figure*}
\centering
\subfloat[UE1 performance at gNB$_1$]{\includegraphics[width=0.5\linewidth]{ 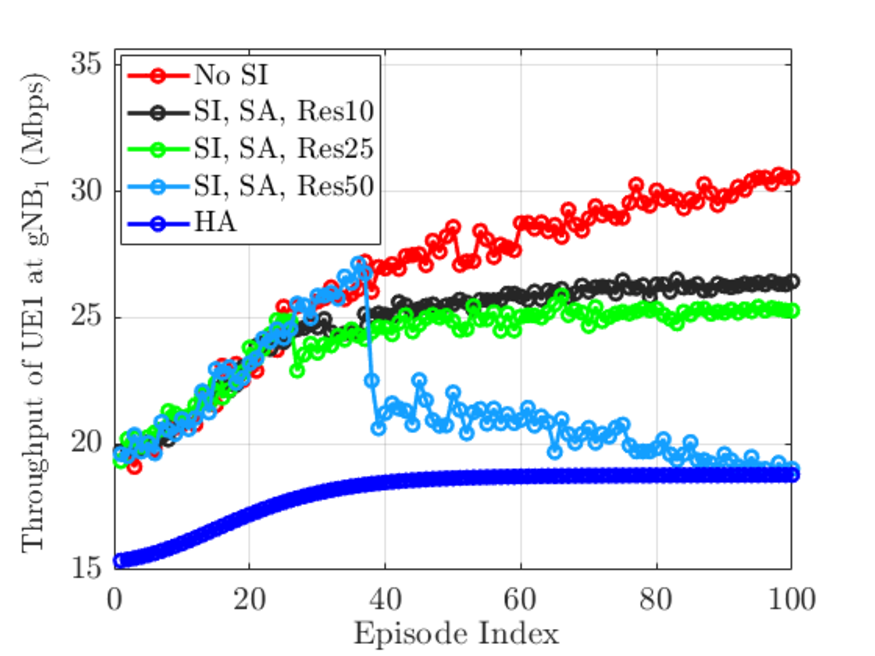}\label{rate_ue1}}
\hfill
\subfloat[UE2 performance at gNB$_2$]{\includegraphics[width=0.5\linewidth]{ 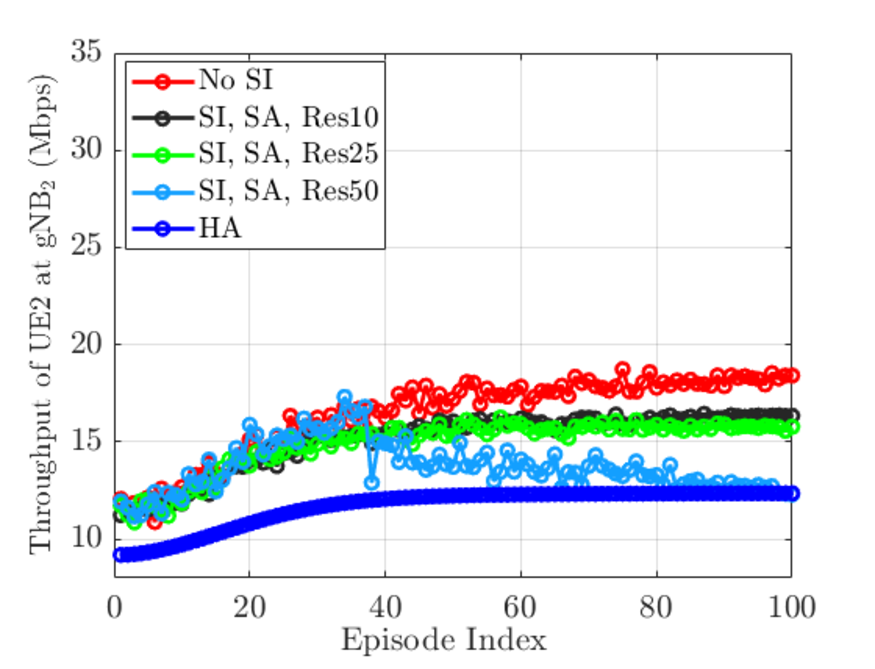}\label{rate_ue2}}
\caption{Throughput performance in a two-cell scenario.} 
\label{Rate@gnb1and2}
\vspace{-6mm}
\end{figure*}

In Figs. \ref{rate_ue1} and \ref{rate_ue2}, we present the throughput achieved by UE1 at gNB$_1$ and by UE2 at gNB$_2$, respectively, as a function of episode index. Comparing Figs. \ref{rate_ue1} and \ref{rate_ue2}, we  observe that UE1 achieves better throughput performance compared to that of UE2. This is because the SINR of UE1 at gNB$_1$ is proportional to $\left(\frac{25}{60}\right)^{-2}$, whereas the SINR of UE2 at gNB$_2$ is proportional to $\left(\frac{40}{75}\right)^{-2}$. We also note that, in the single-cell single-user scenario in Fig. \ref{Single_UE_sumrate}, UE1 achieves a higher throughput of about 40 Mbps with no SI, due to the absence of other-cell interference. However, in the considered two-cell scenario, there is a noticeable decrease in UE1 throughput (about 30 Mbps with no SI in Fig. \ref{rate_ue1}), which is attributed to the other-cell interference.

\begin{figure*}[htbp]
\centering
\subfloat[Sum throughput performance at gNB$_1$]{\includegraphics[width=0.5\linewidth]{ 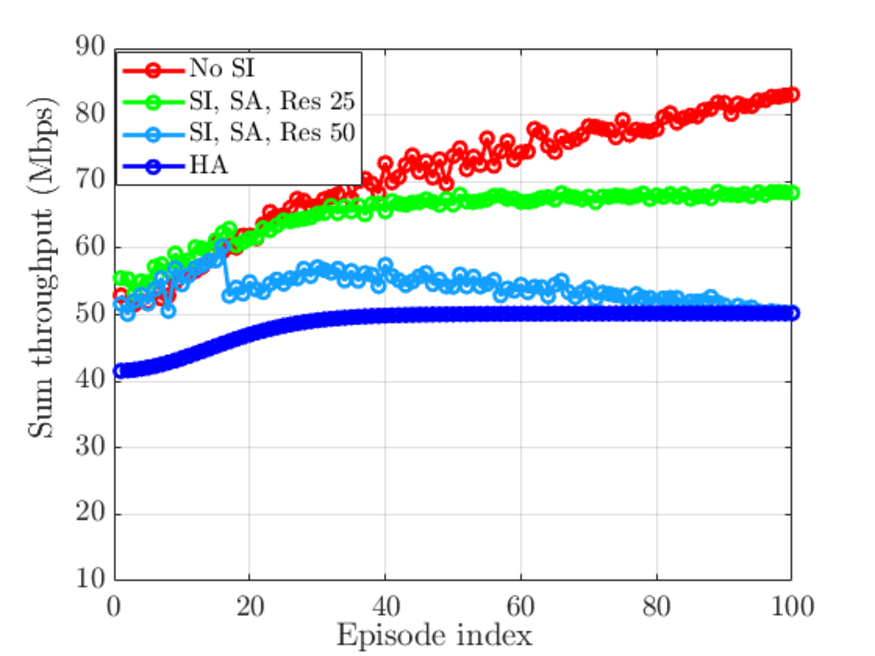}\label{rate_gnb1}}
\hfill
\subfloat[Sum throughput performance at gNB$_2$]{\includegraphics[width=0.5\linewidth]{ 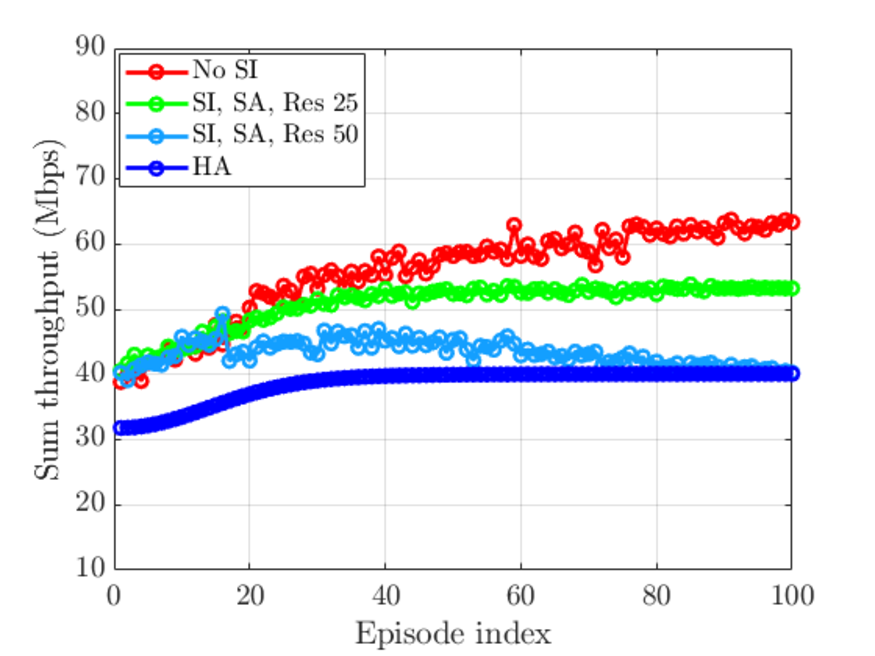}\label{rate_gnb2}}
\caption{Sum throughput performance of a network with two cells and four UEs, with each gNB serving two non-equidistant UEs.} 
\label{Rate@gnb1and2}
\vspace{-2mm}
\end{figure*}

Next, we consider an extended scenario in which the network consists of two gNBs and four UEs. The $\mathrm{gNB}_1$ is located at $(0, 0)$ and the $\mathrm{gNB}_2$ is located at $(100, 0)$. Two UEs, $\mathrm{UE}{1}$ and $\mathrm{UE}{2}$, served by $\mathrm{gNB}_1$, are located at $(10, 0)$ and $(20, 0)$, respectively, corresponding to the distances of 10 m and 20 m from $\mathrm{gNB}_1$. The remaining two UEs, $\mathrm{UE}{3}$ and $\mathrm{UE}{4}$, which are served by $\mathrm{gNB}_2$, are located at $(85, 0)$ and $(70, 0)$, respectively, corresponding to distances of 15 m and 30 m from $\mathrm{gNB}_2$. In this setup, uplink transmissions from UEs associated with one gNB interfere with the reception at the other gNB. $\mathrm{UE}{1}$ and $\mathrm{UE}{2}$ cause interference at $\mathrm{gNB}_2$, and $\mathrm{UE}{3}$ and $\mathrm{UE}{4}$ cause interference at $\mathrm{gNB}_1$.  The penalty function parameters for the UEs are ($\theta^i_1$, $\theta^i_2, \Omega^i$) = (-100 dBm, -95 dBm, $0.625 \times 10^7$), $i=1,2,3,4$. The self-coupling loss at all the UEs is $L_\text{c} = 35$ dB.
For the above setup, Figs. \ref{rate_gnb1} and \ref{rate_gnb2} show the sum throughput performance of $\mathrm{gNB}_1$ and $\mathrm{gNB}_2$, respectively. For $\mathrm{gNB}_1$, it is evident that $\mathrm{UE}{1}$ achieves consistently higher throughput than $\mathrm{UE}{2}$. This difference arises primarily from their relative SINR levels, which are influenced by the positions of interfering UEs and UE locations. Specifically, $\mathrm{UE}{1}$ is located at a distance of 10 m from $\mathrm{gNB}_1$, while the primary interferers $\mathrm{UE}{3}$ and $\mathrm{UE}{4}$ are located at 85 m and 70 m, respectively. Thus, the desired signal power at $\mathrm{gNB}_1$ scales as $10^{-2}$, whereas the interference power roughly scales as $85^{-2} + 70^{-2}$. In contrast, $\mathrm{UE}{2}$ is 20 m away from $\mathrm{gNB}_1$, but experiences interference from $\mathrm{UE}{3}$ and $\mathrm{UE}{4}$ located at 85 m and 70 m, respectively. Consequently, the SINR of $\mathrm{UE}{1}$ at $\mathrm{gNB}_1$ is approximately proportional to $\left(\frac{1}{10^2}\right) / \left(\frac{1}{85^2} + \frac{1}{70^2}\right)$, whereas the SINR of $\mathrm{UE}{2}$ at $\mathrm{gNB}_1$ is proportional to $\left(\frac{1}{20^2}\right) / \left(\frac{1}{70^2} + \frac{1}{85^2}\right)$. These SINR differences explain the throughput advantage of $\mathrm{UE}{1}$. A similar trend is observed for $\mathrm{gNB}_2$. Here, $\mathrm{UE}3$ achieves higher throughput than $\mathrm{UE}4$. $\mathrm{UE}3$ is located at 15 m from $\mathrm{gNB}_2$, while the interfering UEs $\mathrm{UE}1$ and $\mathrm{UE}2$ are at distances of 90 m and 80 m, respectively. Meanwhile, $\mathrm{UE}4$ is 30 m from $\mathrm{gNB}_2$ and experiences interference from the same UEs. Comparing the overall performance of the two gNBs, we find that $\mathrm{gNB}_1$ achieves a higher sum throughput than $\mathrm{gNB}_2$. This is primarily due to its closer user locations (10 m and 20 m), which result in stronger received signal powers and better SINR values, compared to the relatively distant users (70 m and 85 m) of $\mathrm{gNB}_2$.

\begin{table*}
\centering
\begin{tabular}{|l|l|l|}
\hline
\textbf{Component} & \textbf{Operations } & \textbf{Complexity per Step/Episode} \\
\hline \hline
Actor Forward Propagation & Matrix multiplications: $(1024 \times 512) + (512 \times 128)$ & $\sim 656,384$ FLOPs/step \\
\hline
Critic Forward Propagation & Matrix multiplications: $(1024 \times 512) + (512 \times 128)$ & $\sim 656,384$ FLOPs/step \\
\hline
Target Comparison & $1024$-dimensional Q-value comparisons & $\sim 2,048$ FLOPs/step \\
\hline
Actor Backpropagation & Gradients for $128 \times 512$ and $512 \times 1024$ weights & $\mathcal{O}\big(2 \times \text{Batch Size} \times 656,384\big)$ per episode \\
\hline
Critic Backpropagation & Gradients for $128 \times 512$ and $512 \times 1024$ weights & $\mathcal{O}\big(2 \times \text{Batch Size} \times 656,384\big)$ per episode \\
\hline
Replay Buffer size & 500  & $\mathcal{O}(1)$ access time\\
\hline
\end{tabular}

\vspace{0.2cm}
\caption{Computational and storage complexity per episode and step.}
\label{tab:complexity}
\vspace{-4mm}
\end{table*}

\begin{table}
\centering
\begin{tabular}{|l|c|c|c|}
\hline
\textbf{Metric} & \textbf{HA} & \textbf{No SI} & \textbf{SI, SA Res50} \\
\hline \hline
Average iteration time & 11.22s & 11.83s & 11.69s \\
\hline
Data collection time & 11.19s & 11.81s & 11.66s \\
\hline
Training time & 0.027s & 0.020s & 0.022s \\
\hline
\end{tabular}
\vspace{0.2cm}
\caption{{Timing analysis of proposed CA2C algorithm per iteration for different scenarios.}}
\label{tab:iteration_comparison}
\vspace{-4mm}
\end{table}

\subsection{Computational complexity and training time analysis}
In this subsection, we analyze the computational complexity and training time of the   CA2C algorithm. The main operations and their complexities per training step and episode are summarized in Table \ref{tab:complexity}. All timing measurements were conducted on a dedicated benchmarking system with the following specifications: 12\textsuperscript{th} Generation Intel\textsuperscript{\textregistered} Core\textsuperscript{TM} i7-12700 processor (2.1\, GHz base frequency, 12 cores / 20 threads, 25\, MB cache), 32\, GB DDR4 RAM (31.8\, GB usable in dual-channel configuration), and Windows 11 x64 operating system. The software stack included TensorFlow 2.9 with CUDA 11.2 acceleration, Python 3.8, and all measurements were collected using \texttt{time.perf\_counter()} with nanosecond precision. Table \ref{tab:iteration_comparison}  summarizes the timing results from the single-cell experiment involving two equidistant users. These quantitative results demonstrate that the CA2C algorithm maintains reasonable computational complexity while achieving superior performance compared to baseline methods. 

\section{Conclusions}
\label{sec:sec6}
We considered the problem of resource allocation on the uplink with carrier aggregation. The resources for allocation included uplink transmit power at the UE (which is a continuous variable) and the number of CCs and RBs for transmission on the uplink (which are discrete variables). The objective was to maximize the sum throughput. We presented a RL framework to solve the resource allocation problem. Carrier aggregation can result in self interference due to transmit PA non-linearity and self coupling. Considering this SI effect in the RL framework is a new contribution in the paper, which has not been reported before. The way in which we handled the SI in the RL framework through a suitable penalty component in the reward function is novel. The proposed soft avoidance scheme using the composite reward function and fine resolution of RB allocation was shown to achieve increased throughput. The proposed RL framework is applicable for downlink carrier aggregation as well. Future extensions to this work can consider scalability and related issues.

\section{Acknowledgements}
This work was supported by Nokia Solutions and Networks India Pvt Ltd. The authors would like to thank Kim Nielsen, Thomas Jacobsen, Samad Ali, Rajesh Banda, Sheshachalam B S, Vinayak Bellur, Rachit Mahendra, and Srilatha Ramachandran from Nokia for their support and valuable discussions during the course of this project.

\bibliographystyle{IEEEtran}
\bibliography{references}

@article{ddpg2,
  title={Multi-agent actor-critic for mixed cooperative-competitive environments},
  author={Lowe, Ryan and Wu, Yi I and Tamar, Aviv and Harb, Jean and Pieter Abbeel, OpenAI and Mordatch, Igor},
  journal={Proc. NIPS},
  volume={30},
  pages={1-12},
  year={2017}
}

@book{book1,
  title={RF Analog Impairments Modeling for Communication Systems Simulation: Application to OFDM-Based Transceivers},
  author={L. Smaini},
  year={2012},
  publisher={John Wiley \& Sons}
}

@book{book2,
  title={Wireless Receiver Architectures and Design: Antenna, RF, Synthesizers, Mixed Signal and Digital Signal Processing},
  author={T. J. Rouphael},
  year={2014},
  publisher={Academic Press}
}

@ARTICLE{DL_CA_1,
  author={Liao, Hong-Sheng and Chen, Po-Yu and Chen, Wen-Tsuen},
  journal={IEEE Transactions on Mobile Computing}, 
  title={An Efficient Downlink Radio Resource Allocation with Carrier Aggregation in {LTE}-Advanced Networks}, 
  year={2014},
  volume={13},
  number={10},
  pages={2229-2239},
  doi={10.1109/TMC.2013.2297310}}

@INPROCEEDINGS{DL_CA_2,
  author={Tsai, Pei-Ling and Lin, Kate Ching-Ju and Chen, Wen-Tsuen},
  booktitle={IEEE International Conference on Communications}, 
  title={Downlink radio resource allocation with Carrier Aggregation in {MIMO} {LTE}-advanced systems}, 
  year={2014},
  volume={},
  number={},
  pages={2332-2337},
  doi={10.1109/ICC.2014.6883671}}

@INPROCEEDINGS{DL_CA_3,
  author={Shayea, Ibraheem and Ismail, Mahamod and Nordin, Rosdiadee},
  booktitle={2013 IEEE 11th Malaysia International Conference on Communications (MICC)}, 
  title={Downlink spectral efficiency evaluation with carrier aggregation in {LTE}-Advanced system employing Adaptive Modulation and Coding schemes}, 
  year={2013},
  volume={},
  number={},
  pages={98-103},
  doi={10.1109/MICC.2013.6805807}}

@ARTICLE{DL_CA_4,
  author={Yuan, Guangxiang and Zhang, Xiang and Wang, Wenbo and Yang, Yang},
  journal={IEEE Communications Magazine}, 
  title={Carrier aggregation for {LTE}-advanced mobile communication systems}, 
  year={2010},
  volume={48},
  number={2},
  pages={88-93},
  doi={10.1109/MCOM.2010.5402669}}

@ARTICLE{stat_1,
  author={Khoramnejad, Fahime and Joda, Roghayeh and Sediq, Akram Bin and Boudreau, Gary and Erol-Kantarci, Melike},
  journal={IEEE Access}, 
  title={{A}{I}-Enabled Energy-Aware Carrier Aggregation in 5{G} New Radio With Dual Connectivity}, 
  year={2023},
  volume={11},
  number={},
  pages={74768-74783},
  doi={10.1109/ACCESS.2023.3297099}}

@ARTICLE{stat_2,
  author={Hu, Jingzhi and Zhang, Hongliang and Song, Lingyang and Schober, Robert and Poor, H. Vincent},
  journal={IEEE Transactions on Communications}, 
  title={Cooperative Internet of {UAV}s: Distributed Trajectory Design by Multi-Agent Deep Reinforcement Learning}, 
  year={2020},
  volume={68},
  number={11},
  pages={6807-6821},
  doi={10.1109/TCOMM.2020.3013599}}

@ARTICLE{opt_1,
  author={Shen, Kaiming and Yu, Wei},
  journal={IEEE Transactions on Signal Processing}, 
  title={Fractional Programming for Communication Systems—Part II: Uplink Scheduling via Matching}, 
  year={2018},
  volume={66},
  number={10},
  pages={2631-2644},
  doi={10.1109/TSP.2018.2812748}}

@article{opt_2,
  title={Energy efficient power allocation strategy for 5G carrier aggregation scenario},
  author={Gao, Weidong and Ma, Lin and Chuai, Gang},
  journal={EURASIP Journal on Wireless Communications and Networking},
  volume={2017},
  pages={1--10},
  year={2017},
  publisher={Springer}
}

@INPROCEEDINGS{UL_CA_1,
  author={Joda, Roghayeh and Elsayed, Medhat and Abou-Zeid, Hatem and Atawia, Ramy and Sediq, Akram Bin and Boudreau, Gary and Erol-Kantarci, Melike},
  booktitle={IEEE International Conference on Communications}, 
  title={{QoS}-Aware Joint Component Carrier Selection and Resource Allocation for Carrier Aggregation in {5G}}, 
  year={2021},
  volume={},
  number={},
  pages={1-6},
  doi={10.1109/ICC42927.2021.9500923}}

@article{UL_CA_2,
  title={Radio resource management for uplink carrier aggregation in {LTE}-Advanced},
  author={Wang, Hua and Rosa, Claudio and Pedersen, Klaus I},
  journal={EURASIP Journal on Wireless Communications and Networking},
  volume={2015},
  pages={1--15},
  year={2015},
  publisher={Springer}
}

@article{UL_CA8,
    author = {V. Mnih},
    title = {Human-level control through deep reinforcement learning},
    journal = {Nature},
    year = {2015},
    pages ={529-533},
    volume={518},
    number={7540}
}

@article{UL_CA10,
    author = {\color{black} F. Khoramnejad and M. Erol-Kantarci},
    title = {\color{black} {O}n joint offloading and resource allocation: a double deep Q-network approach},
    journal = {\color{black} IEEE Trans. Cognit. Comm. Netw.},
    year = {\color{black} Dec. 2021},
    volume={\color{black} 7},
    number={\color{black} 4},
    pages={\color{black} 1126-1141}
}

@ARTICLE{ref1,
  author={Khoramnejad, Fahime and Joda, Roghayeh and Sediq, Akram Bin and Abou-Zeid, Hatem and Atawia, Ramy and Boudreau, Gary and Erol-Kantarci, Melike},
  journal={IEEE Transactions on Communications}, 
  title={Delay-Aware and Energy-Efficient Carrier Aggregation in {5G} Using Double Deep {Q}-Networks}, 
  year={2022},
  volume={70},
  number={10},
  pages={6615-6629},
  doi={10.1109/TCOMM.2022.3204846}}

@ARTICLE{ref2,
  author={Feriani, Amal and Hossain, Ekram},
  journal={IEEE Communications Surveys \& Tutorials}, 
  title={Single and Multi-Agent Deep Reinforcement Learning for {AI}-Enabled Wireless Networks: A Tutorial}, 
  year={2021},
  volume={23},
  number={2},
  pages={1226-1252},
  doi={10.1109/COMST.2021.3063822}}

@book{ref3,
  title={Reinforcement learning: An introduction},
  author={Sutton, Richard S and Barto, Andrew G},
  year={2018},
  publisher={MIT press}
}

@article{ref4,
  title={Continuous control with deep reinforcement learning},
  author={Lillicrap, TP},
  journal={arXiv preprint arXiv:1509.02971},
  year={2015}
}

@ARTICLE{ce1,
  author={Zhang, Litianyi and She, Changyang and Ying, Kai and Li, Yonghui and Vucetic, Branka},
  journal={IEEE Transactions on Wireless Communications}, 
  title={Unsupervised Learning for Ultra-Reliable and Low-Latency Communications With Practical Channel Estimation}, 
  year={2024},
  volume={23},
  number={4},
  pages={3633-3647},
  doi={10.1109/TWC.2023.3309900}}

@ARTICLE{ra1,
  author={Ji, Zelin and Qin, Zhijin and Tao, Xiaoming and Han, Zhu},
  journal={IEEE Transactions on Wireless Communications}, 
  title={Resource Optimization for Semantic-Aware Networks With Task Offloading}, 
  year={2024},
  volume={23},
  number={9},
  pages={12284-12296},
  doi={10.1109/TWC.2024.3390407}}

@ARTICLE{dd1,
  author={Awan, Daniyal Amir and Cavalcante, Renato Luís Garrido and Yukawa, Masahiro and Stanczak, Slawomir},
  journal={IEEE Transactions on Signal Processing}, 
  title={Robust Online Multiuser Detection: A Hybrid Model-Data Driven Approach}, 
  year={2023},
  volume={71},
  number={},
  pages={2103-2117},
  doi={10.1109/TSP.2023.3282698}}

@article{CA_3GPP,
    author={{3GPP TS 38.101}-1},
    year = {version 17.5.0 Release 17, 2022},
    title = {{User equipment (UE) radio transmission and reception;
    Part 1: Range 1 Standalone}},
    journal = {Third Generation Partnership Project}}

@book{MINLP,
author = {Lee, Jon and Leyffer, Sven},
title = {Mixed Integer Nonlinear Programming},
year = {2011},
isbn = {1461419263},
publisher = {Springer Publishing Company, Incorporated}
}

@INPROCEEDINGS{MINLP_1,
  author={Kibria, Mirza Golam and Lagunas, Eva and Maturo, Nicola and Spano, Danilo and Al-Hraishawi, Hayder and Chatzinotas, Symeon},
  booktitle={2019 IEEE Global Communications Conference (GLOBECOM)}, 
  title={Carrier Aggregation in Multi-Beam High Throughput Satellite Systems}, 
  year={2019},
  volume={},
  number={},
  pages={1-6},
  doi={10.1109/GLOBECOM38437.2019.9014019}}

@article{makantasis2019deep,
  title={A deep reinforcement learning driving policy for autonomous road vehicles},
  author={K. Makantasis and M. Kontorinaki and I. Nikolos},
  journal={arXiv preprint arXiv:1905.09046},
  year={2019}
}

@article{edge2,
  title={Edge artificial intelligence for {6G}: Vision, enabling technologies, and applications},
  author={K. B. Letaief and Y. Shi and J. Lu and J. Lu},
  journal={IEEE J. Sel. Areas in Commun.},
  volume={40},
  number={1},
  pages={5-32},
  year={Jan. 2022},
  publisher={IEEE}
}

@article{edge3,
  title={{\color{black} Federated learning for {6G} communications: Challenges, methods, and future
directions}},
  author={{\color{black} Y. Liu, X. Yuan, Z. Xiong, J. Kang, X. Wang, and D. Niyato}},
  journal={{\color{black} China Commun.}},
  volume={{\color{black} 17}},
  number={{\color{black} 9}},
  pages={{\color{black} 105-118}},
  year={{\color{black} Sep. 2020}},
  publisher={}
}

@article{sinr_est,
  title={{\color{black} SNR estimation in maximum likelihood decoded spatial multiplexing}},
  author={{\color{black} O. Redlich, D. Ezri, and D. Wulich}},
  journal={{\color{black} arXiv preprint 
  arXiv:0909.1209v1 [cs.IT] 7 Sep 2009.}},
}

@article{sefati2022hybrid,
  title={A hybrid service selection and composition for cloud computing using the adaptive penalty function in genetic and artificial bee colony algorithm},
  author={{ S. S. Sefati and S. Halunga}},
  journal={Sensors},
  volume={22},
  number={13},
  pages={4873},
  year={2022},
  publisher={MDPI}
}

@article{xu2023edge,
  title={Edge learning for {B5G} networks with distributed signal processing: semantic communication, edge computing, and wireless sensing},
  author={{ W. Xu, Z. Yang, D. W. K. Ng, M. Levorato, Y. C. Eldar, and M. Debbah}},
  journal={IEEE J. Sel. Topics in Sig. Proc.},
  volume={17},
  number={1},
  pages={9-39},
  year={Jan. 2023},
  publisher={IEEE}
}

\end{document}